\documentclass[fleqn,usenatbib]{mnras}

\usepackage{newtxtext,newtxmath}

\usepackage[T1]{fontenc}

\DeclareRobustCommand{\VAN}[3]{#2}
\let\VANthebibliography\thebibliography
\def\thebibliography{\DeclareRobustCommand{\VAN}[3]{##3}\VANthebibliography}

\usepackage{graphicx}   
\usepackage{amsmath}    
\usepackage{xcolor}     

\title[Merger remnant models with tabulated EOS]{Models of binary neutron star remnants with tabulated equations of state}

\author[P. Iosif and N. Stergioulas]{
Panagiotis Iosif$^{1}$\thanks{E-mail: piosif@auth.gr}
and Nikolaos Stergioulas$^{1}$
\\
$^{1}$Department of Physics, Aristotle University of Thessaloniki, Thessaloniki 54124, Greece
}

\date{Accepted XXX. Received YYY; in original form ZZZ}

\pubyear{2021}

\begin{document}

\label{firstpage}
\pagerange{\pageref{firstpage}--\pageref{lastpage}}
\maketitle

\begin{abstract}
The emergence of novel differential rotation laws that can reproduce the rotational profile of binary neutron star merger remnants has opened the way for the construction of equilibrium models with properties that resemble those of remnants in numerical simulations. We construct models of merger remnants, using a recently introduced 4-parameter differential rotation law and three tabulated, zero-temperature equations of state. The models have angular momenta that are determined by empirical relations, constructed through numerical simulations. After a systematic exploration of the parameter space of merger remnant equilibrium sequences, which includes the determination of turning points along constant angular momentum sequences, we find that a particular rotation law can reproduce the threshold mass to prompt collapse to a black hole with a relative difference of only $\sim 1\%$ with respect to numerical simulations, in all cases considered. Furthermore, our results indicate a possible correlation between the compactness of equilibrium models of remnants at the threshold mass and the compactness of maximum-mass nonrotating models. Another key prediction of binary neutron star merger simulations is a relatively slowly rotating inner region, where the angular velocity $\Omega$ (as measured by an observer at infinity) is mostly due to the frame dragging angular velocity $\omega$. In our investigation of the parameter space of the adopted differential rotation law, we naturally find quasi-spherical (Type A) remnant models with this property. Our investigation clarifies the impact of the differential rotation law and of the equation of state on key properties of binary neutron star remnants and lays the groundwork for including thermal effects in future studies. 
\end{abstract}

\begin{keywords}
stars: neutron -- stars: rotation -- methods: numerical -- relativistic processes -- stars: kinematics and dynamics -- equation of state
\end{keywords}



\section{Introduction}
\label{sec:intro}

The detection of gravitational waves (GW) from the inspiral phase of the GW170817 binary neutron star (BNS) merger event \citep{2017PhRvL.119p1101A,Abbott_2017_multimess} combined with complementary information from its electromagnetic counterpart \citep{Abbott_2017_multimess, GW170817_EM_counterpart, Goldstein_etal_2017} have produced new constraints on the equation of state (EOS), see \citet{Bauswein_etal_2017,GW170817_radii_2018,2018PhRvL.121i1102D,2018PhRvL.120q2702F,2018PhRvL.120z1103M,2019PhRvX...9a1001A,2019PhRvD..99j3009M,2020NatAs...4..625C,2020PhRvD.101l3007L,2020Sci...370.1450D,Breschi_etal_2021} 
and references therein, as well as \citet{2020GReGr..52..109C,2021GReGr..53...27D} for recent reviews. A second likely BNS merger event, GW190425, was reported in \citet{2020ApJ...892L...3A} and more are expected in the next years \citep{2020LRR....23....3A}. Although the sensitivity of the LIGO and Virgo GW detectors was not sufficient to detect the post-merger phase in GW170817 \citep{2017PhRvL.119p1101A,2017ApJ...851L..16A}, such a detection is likely to be achieved in the future, either with upgraded or next-generation detectors, see e.g. \citet{2014PhRvD..90f2004C, 2016CQGra..33h5003C, 2017PhRvD..96l4035C, 2018PhRvL.120c1102B, 2018PhRvD..97b4049Y, 2019PhRvD..99d4014T, 2019PhRvD..99j2004M, 2019MNRAS.485..843O, 2019PhRvD.100d3005E, 2019PhRvD.100d4047T, 2019PhRvD.100j4029B, 2019CQGra..36v5002H, 2020PhRvD.102d3011E, 2020PASA...37...47A, 2020PhRvL.125z1101H, 2020arXiv201112414A, 2021PhRvD.103b2002G,2021CmPhy...4...27P}.

The outcome of a BNS merger is closely tied to the EOS and the total mass $M=m_1 + m_2$ of the system, where $m_1$ and $m_2$ are the binary's components masses, see \citet{Shibata_Hotokezaka_2019, Bernuzzi_2020,2020ARNPS..70...95R, Friedman_Stergioulas_2020} for recent reviews.
If $M<M_\text{thres}$ (the threshold mass for prompt black hole formation), the merger results in a hot, massive and differentially rotating, compact object with a substantial material disk around it.
If, at the same time, $M>M_\text{max,rot}$ (the maximum mass of a cold, uniformly rotating neutron star), the remnant will initially survive several tens of milliseconds (ms) due to the support of differential rotation and thermal pressure. However, the loss of angular momentum, due to GW emission, as well as dissipative effects (e.g. shear viscosity, magnetic breaking and effective viscosity due to the development of the magneto-rotational instability, see \citet{Shibata_Hotokezaka_2019,Ciolfi_2020,Sarin_Lasky_2021, Ruiz_etal_2021_review} for recent reviews and also \citealt{Radice_2020}) will ultimately lead to a delayed collapse to a black hole. A remnant with mass $M_\text{max,rot}>M>M_\text{max}$, where $M_\text{max}$ is the maximum mass of a nonrotating star, will be long-lived, spinning down on the timescale of electromagnetic emission, before reaching the axisymmetric instability limit.
Only if $M<M_\text{max}$, can a stable remnant form. 

During the first few milliseconds after its formation, the remnant is still highly non-axisymmetric, featuring also strong nonlinear oscillations and deformations away from equilibrium. Characteristic nonlinear features are combination tones and spiral deformations \citep{Stergioulas_etal_2011, Bauswein_Stergioulas_2015, Bauswein_etal_2016, Bauswein_Stergioulas_2019}. On a somewhat longer timescale, one can regard the remnant as a quasi-stationary, slowly drifting equilibrium state with the addition of linear oscillations. If one neglects some aspects of the state of the remnant (non-axisymmetric deformations, oscillations, time-dependence and thermal structure), one can construct simplified, stationary axisymmetric models of its structure.

 Merger remnants that survive for more than a few milliseconds before collapsing to black holes have been studied through numerical simulations \citep{Hotokezaka_etal_2011, Sekiguchi_etal_2011, Bauswein_Janka_2012, Bauswein_etal_2012, Hotokezaka_etal_2013,Bernuzzi_etal_2014,Dietrich_etal_2015,DePietri_etal_2016,Radice_etal_2018}. The remnant's structure, including its rotation profile was studied extensively in \citet{Kastaun_Galeazzi_2015,
 Paschalidis_etal_2015,
 Bauswein_Stergioulas_2015, Kastaun_etal_2016,
 East_etal_2016,
 Endrizzi_etal_2016, Kastaun_etal_2017, Ciolfi_etal_2017, Hanauske_etal_2017, Endrizzi_etal_2018, Kiuchi_etal_2018, Ciolfi_etal_2019, East_etal_2019, DePietri_etal_2020, Kastaun_Ohme_2021}. A common finding is that the remnant's rotation profile exhibits a maximum away from the center, which is in sharp disagreement with the differential rotation law by \citet{Komatsu_etal_1989}, hereafter KEH, which was widely used in the context of differentially rotating neutron stars (see \citealt{Ansorg_etal_2009, Espino_Paschalidis_2019, Espino_etal_2019} for different types of equilibrium models that can be constructed with the KEH rotation law). 

A 3-parameter piecewise extension of the KEH rotation law was used in \citet{Bauswein_Stergioulas_2017, Bozzola_etal_2018}, in order to allow the outer regions to rotate more slowly than the core, reaching high masses (typical of remnants) without encountering mass-shedding (see also \citet{Galeazzi_etal_2012,Uryu_etal_2016} for other rotation laws). Two different 3-parameter and 4-parameter rotation laws were proposed by \citet{Uryu_etal_2017}, who presented selected example equilibrium models. 

\citet{Zhou_etal_2019} constructed differentially rotating strange star models, using the 4-parameter rotation law of \citeauthor{Uryu_etal_2017}, whereas \citet{Passamonti_Andersson_2020} and \citet{Xie_etal_2020} studied the onset of the low $T/|W|$ instability \citep{Watts_etal_2005} in models constructed with the 3-parameter rotation law of \citeauthor{Uryu_etal_2017} In \cite{Camelio_etal_2020} models of stationary remnants of a BNS merger at $\sim 10-50$ ms after merger were presented, which were differentially rotating, hot, and baroclinic, using their own, 5-parameter rotation law. The models were constructed with the assumption of spatial conformal flatness (IWM-CFC approximation).

An important aspect of modeling post-merger remnants is to separate the effects of i) the differential rotation law, ii) the cold part of the EOS, and iii) thermal effects on the structure of the remnant and on its dynamical properties (stability to axisymmetric collapse and oscillations). To do so, we embarked on a systematic study of each of these three effects in separation from the other two. Our first step was to present equilibrium sequences of rotating relativistic stars, constructed with the 4-parameter rotation law of \citeauthor{Uryu_etal_2017}, 
adopting a cold, relativistic $N = 1$ polytropic EOS and choosing rotational parameters motivated by simulations of binary neutron star merger remnants \citep{Iosif_Stergioulas_2021}. A distinctive feature of the \citeauthor{Uryu_etal_2017} law is that it allows for the angular velocity to attain a maximum value $\Omega_\mathrm{max}$ away from center (as seen in simulations), which was not possible with the KEH law. We compared the sequences of equilibrium models to published sequences that used the KEH rotation law, revealing only a small influence of the choice of rotation law on the mass of the equilibrium models and a somewhat larger influence on their radius. Both Type A and Type C solutions (in correspondence to the classification of KEH-type models by \citealt{Ansorg_etal_2009}) were found. While our models were highly accurate solutions of the fully general relativistic structure equations, we also demonstrated that for models relevant to merger remnants the IWM-CFC approximation still maintains an acceptable accuracy.

Here, we take a second step in this program and construct sequences of models of post-merger remnants, using the 4-parameter rotation law of \citeauthor{Uryu_etal_2017} and different tabulated EOS. In \citet{Bauswein_Stergioulas_2017} the threshold mass to black hole collapse, as determined in simulations, was reproduced in a semi-analytic way, using equilibrium models obeying a piecewise extension of the KEH rotation law. Following the analysis detailed in that work, our sequences are constructed using an empirical relation between angular momentum at merger and the radius and compactness of the progenitor stars (assuming equal masses). Again, we find both Type A and Type C sequences. For a particular combination of rotation-law parameters, we find that the sequence of merger remnants terminates very near the threshold mass to collapse (as obtained by numerical simulations) for all three representative EOS that we used in this study. For somewhat different combinations of rotation-law parameters, we find sequences of merger remnants with realistic rotation profiles, for which the angular velocity in the core is close to the angular velocity of frame dragging, reproducing a characteristic feature seen in binary neutron star merger simulations. The next step in this program will be the inclusion of thermal effects, which we are planning to present in the future.

The structure of the paper is as follows: in Section~\ref{sec:method} we discuss the theoretical framework and numerical methods. In Section~\ref{sec:results} we present the main results. In Section~\ref{sec:discussion} we discuss our findings.

Throughout the text we set $c = G =1 $ in equations (except for equations where units are explicitly included) and choose appropriate physical units to report numerical results. We also denote with $R_{X}$ the radius of nonrotating neutron stars with gravitational mass $X M_\odot$. E.g. $R_{1.4}$ stands for the radius of a $1.4M_\odot$ star.

\section{Theoretical framework and methods}
\label{sec:method}

\subsection{Spacetime metric and matter description}

Our solutions are fully general relativistic, axisymmetric, and asymptotically flat and we adopt the following form of the line element, in quasi-isotropic coordinates:
\begin{equation}
ds^2 = -e^{\gamma + \rho} dt^2 + e^{\gamma - \rho} r^2 \sin^2 
\theta (d\phi - \omega dt)^2 + e^{2\mu} (dr^2 + r^2 d\theta^2) \label{eq:stationary_axisym_metric} \, , 
\end{equation}
where $ \gamma $, $ \rho $, $ \omega $ and $ \mu $ are metric functions that depend only on the coordinates $r$ and $\theta$. The metric function $\omega$ is the angular velocity of a zero-angular-momentum-observer (ZAMO) and describes the relativistic dragging of inertial frames due to rotation.

The matter is described as a perfect fluid with a stress-energy tensor of the form
\begin{equation}
T^{\alpha \beta} = (\epsilon + P) u^\alpha u^\beta + P g^{\alpha \beta} \label{eq:stress_energy_tensor} \, , 
\end{equation}
where $\epsilon$ is the energy density, $P$ is the pressure, $u^\alpha$ is the four-velocity of the fluid and $g_{\alpha\beta}$ is the metric tensor. Further details on the basic equations and concepts can be found in \citet{Friedman_Stergioulas_2013, Paschalidis_Stergioulas_2017}.

\subsection{Rotation law}

The 1-parameter rotation law of \citet{Komatsu_etal_1989} (suitable for rotating proto-neutron stars formed after core-collapse) is
\begin{equation}
F(\Omega) = A^2 (\Omega_c - \Omega),  \label{eq:KEH_rotlaw}
\end{equation}
where $F \equiv u^t u_{\phi} $ is the gravitationally redshifted angular momentum per unit rest mass and enthalpy and $ \Omega_c $ is the angular velocity at the center of the star. In (\ref{eq:KEH_rotlaw}) the parameter $ A $ determines the length scale over which the angular velocity $\Omega$ changes. 

In contrast, the 4-parameter rotation law of \cite{Uryu_etal_2017} reads
\begin{equation}
\Omega = \Omega_c \frac{1 + \left( \dfrac{F}{B^2 \Omega_c} \right)^p}{1 + \left( \dfrac{F}{A^2 \Omega_c} \right)^{q+p}} \, , \label{eq:Uryuetal_rotlaw8}
\end{equation}
(hereafter Uryu+ law). As in \cite{Iosif_Stergioulas_2021}, here we also fix two of the four parameters to the specific values $p=1$ and $q=3$. On one hand, setting integer values for $p$ and $q$ allows us to obtain an algebraic expression for the first integral of the hydrostationary equilibrium equation. On the other hand, fixing two of the four parameters allows us to investigate in detail a more manageable two-parameter space. The choice of $q=3$ is motivated by the fact that for this value the rotation law tends to the Keplerian law at a large distance from the center, in the Newtonian limit.

As in \citet{Uryu_etal_2017} and \citet{Zhou_etal_2019}, instead of investigating different values for the parameters $A$ and $B$, we choose to work with the parameters 
\begin{equation}
\lambda_1 \equiv \frac{\Omega_\text{max}}{\Omega_c}  \label{eq:lambda1},
\end{equation}
\begin{equation}
\lambda_2 \equiv \frac{\Omega_e}{\Omega_c},   \label{eq:lambda2}
\end{equation}
where $\Omega_e$ is the angular velocity at the equator. This has the advantage of choosing parameter values that can be directly set by inspecting the ratios of $\Omega_\text{max}$ and $\Omega_e$ with respect to $\Omega_c$, as obtained in numerical simulations of BNS remnants. Note that all angular velocities are defined with respect to an observer at infinity.

\subsection{Numerical scheme}

In order to build our equilibrium configurations we use an extended version of the \textsc{rns} code \citep{Stergioulas_Friedman_1995}, which implements the iterative \citet{Komatsu_etal_1989} scheme with improvements by \citet{Cook_etal_1992}. The initial \textsc{rns} code was updated to tackle differential rotation in \citet{Stergioulas_etal_2004} and further extended for the 3-parameter, piecewise KEH law in \citet{Bauswein_Stergioulas_2017}. In \citet{Iosif_Stergioulas_2021}, we extended \textsc{rns} with the implementation of the 4-parameter Uryu+ rotation law of Eq. \eqref{eq:Uryuetal_rotlaw8}. This allowed for the construction of models with realistic rotation profiles for BNS merger remnants that have off-center maxima in the angular velocity profile. The solutions were shown to be highly accurate and converging at second order with an increasing number of grid points. A standard resolution was chosen that yields solutions with 3-dimensional virial theorem index (GRV3) of order $10^{-5}$. In the present study, we employ a grid size of $\text{SDIV} \times \text{MDIV}=401 \times 201$ (compactified radial times angular) for all models. We refer to \citet{Iosif_Stergioulas_2021} for further details on the numerical scheme.

\subsection{Equations of state}
\label{sec:eos}

\begin{figure}
    \includegraphics[width=0.93\columnwidth]{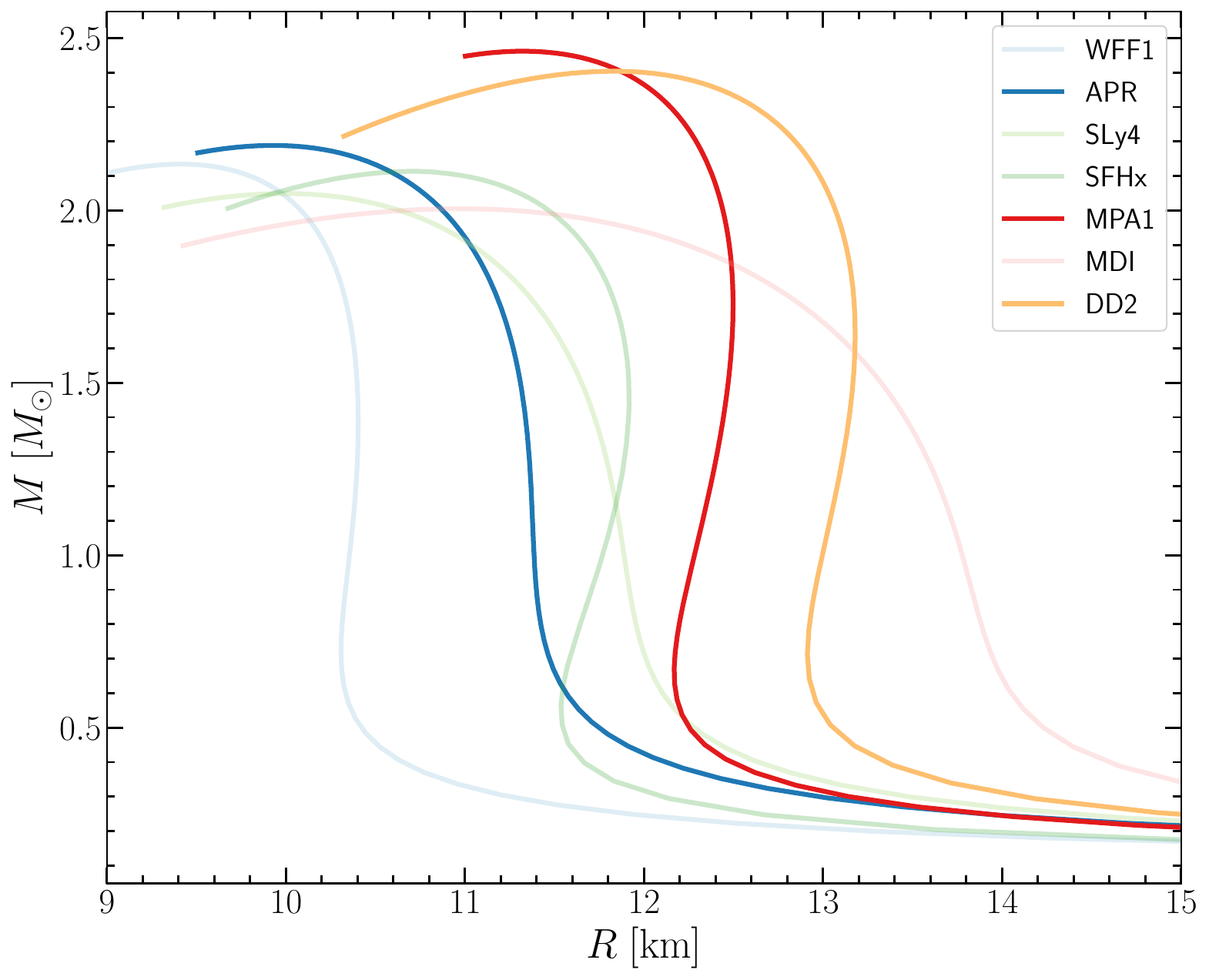}
    \caption{Gravitational mass $M$ vs. circumferential radius $R$ of nonrotating models for different EOS.}
    \label{fig:M_R_plot}
\end{figure}

Considerable uncertainty still exists in the determination of the EOS of dense nuclear matter. Figure~\ref{fig:M_R_plot} shows the gravitational mass $M$ vs. the circumferential radius $R$ for nonrotating models constructed with several different hadronic EOS that cover the large uncertainty range that existed before the historic detection of gravitational waves from the source GW170817. The initial analysis of GW170817 resulted in a constraint on neutron star radii $R=11.9_{-1.4}^{+1.4}$ km
\citep{GW170817_radii_2018} for both stars involved in the merger, at the 90\% credible level. In the meantime, a large number of studies presented multi-messenger constraints on the neutron star radius, taking into account observations in the electromagnetic spectrum as well as nuclear-theory computations using chiral effective field theory. 
Recent studies provide predictions for the radius of a $1.4 M_{\odot}$ neutron star with an uncertainty range of 
$R_{1.4}=12.32_{-1.47}^{+1.09}$ km (90\% credible level) \citep{2020PhRvD.101l3007L}, 
$R_{1.4}=11.0_{-0.6}^{+0.9}$ km
 (90\% credible level) \citep{2020NatAs...4..625C},
 $R_{1.4}=11.75_{-0.81}^{+0.86}$ km
(90\% credible level) \citep{2020Sci...370.1450D}, 
$R_{1.4}=12.2_{-0.5}^{+0.5}$ km (1$\sigma$ level) \citep{Breschi_etal_2021} and $R_{1.4}=11.94_{-0.87}^{+0.76}$ km (90\% credible level) \citep{Pang_etal_2021}. Furthermore, the Neutron Star Interior Composition Explorer (NICER, \citealt{NICER_instr_ref}) measurements of PSR J0740+6620 have yielded radius estimates of $R_{1.4}=12.45_{-0.65}^{+0.65}$ km (1$\sigma$ level) \citep{Miller_etal_2021_NICER}, $R_{1.4}=12.33_{-0.81}^{+0.76}$ km (95\% credible level) and $R_{1.4}=12.18_{-0.79}^{+0.56}$ km (95\% credible level) for different high-density EOS parameterizations \citep{Riley_etal_2021_NICER, Raaijmakers_etal_2021_NICER}.

Taking into account the above range of radii uncertainties, we selected three tabulated, zero-temperature, hadronic EOS that correspond to typical neutron star radii between 11 and 13 km. These are APR \citep{APR_1998, BPS_1971, Douchin_Haensel_2001}, DD2 \citep{HEMPEL2010210_DD2, Typel_etal_2010_DD2, MOLLER1997131_DD2} and MPA1 \citep{MPA1_1987}, shown with darker colors in Figure~\ref{fig:M_R_plot}. All three EOS satisfy the current constraints for the maximum neutron star mass \citep{Demorest_etal_2010, Antoniadis_etal_2013, Cromartie_etal_2020} as well as the minimum radius constraint, when combining causality and GW170817, of $R_{1.6} \geq 10.68$ km \citep{Bauswein_etal_2017}. EOS with strong phase transitions are not included in the present study.

\subsection{Construction of merger remnant sequences}
\label{sec:remn_seq}

Our aim is to construct sequences of equilibrium models that mimic characteristic properties of post-merger remnants and reach the threshold mass to prompt collapse.

In \cite{Bauswein_Stergioulas_2017} an empirical, EOS-insensitive relation that connects the angular momentum at merger $J_{\rm merger}$ to the total mass of a binary neutron star system $M_{\rm tot}$ was constructed: 
\begin{equation}
\frac{cJ_{\text {merger }}}{G M_\odot^2} \simeq a \frac{M_{\text {tot }}}{M_\odot}- \left( b + \frac{R_{1.5}-R_{1.5}^{\mathrm{DD} 2}}{10 \mathrm{~km}}\right),
\label{BS17-Jempirical}
\end{equation}
where $a = 4.041$ and $b = 4.658$. An alternative relation was constructed by \citet{Lucca_etal_2021}, who expressed $J_{\rm merger}$ as a function of the radius $R_{\rm NS}$ and compactness $C_{\rm NS} = GM_{\rm NS}/c^2 R_{\rm NS}$ of a nonrotating neutron star with mass $M_{\rm NS}=M_{\rm tot}/2$
\begin{equation}
    \frac{GJ_{\text{merger}}}{c^3 R^2_{\text{NS}}} = a_1 C_{\text{NS}} + a_2 \, ,
    \label{eq:LSGF20eqn1}
\end{equation}
where $a_1 = 0.8765 \pm 0.0051$ and $a_2 = -\left(5.209 \pm 0.077 \right) \times 10^{-2}$ ($1\sigma$ credible level). We find that the two empirical relations are in good agreement with each other (see Appendix~\ref{appendix} for a comparison of the two relations' predictions). Note that both \eqref{BS17-Jempirical} and \eqref{eq:LSGF20eqn1} have been derived by analyzing different sets of numerical simulations assuming equal mass binaries.

We construct sequences of models of merger remnants with different combinations of $\{\lambda_1, \lambda_2\}$ and with remnant masses of $M_\text{tot} = \{2.2, 2.3, 2.4, 2.5, \dots \}M_\odot$. We continue to larger values with a step of $0.1 M_\odot$ up to the maximum possible $M_\text{tot}$ for which we can construct an equilibrium sequence for the particular rotation law and EOS. For each value of $M_\text{tot}$, we compute the corresponding $R_{\rm NS}$ and $M_{\rm NS}$ of a nonrotating star. From (\ref{eq:LSGF20eqn1}) we compute the corresponding angular momentum of the remnant. The detailed properties of the equilibrium models of merger remnant sequences are reported in Tables~\ref{tab:APR_remnants_physical_quantities}, \ref{tab:DD2_remnants_physical_quantities} and \ref{tab:MPA1_remnants_physical_quantities} and they are discussed in detail in Section \ref{sec:results}.

\subsection{Constant angular momentum sequences and turning points}

For each merger remnant model, we also construct the corresponding sequence of equilibrium models with the same rotation law and fixed $J_{\rm merger}$. In the case of uniform rotation, the line connecting the turning points of each constant angular momentum sequence, i.e. the points where
\begin{equation}
	\frac{d M}{d \epsilon_\text{max}} \biggm|_{J=\text{const.}} = 0,
	\label{eq:turn_points}
\end{equation}
(where $\epsilon_\text{max}$ is the maximum value of the energy density in the star) defines the secular instability limit to axisymmetric perturbations \citep{Friedman_etal_1988}. Note that the turning point criterion is only a sufficient condition for secular instability of uniformly rotating stars.

The dynamical instability to prompt collapse to a black hole is detected through numerical simulations or by finding the models for which the frequency of the fundamental quasi-radial mode vanishes. For uniform rotation, the dynamical instability limit for prompt collapse is very close to the secular instability limit and it occurs slightly earlier (see \citet{Friedman_Stergioulas_2013} for a detailed discussion). In the case of differential rotation, \citet{Weih_etal_2018} demonstrated (through numerical simulations) that for particular choices of the KEH law the dynamical instability also sets in very close to the secular instability limit (the central density of dynamically unstable models was at most several percent smaller than the central density at the turning points).

Given the above findings for uniformly rotating as well as differentially rotating models with the KEH law and since we don't yet have dynamical or perturbative calculations for models constructed with the Uryu+ law, we adopt the line connecting the turning points of constant angular momentum sequences as a reasonably \textit{approximate} indicator of dynamical instability.

\section{Main Results}
\label{sec:results}

\subsection{Sequences of Type C models and threshold mass to prompt collapse}

Initially, we focus on two combinations of $\{\lambda_1, \lambda_2\}$ that were shown \citep{Iosif_Stergioulas_2021} to yield Type C solutions according to the classification of \citet{Ansorg_etal_2009}. These are sequences along which the models transition continuously from quasi-spherical to quasi-toroidal configurations, as the polar to equatorial axis ratio $r_p/r_e$ is decreased. We find that setting parameters $\{\lambda_1, \lambda_2\}$ equal to the pairs of values $\{2.0, 0.5\}$ and $\{1.5, 0.5\}$, continues to result in Type C solutions for the tabulated EOS we examined, as was the case for the $N=1$ polytropes in \citet{Iosif_Stergioulas_2021}.

\begin{figure}
        \includegraphics[width=0.93\columnwidth]{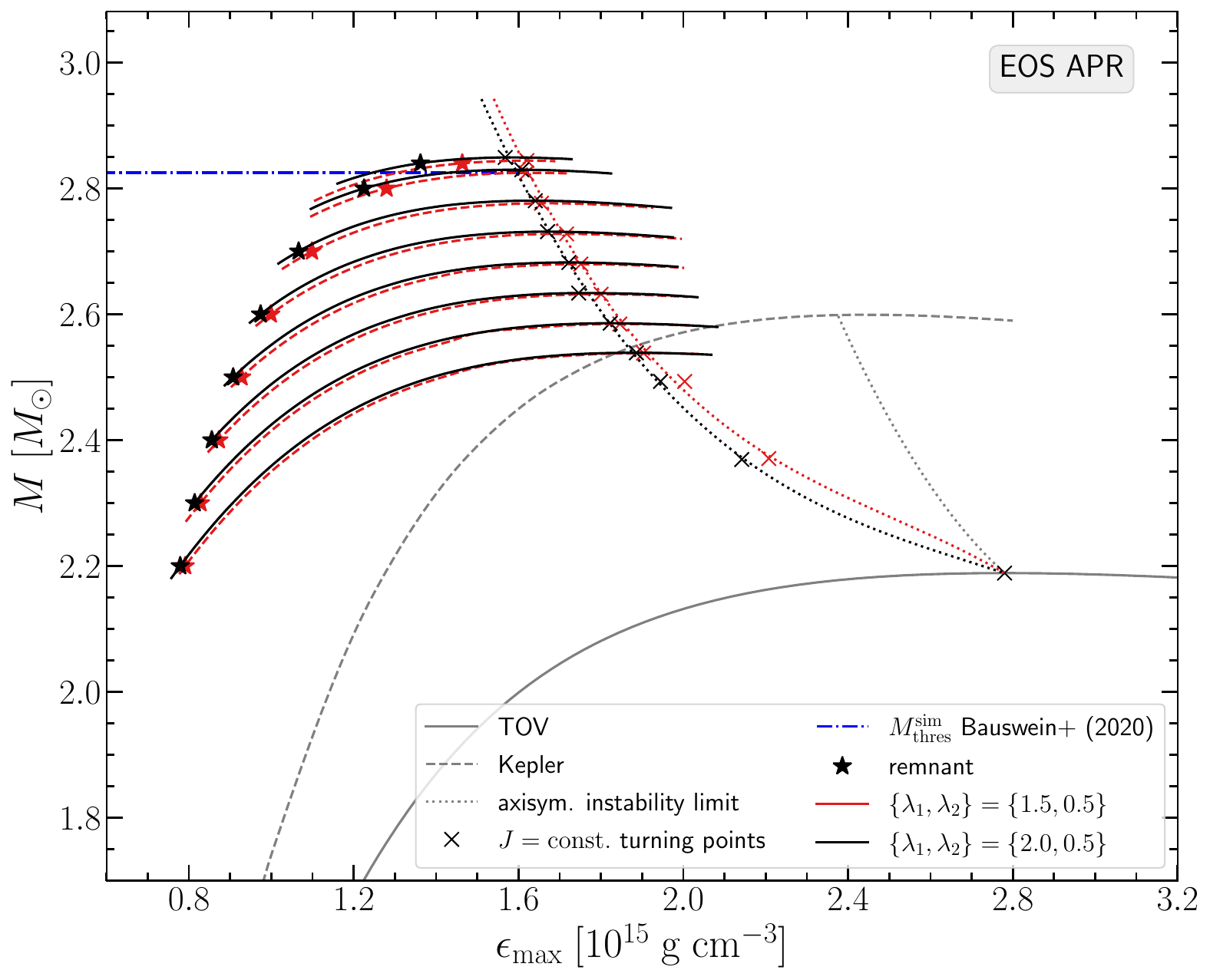}
    \caption{Gravitational mass $M$ vs. maximum energy density $\epsilon_\mathrm{max}$ for the APR EOS. Two choices of rotation law parameters yielding Type C solutions are shown. The nonrotating (TOV) sequence (grey solid line), the mass-shedding (Kepler) limit for uniform rotation (grey dashed line) and the axisymmetric instability limit for uniform rotation (grey dotted line) are shown as reference.}
    \label{fig:APR_Mass_emax_TypeC}
\end{figure}

\begin{figure}
        \includegraphics[width=0.93\columnwidth]{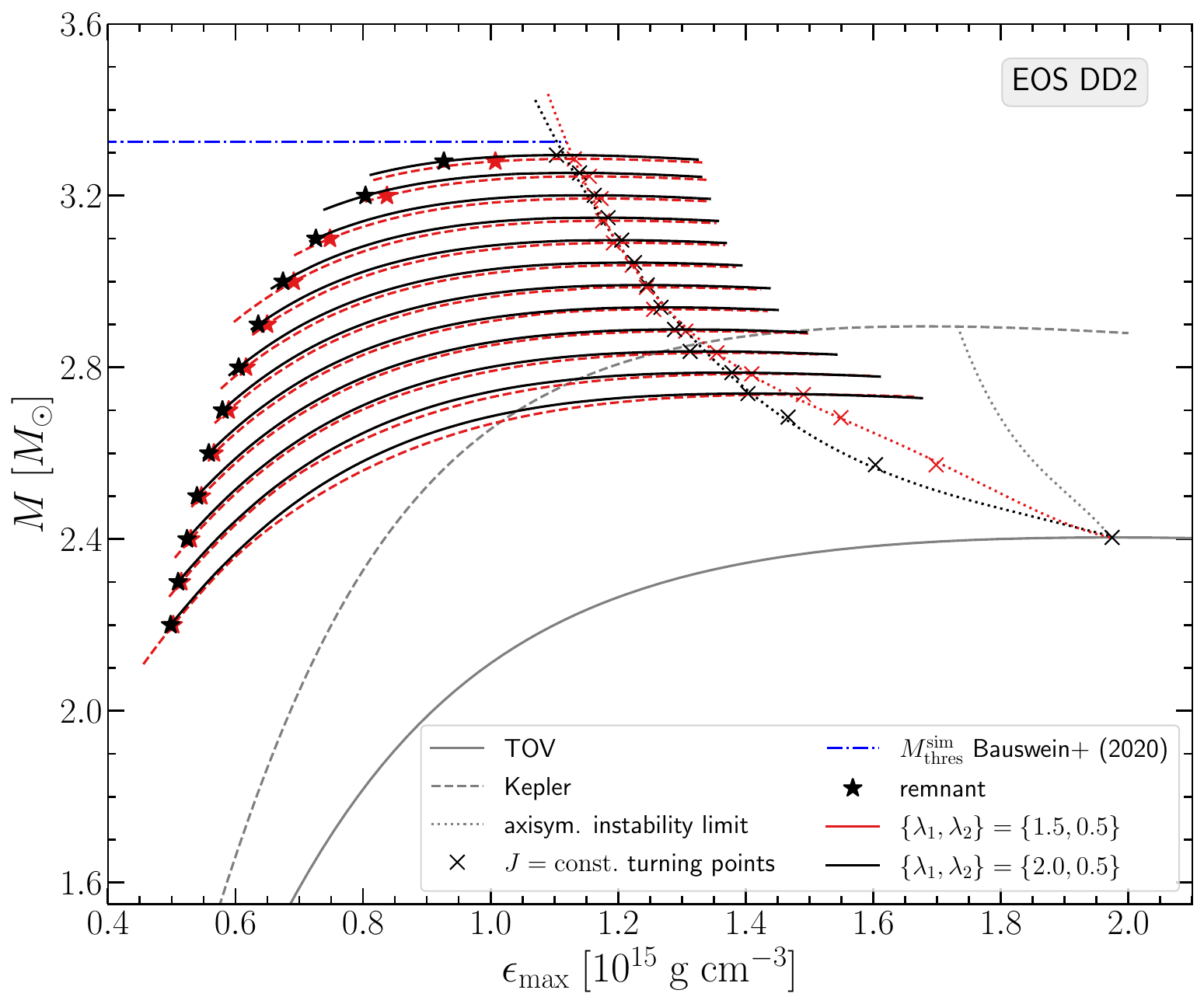}
    \caption{Same as Figure~\ref{fig:APR_Mass_emax_TypeC} for the DD2 EOS.}
    \label{fig:DD2_Mass_emax_TypeC}
\end{figure}

\begin{figure}
        \includegraphics[width=0.93\columnwidth]{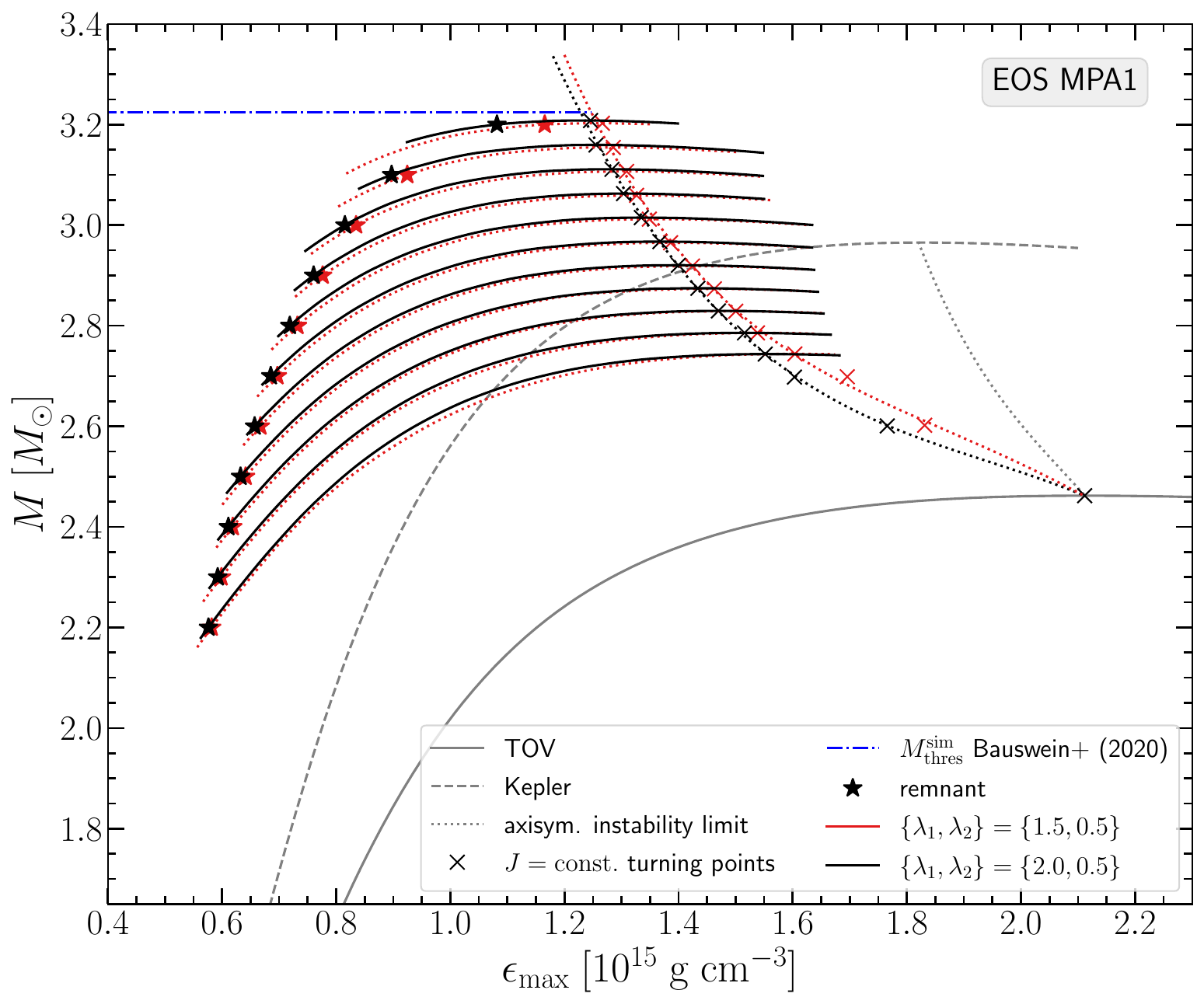}
    \caption{Same as Figure~\ref{fig:APR_Mass_emax_TypeC} for the MPA1 EOS.}
    \label{fig:MPA1_Mass_emax_TypeC}
\end{figure}

As can be seen in Figure~\ref{fig:APR_Mass_emax_TypeC} for the APR EOS, the $J$-constant curves for both of these rotation laws are overlapping and the turning points we locate are quite close as well. This is to be expected, as these two particular rotation laws correspond to similar $\Omega(r)$ rotational profiles \citep[Figure 8]{Iosif_Stergioulas_2021}. Asterisks in black and red denote the remnant models found for the respective $J_\text{merger}$ values predicted by the empirical relation \eqref{eq:LSGF20eqn1}. The gravitational masses of these configurations for the APR EOS start at $2.2 M_\odot$ for the least massive model and end at $2.84 M_\odot$ for the most massive model (Table~\ref{tab:APR_remnants_physical_quantities}). The picture is similar for the other two EOS we consider, DD2 (see Figure~\ref{fig:DD2_Mass_emax_TypeC} and Table~\ref{tab:DD2_remnants_physical_quantities}) and MPA1 (see Figure~\ref{fig:MPA1_Mass_emax_TypeC} and Table~\ref{tab:MPA1_remnants_physical_quantities}). The only difference is that higher masses (as well as higher angular momentum values) are reached for the most massive remnant models at $3.28 M_\odot$ for DD2 and at $3.2 M_\odot$ for MPA1.

\begin{figure}
        \includegraphics[width=0.93\columnwidth]{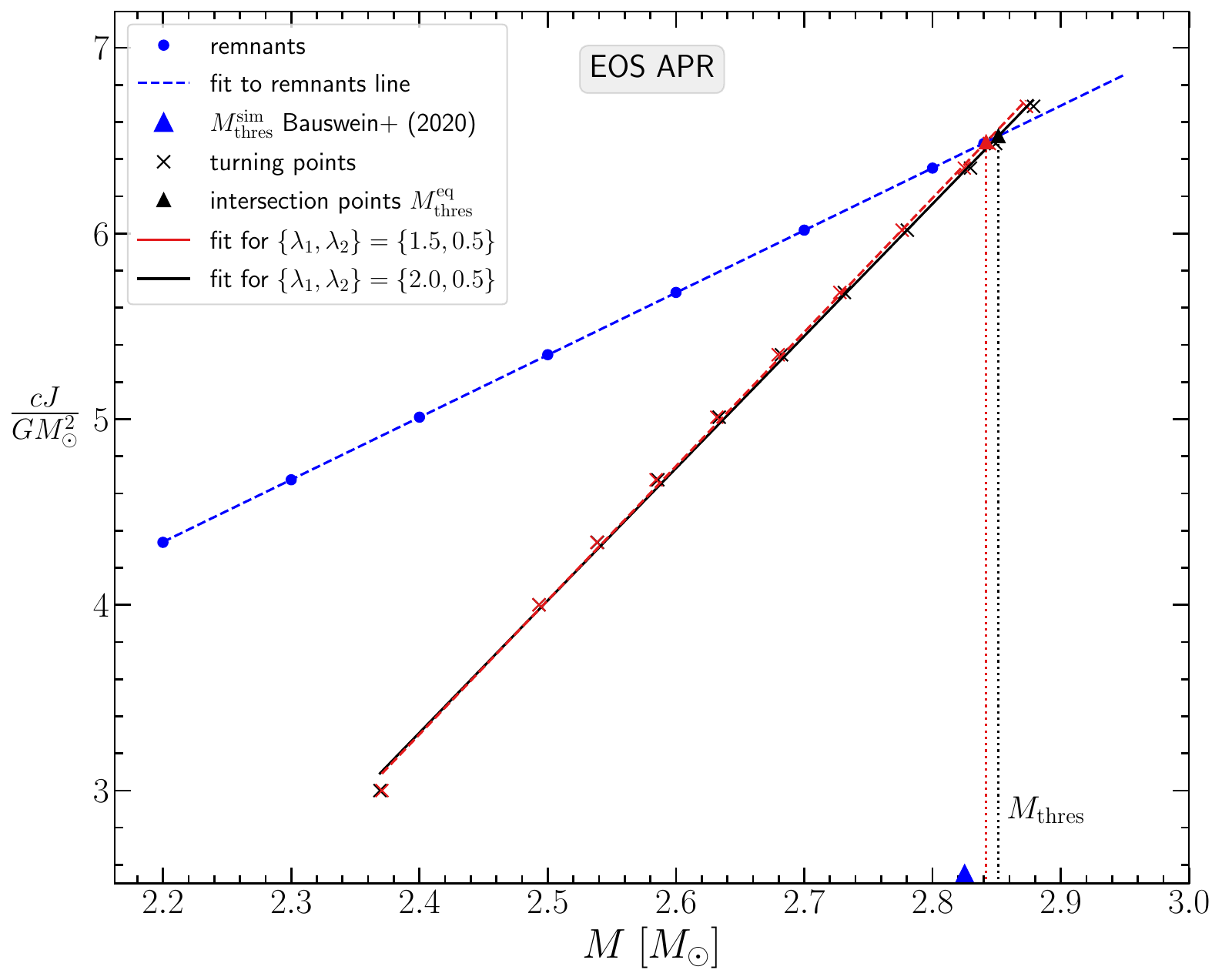}
    \caption{Angular momentum $J$ vs. gravitational mass $M$ for the APR EOS. The intersection of the remnants' and the turning points' fitted lines determines the threshold mass to collapse calculated from our equilibrium models, $M_\mathrm{thres}^\mathrm{eq}$.}
    \label{fig:Mthres_APR}
\end{figure}

\begin{figure}
        \includegraphics[width=0.93\columnwidth]{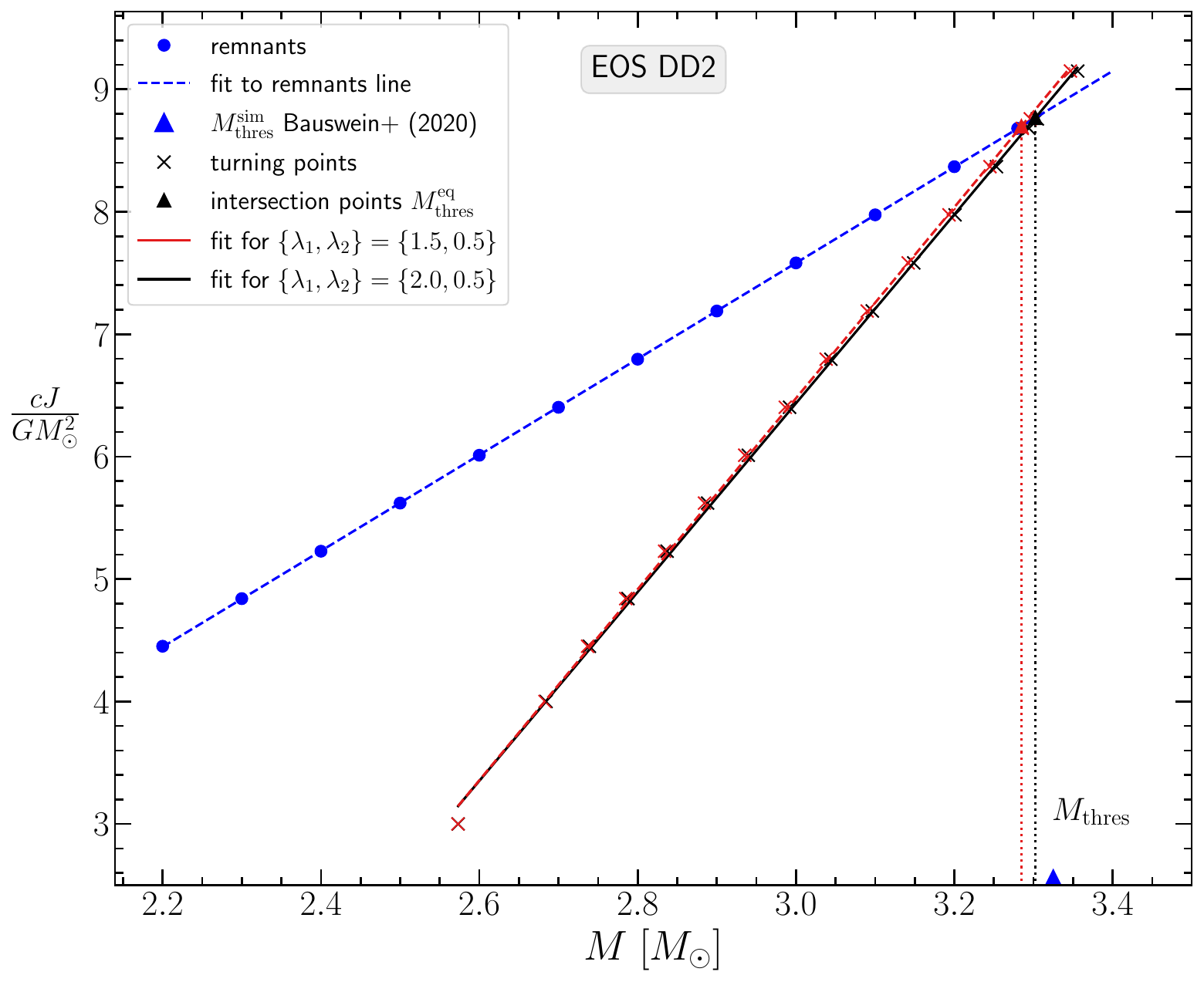}
    \caption{Same as Figure~\ref{fig:Mthres_APR} for the DD2 EOS.}
    \label{fig:Mthres_DD2}
\end{figure}

\begin{figure}
        \includegraphics[width=0.93\columnwidth]{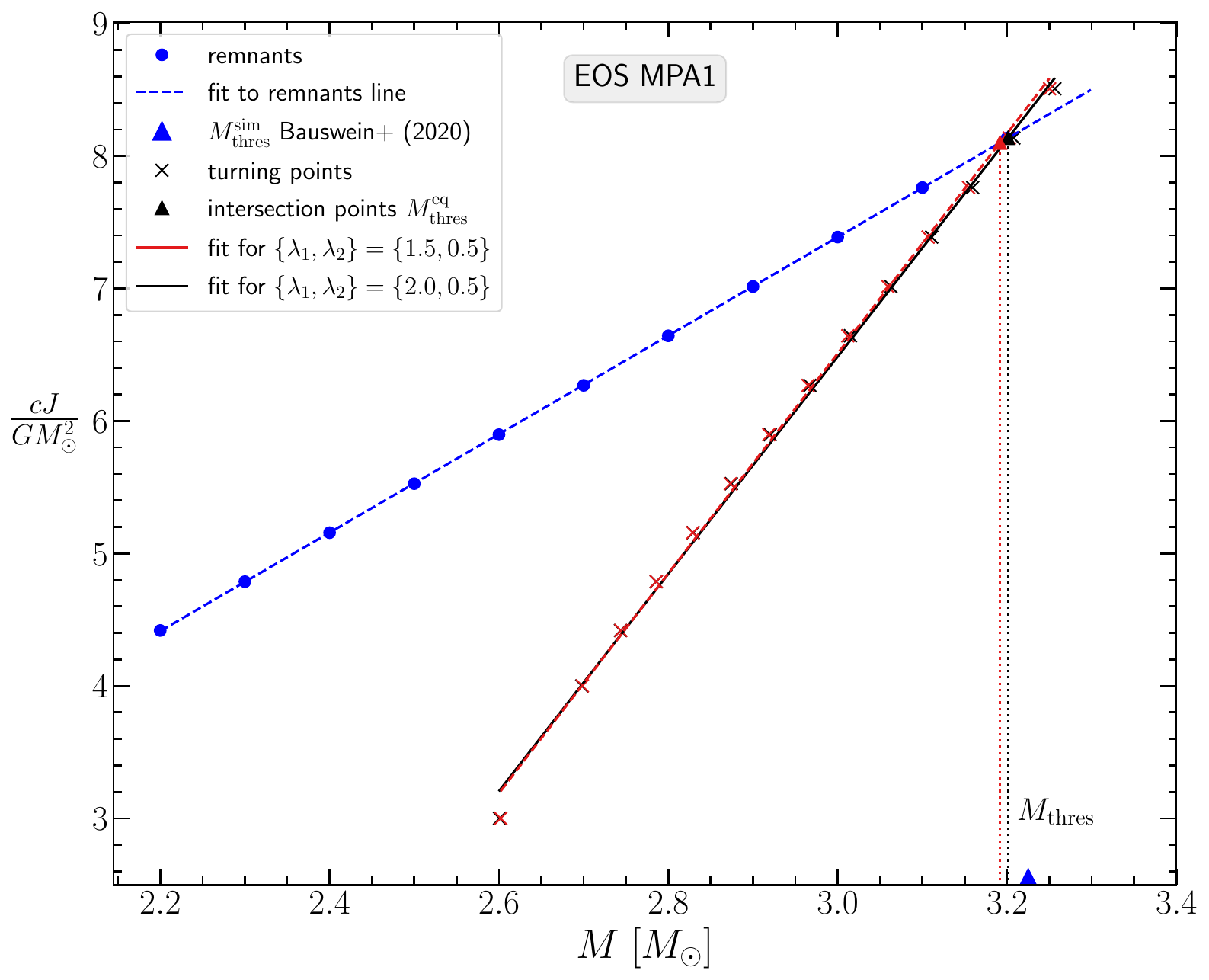}
    \caption{Same as Figure~\ref{fig:Mthres_APR} for the MPA1 EOS.}
    \label{fig:Mthres_MPA1}
\end{figure}

Qualitatively, the remnant sequence rises to larger masses at an almost constant (steep) slope that after a point abruptly diminishes, allowing the remnant sequence to intersect with the turning point line. As in \cite{Bauswein_Stergioulas_2017}, we find that this intersection is related to the threshold mass for prompt collapse, $M_\text{thres}^\text{sim}$ (as determined by numerical simulations in \citealt{Bauswein_etal_2020}). This mass is indicated by a blue horizontal dash-dotted line in Figures~\ref{fig:APR_Mass_emax_TypeC}, \ref{fig:DD2_Mass_emax_TypeC}, \ref{fig:MPA1_Mass_emax_TypeC}, \ref{fig:APR_Mass_emax_TypeA}, \ref{fig:DD2_Mass_emax_TypeA} and \ref{fig:MPA1_Mass_emax_TypeA}. The intersection can also be determined in Figures~\ref{fig:Mthres_APR}, \ref{fig:Mthres_DD2} and \ref{fig:Mthres_MPA1}, which show the angular momentum $J$ as a function of the gravitational mass $M$ of the remnant sequence (blue line) and the line connecting the turning points of $J$-constant sequences for $\{\lambda_1, \lambda_2\} = \{2.0, 0.5\}$ (black line) and $\{\lambda_1, \lambda_2\} = \{1.5, 0.5\}$ (red line), for the three EOS. Note that the data for the remnant sequences in Figures~\ref{fig:Mthres_APR}, \ref{fig:Mthres_DD2} and \ref{fig:Mthres_MPA1}, as obtained from the empirical relation (\ref{eq:LSGF20eqn1}), form a straight line, in agreement with the form of the original empirical relation 
(\ref{BS17-Jempirical}).

To locate the intersection point of the remnant sequence with the turning points line, we calculate linear fits for the remnant models and the turning points. These linear relations have the form
\begin{equation}
J = a M - b.
\label{eq:linear_fits_RL_TP}
\end{equation}
For the remnant sequences, this also facilitates a direct comparison between the $J_\text{merger}$ empirical relations (\ref{BS17-Jempirical}) and (\ref{eq:LSGF20eqn1}). In essence, \eqref{eq:linear_fits_RL_TP}
casts (\ref{eq:LSGF20eqn1}) in the form of (\ref{BS17-Jempirical}).
The coefficients $a$ and $b$ of these linear fits are reported in Table~\ref{tab:linear_fits_coeffs_RL_TP}, together with their respective errors $\delta a$ and $\delta b$.

The values of $M_\text{thres}^\text{eq}$ determined from the intersections of the linear fits for the remnant sequence and each turning point sequence in Figure~\ref{fig:Mthres_APR}, \ref{fig:Mthres_DD2} and \ref{fig:Mthres_MPA1} are in excellent agreement with published values for the respective quantity $M_\text{thres}^\text{sim}$ from numerical simulations \citep[Supplementary Material]{Bauswein_etal_2020}\footnote{We note that in the simulations of \citet{Bauswein_etal_2020} the EOS table used for DD2 provided temperature dependence, whereas the APR and MPA1 were used in a hybrid form (zero-temperature EOS supplemented by an approximate thermal part).}. For the case $\{\lambda_1, \lambda_2\} = \{2.0, 0.5\}$ we find $M_\text{thres}^\text{eq} = \{2.851, 3.302, 3.201\}$, which is to be compared to $M_\text{thres}^\text{sim}= \{2.825, 3.325, 3.225\}$, for the EOS APR, DD2 and MPA1 correspondingly (see Table~\ref{tab:Mthres_eq_comparison}). We also report the relative difference between $M_\text{thres}^\text{eq}$ and $M_\text{thres}^\text{sim}$ for each EOS and rotation law as
\begin{equation}
\delta M_\text{thres} = \biggm| \frac{M_\text{thres}^\text{sim} - M_\text{thres}^\text{eq}}{M_\text{thres}^\text{sim}} \biggm|.
\label{eq:delta_rel_error}
\end{equation}
The agreement with the threshold mass values from numerical simulations is at the 1\% level, which is quite remarkable, considering that we only use \textit{zero-temperature} EOS in constructing our models. A possible explanation is that in the case of prompt collapse, the kinetic energy of the collision does not have enough time to be transformed into thermal energy, through shock heating. Therefore, it would be the cold part of the EOS that mainly determines the properties of prompt collapse. We elaborate more on this topic on Section~\ref{sec:discussion}. 

Also, the agreement at the level of 1\% between our current models and numerical simulations is a significant improvement with respect to the agreement at the level of 3\% - 7\% in \citet{Bauswein_Stergioulas_2017}, where a particular 3-parameter piecewise KEH-type rotation law was used. This is achieved without a direct reconstruction of particular merger remnants (i.e. without trying to match all properties of a remnant, as extracted from a numerical simulation), but using equilibrium models for particular $\{\lambda_1, \lambda_2\}$ values, in conjunction with the empirical relation for the angular momentum at merger. 

The above findings indicate that the Uryu+ law is not simply qualitatively appropriate for merger remnants, in the sense that it allows for the maximum angular velocity to appear off-center, but it can also yield precise numerical results, at least for certain properties of the remnants.

\begin{table}
        \centering
        \caption{Coefficients of the linear fits $J = a M - b$ and their respective errors, that determine the intersection of the remnants sequence and the turning points line for each EOS (Figures~\ref{fig:Mthres_APR}, \ref{fig:Mthres_DD2} and \ref{fig:Mthres_MPA1}). The abbreviations RL, TP20 and TP15 stand for "remnant line" and "turning point line" with $\{\lambda_1, \lambda_2\}=\{2.0, 0.5\}$ and $\{\lambda_1, \lambda_2\}=\{1.5, 0.5\}$ respectively. The errors in the coefficients of the linear fits, $\delta a$ and $\delta b$, are calculated with the standard formulas of simple linear regression and correspond to uncertainties at the $1\sigma$ level.}
        \label{tab:linear_fits_coeffs_RL_TP}
        \begin{tabular}{cccccc}
                \hline
                 EOS & line & $a$ & b & $\delta a$ & $\delta b$\\
                \hline
                APR & RL & 3.3562 & 3.0453 & 0.0027 & 0.0069\\
				& TP20 & 7.1189 & 13.7739 & 0.0823 & 0.2220\\
				& TP15 & 7.2199 & 14.0254 & 0.0827 & 0.2209\\
                DD2 & RL & 3.9190 & 4.1758 & 0.0022 & 0.0062\\
				& TP20 & 7.7087 & 16.6908 & 0.0523 & 0.1580\\
				& TP15 & 7.8003 & 16.9247 & 0.0530 & 0.1597\\
				MPA1 & RL & 3.7183 & 3.7673 & 0.0029 & 0.0079\\
				& TP20 & 8.2014 & 18.1185 & 0.1116 & 0.3294\\
				& TP15 & 8.3019 & 18.3962 & 0.1103 & 0.3253\\ 			             
                \hline        
        \end{tabular}
\end{table}

\begin{table}
        \centering
        \caption{Comparison of the threshold mass deduced from equilibrium models $M_\text{thres}^\text{eq}$ with the respective quantity $M_\text{thres}^\text{sim}$ from the numerical simulations of \citet{Bauswein_etal_2020}. The angular momentum value we find at the intersection point, $J_\text{thres}^\text{eq}$, is also reported. The last column lists the absolute value of the relative difference $\delta M_\text{thres}$ calculated via \eqref{eq:delta_rel_error}.}
        \label{tab:Mthres_eq_comparison}
        \begin{tabular}{ccccc}
                \hline
                 EOS & $M_\text{thres}^\text{eq}$ & $J_\text{thres}^\text{eq}$  & $M_\text{thres}^\text{sim}$ & $\delta M_\text{thres}$\\
                 \{$\lambda_1$, $\lambda_2$\} & $[M_\odot]$ & $[\frac{G M_{\odot}^2} {c}]$ & $[M_\odot]$ & [\%] \\
                \hline
                APR & & & 2.825\\
                \{2.0, 0.5\} & 2.851 & 6.524 & & 0.92 \\
				\{1.5, 0.5\} & 2.842 & 6.492 & & 0.60\\
				DD2 & & & 3.325\\
                \{2.0, 0.5\} & 3.302 & 8.766 & & 0.69 \\
				\{1.5, 0.5\} & 3.285 & 8.697 & & 1.20\\
				MPA1 & & & 3.225\\
                \{2.0, 0.5\} & 3.201 & 8.136 & & 0.74\\
				\{1.5, 0.5\} & 3.192 & 8.100 & & 1.02\\ 			             
                \hline
        \end{tabular}
\end{table}

\subsection{Domain of Type A solutions}
\label{sec:confirmations}
Merger remnants that do not collapse promptly, can evolve towards nearly axisymmetric, quasi-stationary configurations (at least before a possible delayed collapse sets in) that can be approximated with suitable equilibrium models. This involves Type A solutions\footnote{We use the same nomenclature as the classification of \citet{Ansorg_etal_2009} for models constructed with the KEH rotation law.}, i.e. sequences of models that remain quasi-spherical (the maximum density is always at the center) as the axis ratio $r_p/r_e$ is decreased (i.e. the rotation rate increases) until the mass-shedding limit is reached.

In our recent work \citet{Iosif_Stergioulas_2021}, we found that the Uryu+ rotation law with $\{\lambda_1, \lambda_2\} = \{2.0, 1.0\}$ and with $\{\lambda_1, \lambda_2\} = \{1.5, 1.0\}$ yields Type A solutions for the $N=1$ polytropic EOS. In addition, we highlighted the fact that according to recent numerical simulations \citep{Hanauske_etal_2017, DePietri_etal_2020}, a value of $\lambda_2 = 1$ seems to be favoured over $\lambda_2 = 0.5$, for the case of compact remnants from BNS mergers, while $\lambda_1$ ranges between 1.7-1.9 for realistic EOS. We therefore probe in more detail models with this range of parameters. To that end, we set the value of $\lambda_2$ to 1.0 and explore the range $\lambda_1 \in [1.5, 2.0]$ with a step of 0.1.

A vertical "scan" of the mass vs. $\epsilon_\text{max}$ parameter space for specific Type A $\{\lambda_1, \lambda_2\}$ pairs (fixing the value of the maximum energy density and gradually decreasing the axis ratio $r_p/r_e$), revealed that these sequences reached a point, after which it was not possible to further construct equilibrium solutions as the maximum density was increased. This behavior is much more stark for the case of $\{\lambda_1, \lambda_2\} = \{2.0, 1.0\}$ than for $\{\lambda_1, \lambda_2\} = \{1.5, 1.0\}$. We note that the terminal models encountered for each choice of $\{\lambda_1, \lambda_2\}$ pairs are not close to mass-shedding. For the highest $\{\lambda_1, \lambda_2\} = \{2.0, 1.0\}$, the $\Omega_e / \Omega_K$ ratio of the terminal models ranges between 0.6-0.7 as the angular momentum increases, whereas for $\{\lambda_1, \lambda_2\} = \{1.5, 1.0\}$ it ranges between 0.6-0.8. However, the maximum density remained at the center of the configuration even at the highest rotation rates that were achieved and we classify these models as Type A.

\begin{figure}
        \includegraphics[width=0.93\columnwidth]{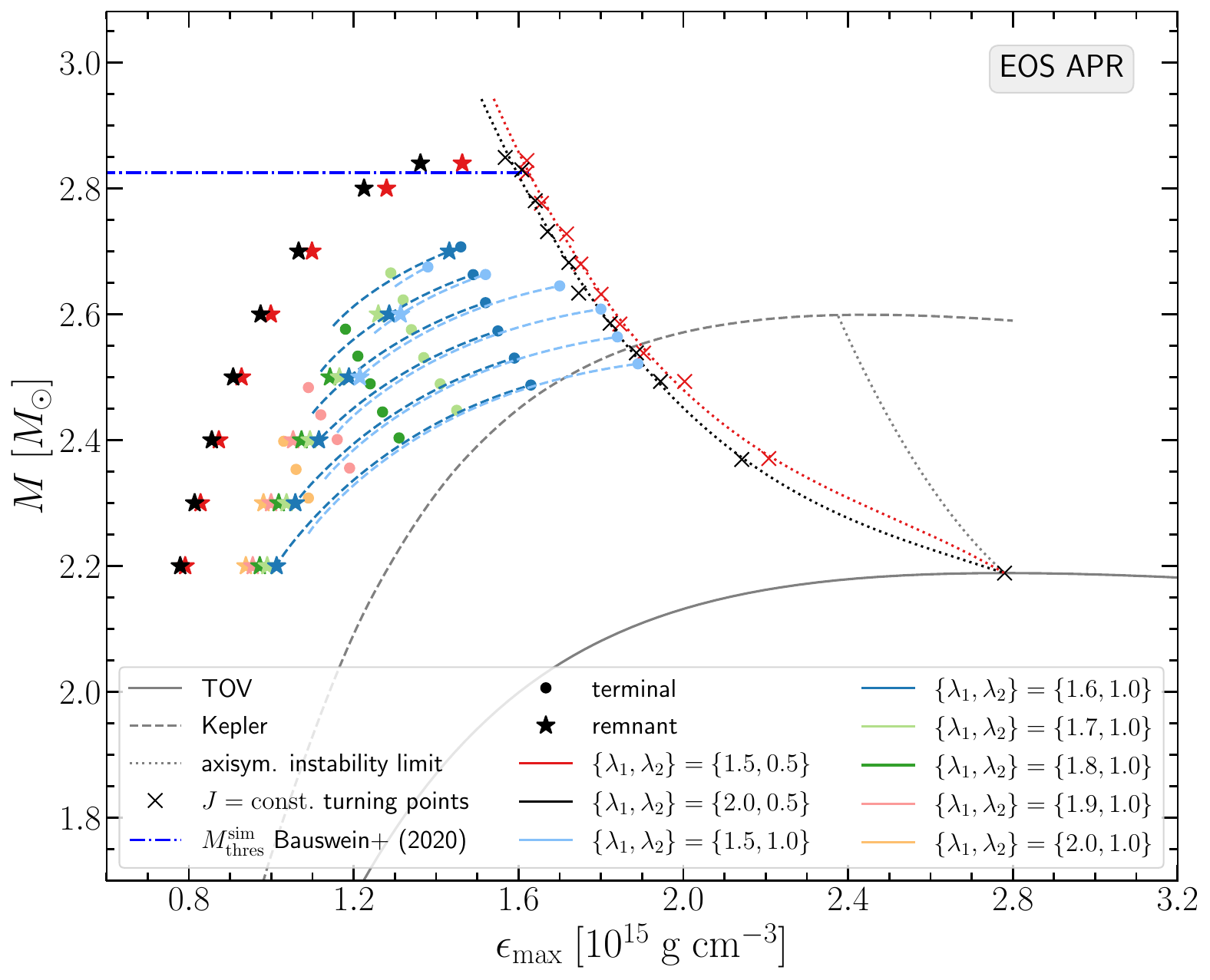}
    \caption{Gravitational mass $M$ vs. maximum energy density $\epsilon_\mathrm{max}$ for the APR EOS. Six choices of rotation law parameters yielding Type A solutions are presented ($\lambda_1$ varies from 1.5 to 2 while $\lambda_2$ is held fixed and equal to 1). The nonrotating (TOV) sequence (grey solid line), the mass-shedding (Kepler) limit for uniform rotation (grey dashed line) and the axisymmetric instability limit for uniform rotation (grey dotted line), together with the data corresponding to Type C solutions are shown as reference.}
    \label{fig:APR_Mass_emax_TypeA}
\end{figure}

Figure~\ref{fig:APR_Mass_emax_TypeA} displays the six Type A remnant sequences corresponding to the six $\{\lambda_1, \lambda_2\}$ pairs that we investigated\footnote{As a related remark, see also \citet[Figure 1]{Tsokaros_etal_2020} for the behavior of the turning points for different parameter choices with the KEH rotation law.} for the APR EOS. Remnant models are shown as asterisks (different colors correspond to different $\{\lambda_1, \lambda_2\}$ values). The terminal models along each $J$-constant sequence for each $\{\lambda_1, \lambda_2\}$ pair are shown as dots of matching color. We find that the choice of $\lambda_1=1.6$ allows for the most massive Type A remnant model for this EOS, with a gravitational mass of $M_\text{tot}=2.7 M_\odot$. 

\begin{figure}
        \includegraphics[width=0.93\columnwidth]{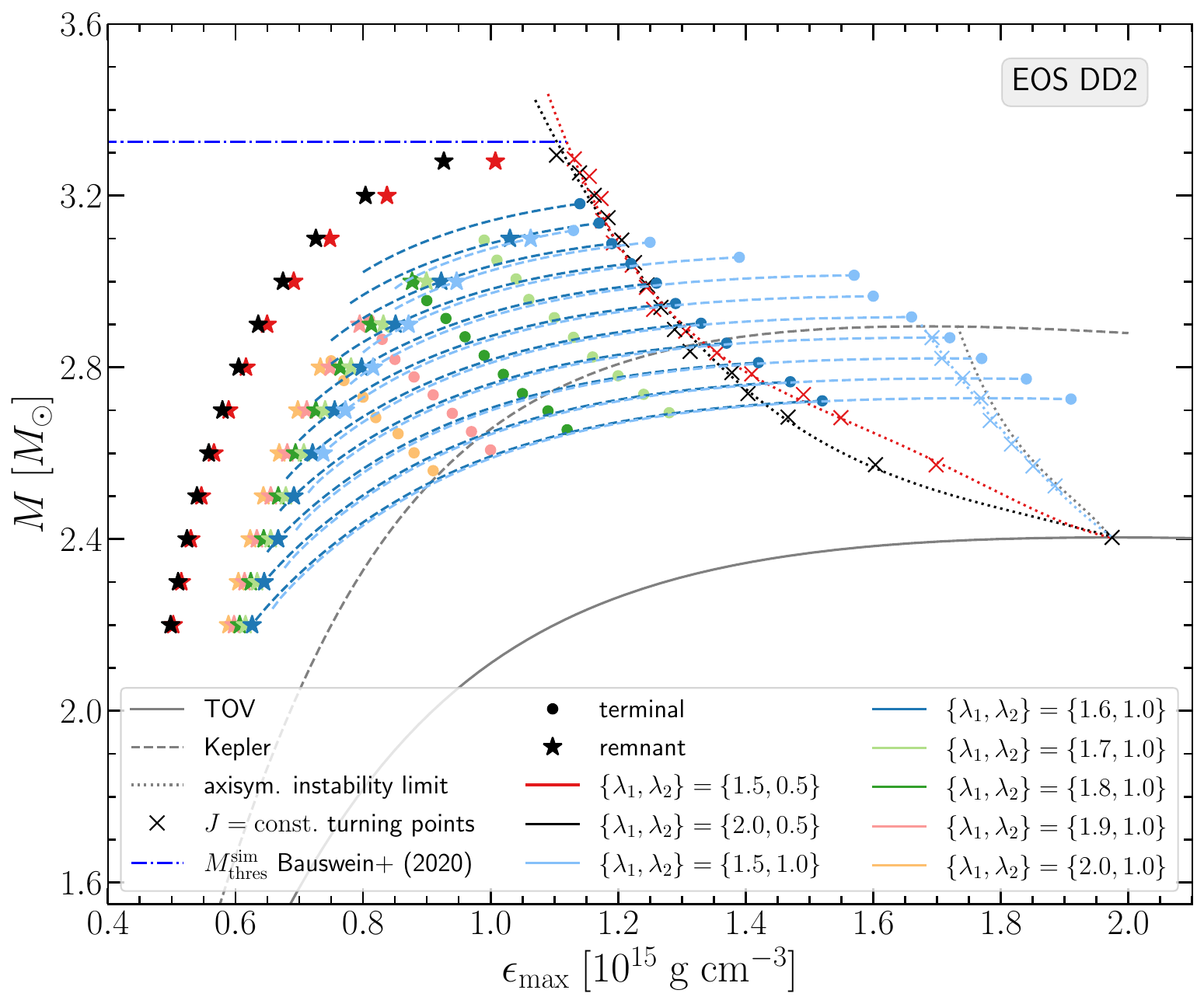}
    \caption{Same as Figure~\ref{fig:APR_Mass_emax_TypeA} for the DD2 EOS.}
    \label{fig:DD2_Mass_emax_TypeA}
\end{figure}

For the $\{\lambda_1, \lambda_2\} = \{1.5, 1.0\}$ and $\{\lambda_1, \lambda_2\} = \{1.6, 1.0\}$ cases we explicitly show the constant angular momentum sequences (as light blue and dark blue dashed lines respectively). We note that the $\{\lambda_1, \lambda_2\} = \{1.6, 1.0\}$ $J$-constant lines in Figure~\ref{fig:APR_Mass_emax_TypeA} nearly merge into those of $\{\lambda_1, \lambda_2\} = \{1.5, 1.0\}$ as the maximum density increases. $J$-constant lines constructed with other pairs of $\{\lambda_1, \lambda_2\}$ also tend to merge with those of $\{\lambda_1, \lambda_2\} = \{1.5, 1.0\}$, but for clarity, we omit the $J$-constant lines for $\lambda_1= \{1.7, 1.8, 1.9, 2.0\}$ in Figure~\ref{fig:APR_Mass_emax_TypeA}. Moreover, as a rule of thumb the $J$-constant lines for $\{\lambda_1, \lambda_2\} = \{1.5, 1.0\}$ reach higher maximum densities than the $J$-constant lines for $\{\lambda_1, \lambda_2\} = \{1.6, 1.0\}$. Note though, that for the highest $J$-constant sequence for the APR and DD2 EOS (Figures~\ref{fig:APR_Mass_emax_TypeA} and \ref{fig:DD2_Mass_emax_TypeA} respectively), this trend is reversed, i.e. it is the highest $J$-constant line for $\{\lambda_1, \lambda_2\} = \{1.6, 1.0\}$ that actually reaches higher maximum densities. In our investigation of this behaviour, we found that an increasingly smaller step size in axis ratio was required (reaching as low as $10^{-4}$) to locate equilibrium solutions for $\{\lambda_1, \lambda_2\} = \{1.5, 1.0\}$ (which yields the weakest differential rotation we consider) in the parameter space of increasing energy density values and high angular momentum values. This effect is indicating numerical stiffness. However, one would have to perform similar trials with other equilibrium codes to clarify whether this issue: (i) has some physical origin (e.g. it could be possible that weaker differential rotation cannot produce \textit{high enough} angular velocities to accommodate \textit{arbitrarily high} angular momentum values), (ii) or if it is due to numerical stiffness of the problem, (iii) or if it is due to the existence of different types of differentially rotating solutions that have the same central density and axis ratio\footnote{We note that the numerical method used in the \textsc{rns} code does not discriminate between different types of models with same axis ratio, central or maximum density and differential rotation law, so that the numerical iterations can alternate between two different solutions, instead of converging to a specific one.}.

\begin{figure}
        \includegraphics[width=0.93\columnwidth]{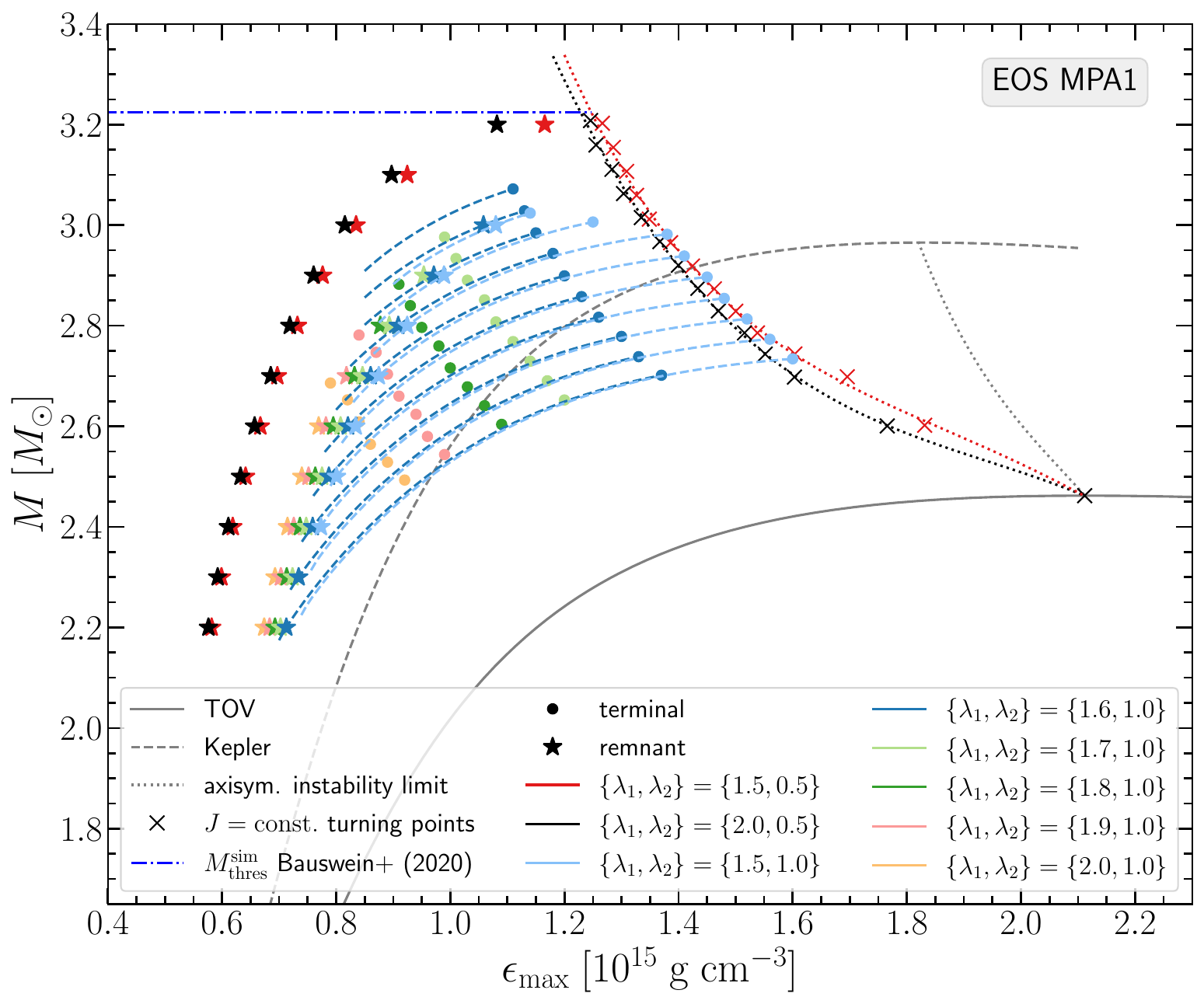}
    \caption{Same as Figure~\ref{fig:APR_Mass_emax_TypeA} for the MPA1 EOS.}
    \label{fig:MPA1_Mass_emax_TypeA}
\end{figure}

Gathering all the evidence, some interesting observations can be made in connection to earlier works in the literature, where the KEH rotation law was used. First of all, we note that Type C remnant models are able to reach higher masses than Type A models, in agreement with findings in \citet{GondekRosinska_etal_2017, Studzinska_etal_2016, Espino_Paschalidis_2019} for the KEH rotation law. Concerning the terminal models encountered for the Type A $J$-constant sequences, they can be interpreted as a confirmation that the domain of Type A solutions shrinks for higher densities and stronger differential rotation. Specifically, with stronger differential rotation we do not find Type A solutions above a certain maximum energy density, whereas we can still find Type C solutions\footnote {Note that we do not discuss Type B and Type D solutions of \citet{Ansorg_etal_2009} in this study. Depending on the choice of parameters, Type B solutions can co-exist with those of Type A and Type D with those of Type C.} (or Type A solutions with a weak differential rotation). 

\begin{figure*}
        \includegraphics[width=0.945\textwidth]{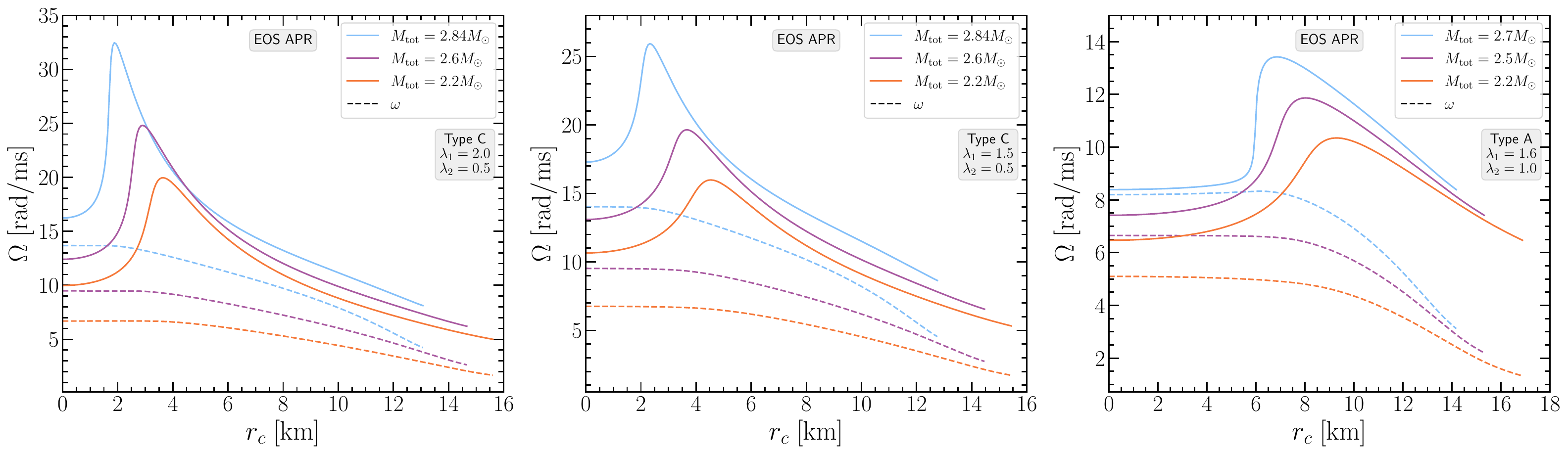}
    \caption{Angular velocity profiles $\Omega$ vs. the circumferential radial coordinate $r_c$, in the equatorial plane for the APR EOS. \textit{Left panel:} Type C models with $\{\lambda_1, \lambda_2\}= \{2.0, 0.5\}$. \textit{Middle panel:} Type C models with $\{\lambda_1, \lambda_2\}= \{1.5, 0.5\}$. \textit{Right panel:} Type A models with $\{\lambda_1, \lambda_2\}= \{1.6, 1.0\}$. In each panel the profiles for the most massive, the least massive and an intermediate mass remnant model are shown. The dashed lines correspond to the frame dragging angular velocity $\omega(r_c)$ in the equatorial plane for each different model. For the most massive Type A model, we observe that $\Omega \simeq \omega$ in the core, as has been reported in BNS merger simulations.}
    \label{fig:rot_profiles_APR}
\end{figure*}

\begin{figure*}
        \includegraphics[width=0.945\textwidth]{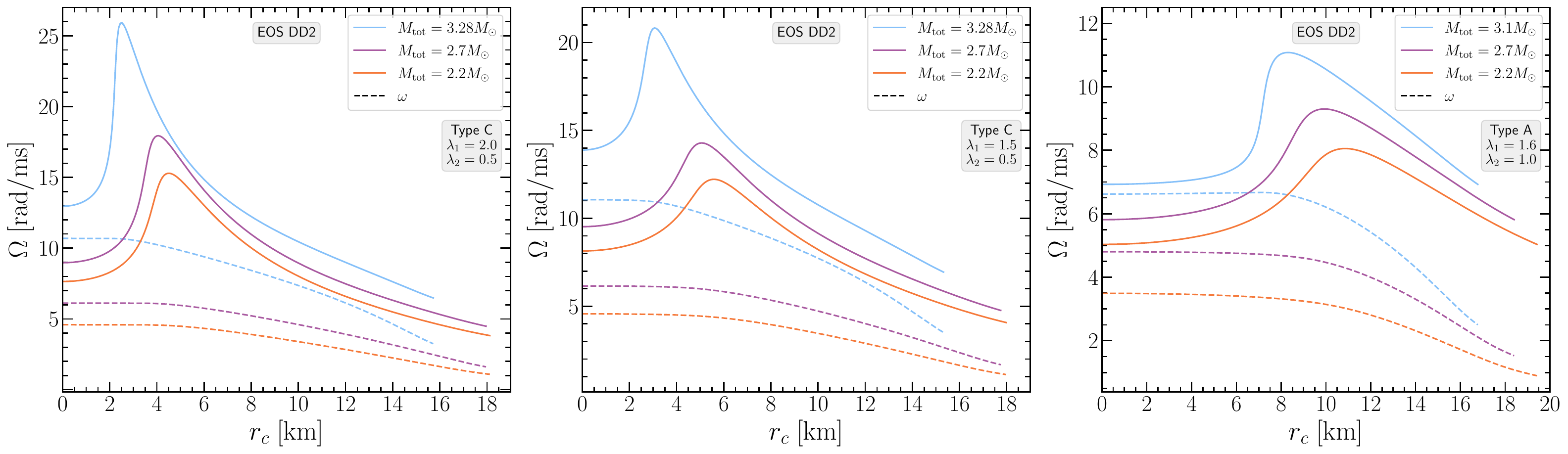}
    \caption{Same as Figure~\ref{fig:rot_profiles_APR} for the DD2 EOS.}
    \label{fig:rot_profiles_DD2}
\end{figure*}

\begin{figure*}
        \includegraphics[width=0.945\textwidth]{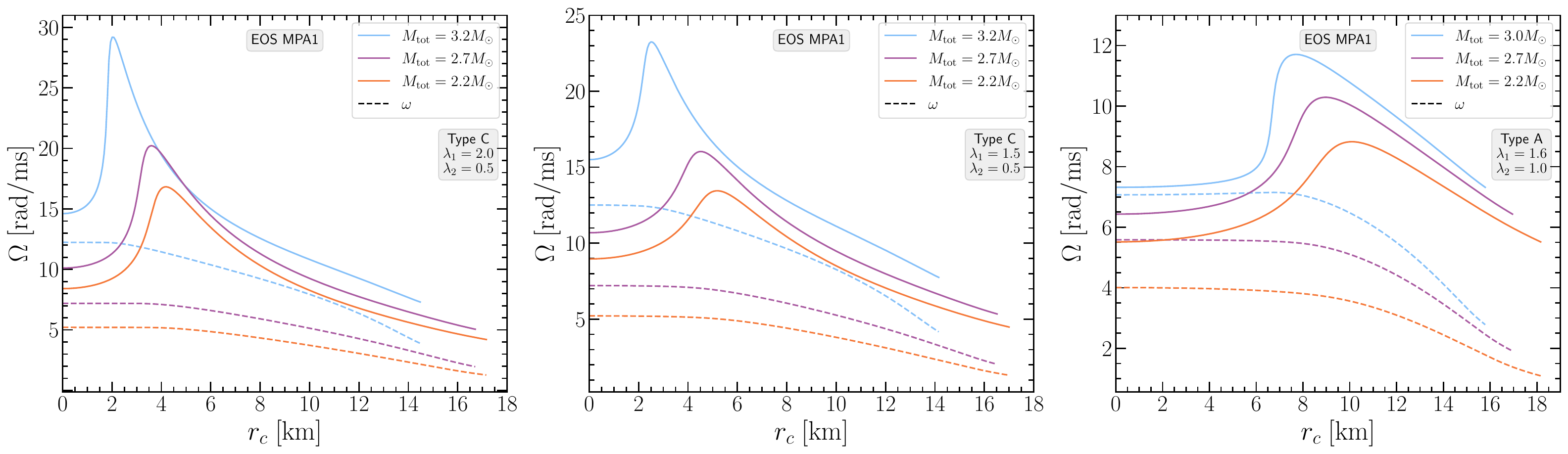}
    \caption{Same as Figure~\ref{fig:rot_profiles_APR} for the MPA1 EOS.}
    \label{fig:rot_profiles_MPA1}
\end{figure*}

The above property of differentially rotating models was highlighted for the KEH rotation law in  \citet{Studzinska_etal_2016, GondekRosinska_etal_2017} for polytropes, \citet{Espino_Paschalidis_2019} for realistic EOS and \citet{Szkudlarek_etal_2019} for strange quark stars. From our findings for the Uryu+ rotation law, it seems that the different types of solutions are not tied to the particular KEH rotation law (for which they were originally discovered), but appear also for other, more general rotation laws, such as the one by \citeauthor{Uryu_etal_2017} considered here. 

\begin{figure*}
        \includegraphics[width=0.945\textwidth]{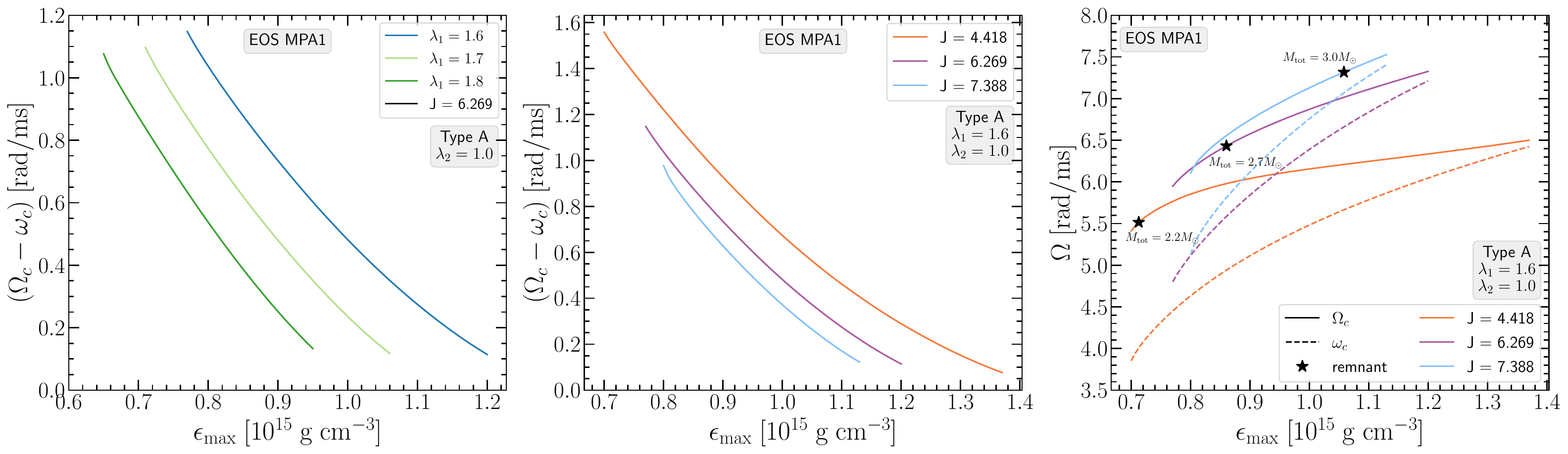}
    \caption{Investigation of the frame dragging contribution to rotation for the MPA1 EOS. \textit{Left panel:} Difference $(\Omega_c - \omega_c)$ between the angular velocity and the frame dragging values at the center of the configuration vs. the maximum energy density. The solid lines correspond to $J$-constant sequences with $J=6.269$, along which remnant models with $M_\text{tot}=2.7 M_\odot$ are constructed. The colors represent different options for $\lambda_1$ yielding different strengths of differential rotation and are matching the colors of the respective options in Figure~\ref{fig:MPA1_Mass_emax_TypeA} for ease of reference. \textit{Middle panel:} Same quantities plotted as in the left panel, but for a single strength of differential rotation $\{\lambda_1, \lambda_2\}= \{1.6, 1.0\}$. The solid lines are $J$-constant sequences corresponding to the most massive ($M_\text{tot}=3.0 M_\odot$), least massive ($M_\text{tot}=2.2 M_\odot$) and an intermediate mass $M_\text{tot}=2.7 M_\odot$ remnant model. \textit{Right panel:} Central angular velocity $\Omega_c$ (solid lines) and frame dragging at the center $\omega_c$ (dashed lines) vs. the maximum energy density for $\{\lambda_1, \lambda_2\}= \{1.6, 1.0\}$. The different colors correspond to the same $J$-constant sequences as in the middle panel. The asterisks represent the respective remnant models. The colors selected in the middle and right panels match those of the right panel of Figure~\ref{fig:rot_profiles_MPA1} for ease of reference.}
    \label{fig:Omom_MPA1}
\end{figure*}

Figures~\ref{fig:DD2_Mass_emax_TypeA} and \ref{fig:MPA1_Mass_emax_TypeA} show the same investigation of Type A models as in Figure~\ref{fig:APR_Mass_emax_TypeA}, but for the EOS DD2 and MPA1. We note that also for these EOS, the choice of $\lambda_1 =1.6$ allows for the construction of the most massive Type A remnant model and also for a full remnant sequence (i.e. starting from the lowest remnant mass of $M_\text{tot}=2.2 M_\odot$ we consider). Within the parameter range that we investigated, the highest mass Type A remnant model reached with EOS DD2 was $3.1 M_\odot$, while for EOS MPA1 it was $3 M_\odot$. Physical quantities of Type A remnant models are reported in Tables~\ref{tab:APR_remnants_physical_quantities}, \ref{tab:DD2_remnants_physical_quantities} and \ref{tab:MPA1_remnants_physical_quantities} for EOS APR, DD2 and MPA1 respectively.

In \citet[Figure 8]{Iosif_Stergioulas_2021} we showed that for $\{\lambda_1, \lambda_2\} = \{1.5, 1.0\}$ the differential rotation is quite weak, compared to other values of $\lambda_1$ in the range we consider here. For the stiffest EOS we consider (DD2) this leads to Type A models for which we can find turning points along the $J$-constant sequences (light blue dotted curve in Figure~\ref{fig:DD2_Mass_emax_TypeA}). Moreover, these turning points are close to the axisymmetric instability limit for uniform rotation. 

\subsection{Frame dragging contribution to rotation}
\label{sec:Omega_sim_omega}

Figures~\ref{fig:rot_profiles_APR}, \ref{fig:rot_profiles_DD2} and \ref{fig:rot_profiles_MPA1} show the angular velocity $\Omega(r_c)$ rotational profiles versus the circumferential radial coordinate $r_c$ at the equatorial plane, for the three EOS under study, APR, DD2 and MPA1. For each EOS, a triplet of panels is shown, with each left panel corresponding to rotation laws $\{\lambda_1, \lambda_2\}=\{2.0, 0.5\}$, each middle panel to $\{\lambda_1, \lambda_2\}=\{1.5, 0.5\}$ and each right panel to $\{\lambda_1, \lambda_2\}=\{1.6, 1.0\}$. Every individual panel shows $\Omega(r_c)$, as well as the frame dragging angular velocity $\omega(r_c)$ for the most massive, the least massive and for an intermediate mass remnant model constructed with the particular choice of parameters $\{\lambda_1, \lambda_2\}$. A common finding shared among the three EOS explored, is that models with $\{\lambda_1, \lambda_2\}=\{2.0, 0.5\}$ reach the highest angular velocity compared to the other two options and models with $\{\lambda_1, \lambda_2\}=\{1.6, 1.0\}$ reach the smallest angular velocity peaks in their profile. This is in agreement with corresponding rotational profiles for these parameter values for polytropic equilibrium models with $N=1$ \citep[Figure 8]{Iosif_Stergioulas_2021}.

\begin{figure*}
        \includegraphics[width=0.945\textwidth]{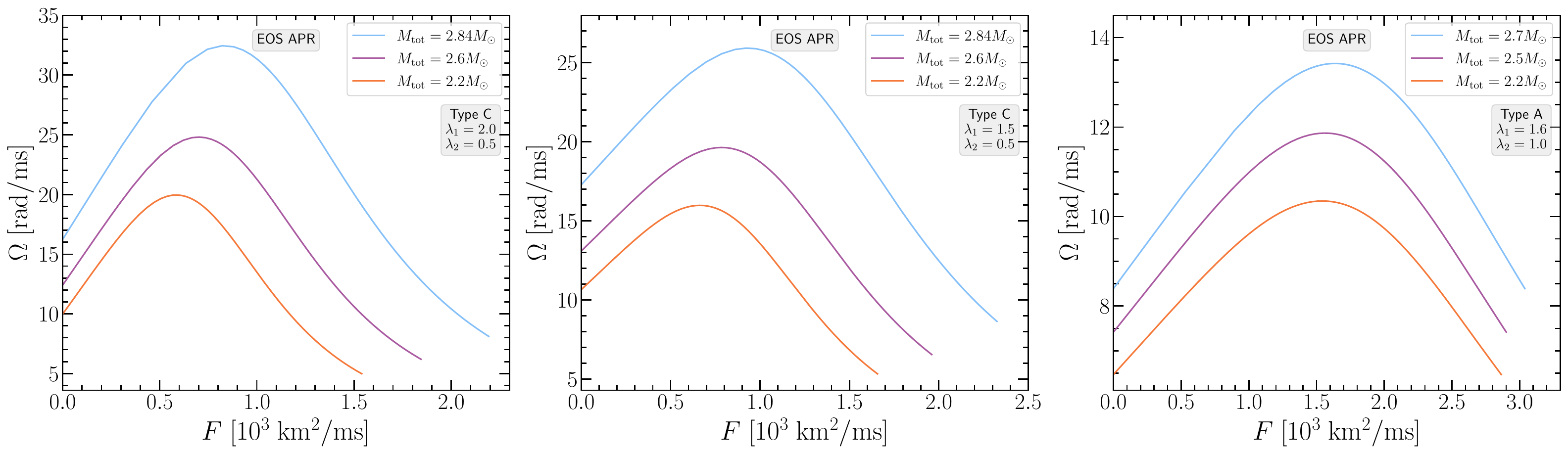}
    \caption{Angular velocity profiles $\Omega$ vs. the gravitationally redshifted angular momentum per unit rest mass and enthalpy $F$, in the equatorial plane for the APR EOS. \textit{Left panel:} Type C models with $\{\lambda_1, \lambda_2\}= \{2.0, 0.5\}$. \textit{Middle panel:} Type C models with $\{\lambda_1, \lambda_2\}= \{1.5, 0.5\}$. \textit{Right panel:} Type A models with $\{\lambda_1, \lambda_2\}= \{1.6, 1.0\}$. In each panel the profiles for the most massive, the least massive and an intermediate mass remnant model are shown.}
    \label{fig:Omega_F_profiles_APR}
\end{figure*}

\begin{figure*}
        \includegraphics[width=0.945\textwidth]{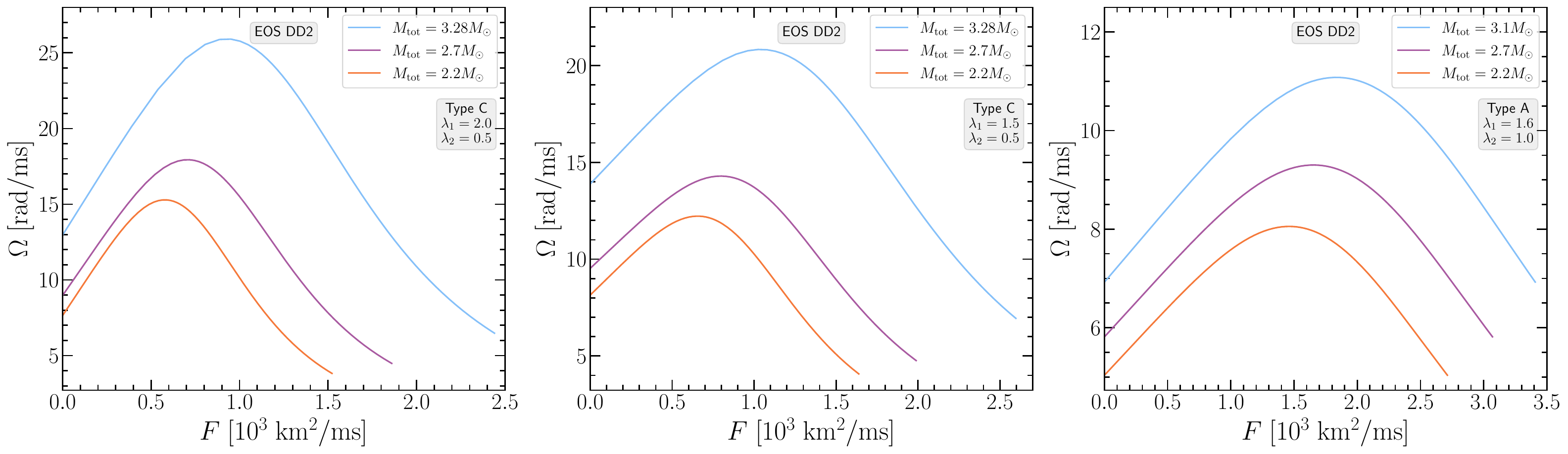}
    \caption{Same as Figure~\ref{fig:Omega_F_profiles_APR} for the DD2 EOS.}
    \label{fig:Omega_F_profiles_DD2}
\end{figure*}

\begin{figure*}
        \includegraphics[width=0.945\textwidth]{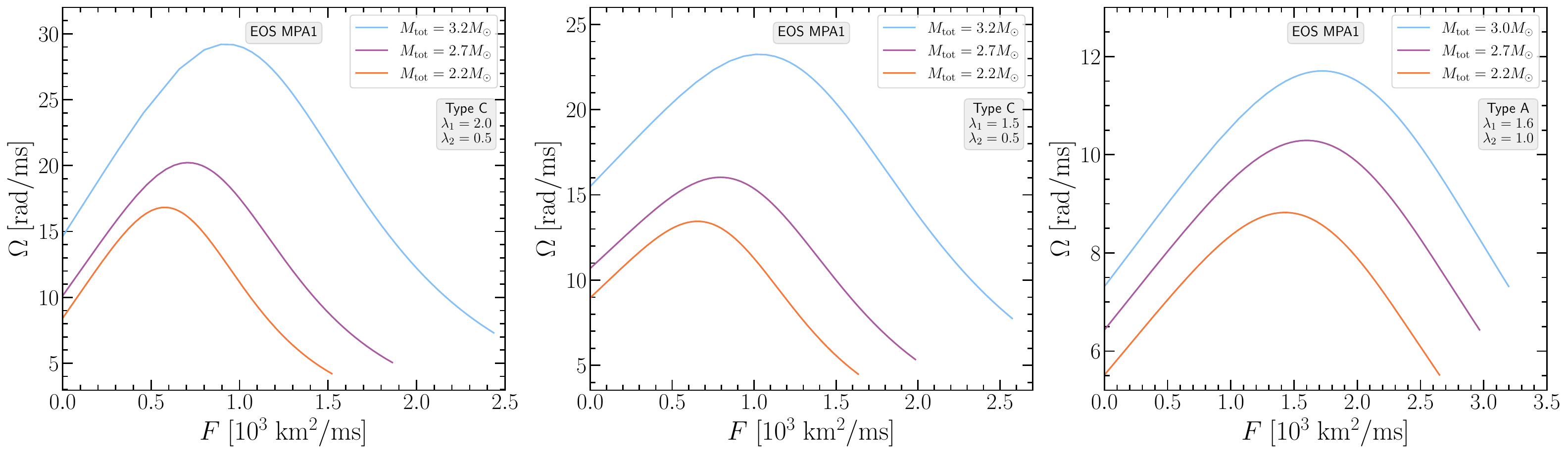}
    \caption{Same as Figure~\ref{fig:Omega_F_profiles_APR} for the MPA1 EOS.}
    \label{fig:Omega_F_profiles_MPA1}
\end{figure*}

Examining Type A remnants we observe that for the most massive models the central part of the configuration (i.e. approximately up to $r_c \sim 5 \, \text{km} $) rotates slowly compared to the rest of the configuration. However, this rotation rate is mostly due to the contribution of the frame dragging $\omega$, which means that with respect to a ZAMO this part of the model is \textit{almost nonrotating}. To our knowledge, this is the first time that a differential rotation law has been shown to reproduce this feature, already known from numerical simulations: similar rotation profiles have been presented and analyzed in \citet{Kastaun_Galeazzi_2015, Endrizzi_etal_2016, Kastaun_etal_2016, Kastaun_etal_2017, Ciolfi_etal_2017}. 

Note that we did not observe the $\Omega \sim \omega$ behaviour near the center of the star \textit{only} for the most massive Type A models (i.e. the rightmost dark blue asterisks of the remnant sequences for $\{\lambda_1, \lambda_2\}=\{1.6, 1.0\}$ in Figures~\ref{fig:APR_Mass_emax_TypeA}, \ref{fig:DD2_Mass_emax_TypeA} and \ref{fig:MPA1_Mass_emax_TypeA} corresponding to the light blue curves of the right panels in Figures~\ref{fig:rot_profiles_APR}, \ref{fig:rot_profiles_DD2} and \ref{fig:rot_profiles_MPA1} for each EOS considered). This feature also appears in less massive models that have higher central densities (i.e. models neighbouring the terminal models shown as dark blue dots for $\{\lambda_1, \lambda_2\}=\{1.6, 1.0\}$ in Figures~\ref{fig:APR_Mass_emax_TypeA}, \ref{fig:DD2_Mass_emax_TypeA} and \ref{fig:MPA1_Mass_emax_TypeA}). Specifically, for \textit{all} the rotation law parameters yielding Type A configurations that we considered (i.e. the six $\{\lambda_1, \lambda_2\}$ pairs with $\lambda_2=1.0$ in Figures~\ref{fig:APR_Mass_emax_TypeA}, \ref{fig:DD2_Mass_emax_TypeA} and \ref{fig:MPA1_Mass_emax_TypeA}), we found that $\omega$ approaches $\Omega$ as the maximum density increases along a $J$-constant sequence, until the $\Omega \sim \omega$ feature appears as we reach the terminal model. To clarify even further, we found that 
\begin{itemize}
    \item along a $J$-constant sequence, the $\Omega \sim \omega$ behaviour appears at lower central densities as the strength of differential rotation is increased\footnote{This holds as a rule of thumb, barring a possible numerical stiffness encountered for high $J$ values and weak differential rotation (see discussion in Section~\ref{sec:confirmations}).} (see left panel of Figure~\ref{fig:Omom_MPA1})
    \item for a specified strength of differential rotation (i.e. fixing the differential rotation law's parameters at certain values), the $\Omega \sim \omega$ behaviour appears at lower central densities as the angular momentum $J$ is increased (see middle panel of Figure~\ref{fig:Omom_MPA1}).
    \item as the maximum energy density is increased along a $J$-constant sequence, the frame dragging at the center $\omega_c$ \textit{increases faster} than the central angular velocity $\Omega_c$, thus giving rise to the $\Omega \sim \omega$ feature (see right panel of Figure~\ref{fig:Omom_MPA1}).
\end{itemize}
Note that while Figure~\ref{fig:Omom_MPA1} demonstrates the above for the MPA1 EOS, similar behaviour is observed for the other two EOS that we studied.

\begin{figure*}
        \includegraphics[width=0.945\textwidth]{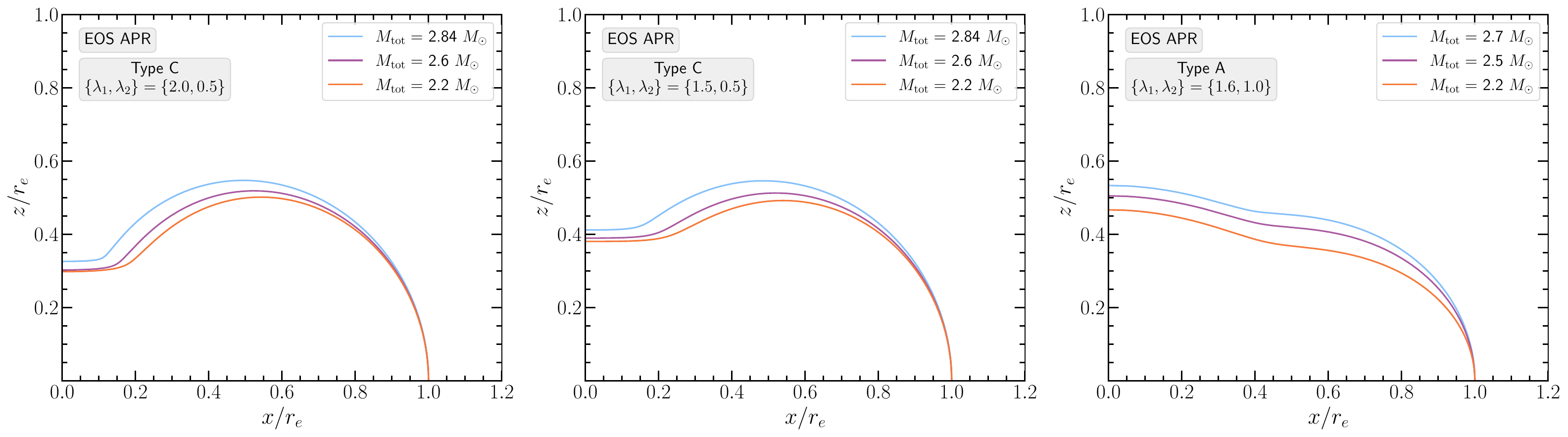}
    \caption{Stellar surfaces for the APR EOS. \textit{Left panel:} Type C models with $\{\lambda_1, \lambda_2\}= \{2.0, 0.5\}$. \textit{Middle panel:} Type C models with $\{\lambda_1, \lambda_2\}= \{1.5, 0.5\}$. \textit{Right panel:} Type A models with $\{\lambda_1, \lambda_2\}= \{1.6, 1.0\}$. In each panel the surfaces of the most massive, the least massive and an intermediate mass remnant model are shown.}
    \label{fig:surfaces_APR}
\end{figure*}

\begin{figure*}
        \includegraphics[width=0.945\textwidth]{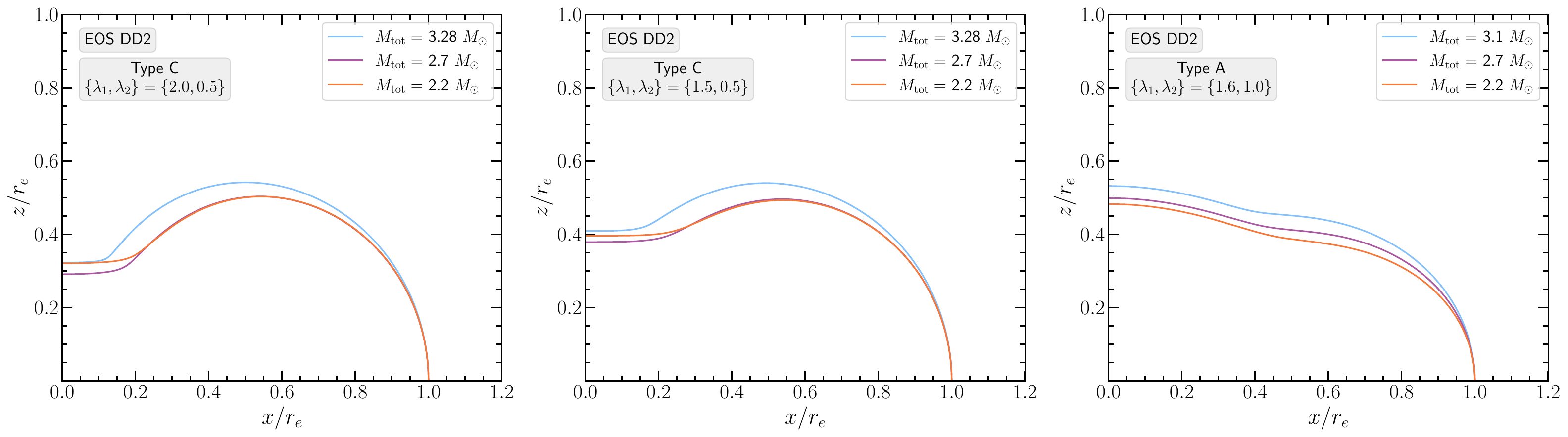}
    \caption{Same as Figure~\ref{fig:surfaces_APR} for the DD2 EOS.}
    \label{fig:surfaces_DD2}
\end{figure*}

\begin{figure*}
        \includegraphics[width=0.945\textwidth]{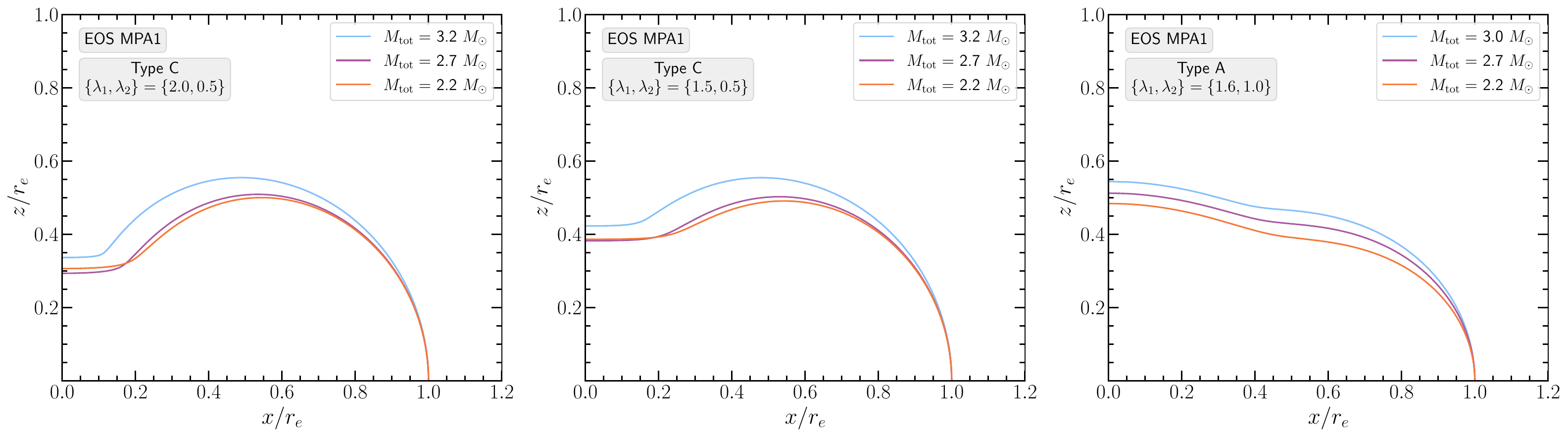}
    \caption{Same as Figure~\ref{fig:surfaces_APR} for the MPA1 EOS.}
    \label{fig:surfaces_MPA1}
\end{figure*}

\begin{figure*}
        \includegraphics[width=0.945\textwidth]{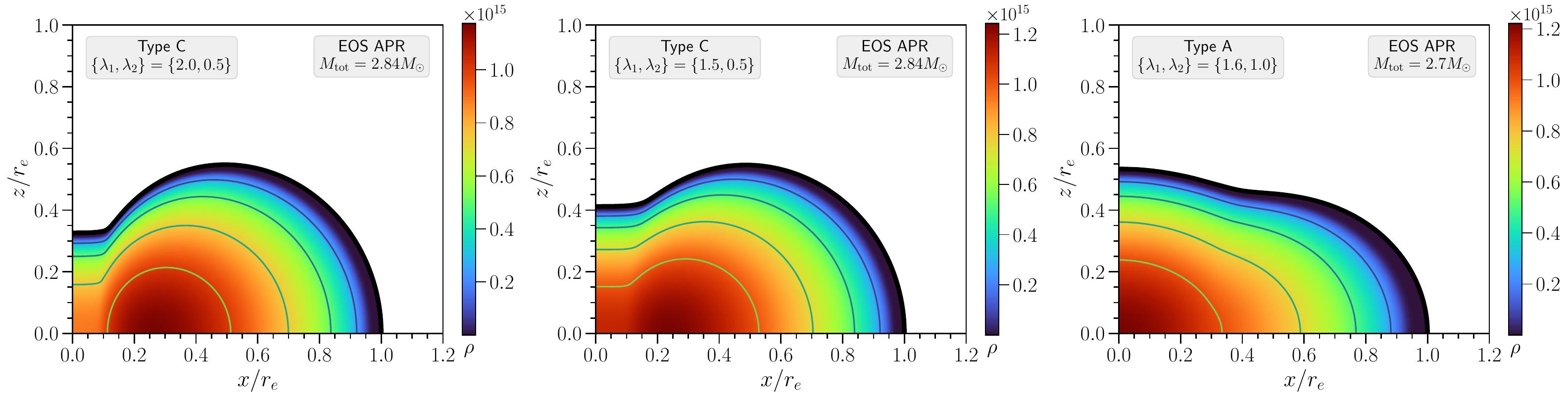}
    \caption{Two-dimensional rest mass density distribution $\rho \, [\mathrm{g \, cm^{-3}}]$ of the most massive remnant models for the APR EOS. \textit{Left panel:} Type C model with $\{\lambda_1, \lambda_2\}= \{2.0, 0.5\}$. \textit{Middle panel:} Type C model with $\{\lambda_1, \lambda_2\}= \{1.5, 0.5\}$. \textit{Right panel:} Type A model with $\{\lambda_1, \lambda_2\}= \{1.6, 1.0\}$.}
    \label{fig:2Drestmass_APR}
\end{figure*}

\begin{figure*}
        \includegraphics[width=0.945\textwidth]{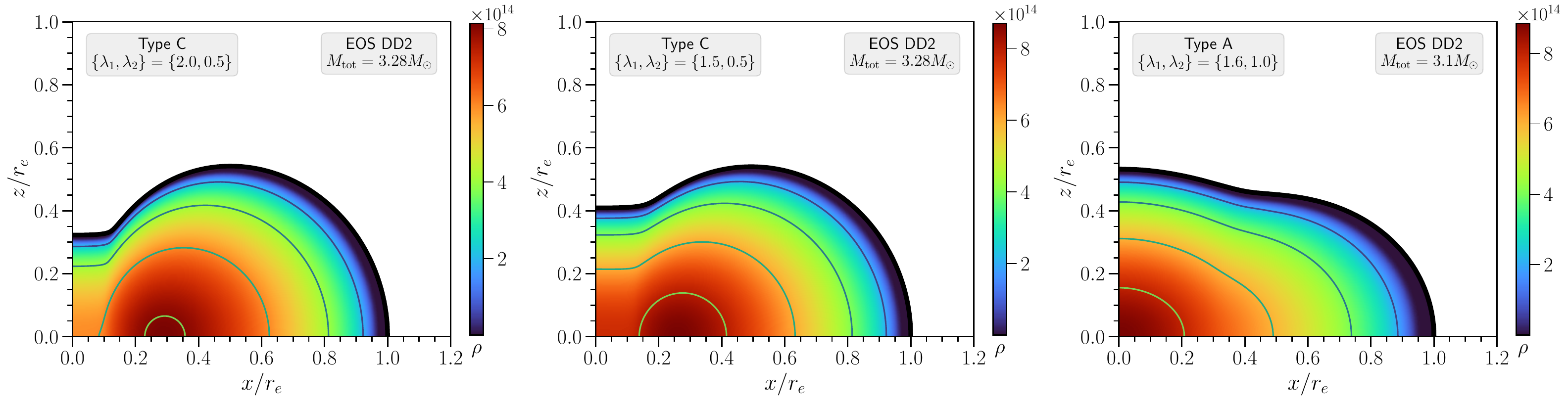}
    \caption{Same as Figure~\ref{fig:2Drestmass_APR} for the DD2 EOS.}
    \label{fig:2Drestmass_DD2}
\end{figure*}

\begin{figure*}
        \includegraphics[width=0.945\textwidth]{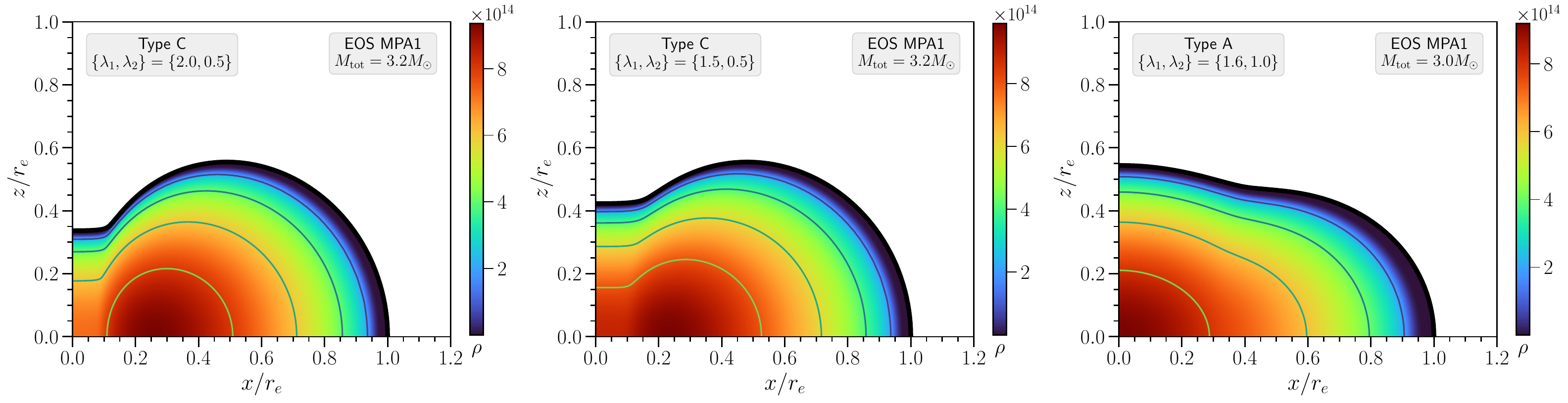}
    \caption{Same as Figure~\ref{fig:2Drestmass_APR} for the MPA1 EOS.}
    \label{fig:2Drestmass_MPA1}
\end{figure*}

\subsection{$\Omega(F)$ profiles}

Figures~\ref{fig:Omega_F_profiles_APR}, \ref{fig:Omega_F_profiles_DD2} and \ref{fig:Omega_F_profiles_MPA1} follow the same organization as corresponding figures for the rotation profiles $\Omega(r_c)$ and present the angular velocity $\Omega(F)$ profiles that define each rotation law. We find that for each choice of rotation law parameters (e.g. if one observes the left panels of Figures~\ref{fig:Omega_F_profiles_APR}, \ref{fig:Omega_F_profiles_DD2} and \ref{fig:Omega_F_profiles_MPA1} "vertically", etc.) the resulting $\Omega(F)$ profiles are qualitatively similar between the three EOS that we consider. This is in agreement with analogous behaviour of the corresponding $\Omega(r_c)$ profiles in Figures~\ref{fig:rot_profiles_APR}, \ref{fig:rot_profiles_DD2} and \ref{fig:rot_profiles_MPA1}.

Note that in all cases, the inverse profile, $F(\Omega)$, would not be a one-to-one function. In the case of the KEH rotation law, one can simply integrate $F(\Omega)$ in the equation of hydrostationary equilibrium. However, for the Uryu+ rotation law, one needs to express the equation of stationary equilibrium in terms of an integral $\Omega(F)$ (see \cite{Iosif_Stergioulas_2021} for details).

\subsection{Structure of the remnants: surface and density distribution}

Stellar surfaces in the $x-z$ plane (with $x$ and $z$ normalized with the equatorial coordinate radius $r_e$) for the most massive, least massive and an intermediate mass remnant model are shown in Figures~\ref{fig:surfaces_APR}, \ref{fig:surfaces_DD2} and \ref{fig:surfaces_MPA1}, for the three EOS employed and for three different choices of parameters $\{\lambda_1, \lambda_2\}$ (similar to the corresponding figures of the rotation profiles). We note that for tabulated EOS we always have a finite minimum pressure value $P_\text{min}$ at the minimum density $\rho_\text{min}$. For the three EOS employed, the range of $\rho_\text{min}$ is sub-grid varying from $\sim 8 \, \mathrm{g \, cm^{-3}}$ to $\sim 10^3 \, \mathrm{g \, cm^{-3}}$, i.e. so small that not a single grid point is removed from the model. Furthermore, since the \textsc{rns} code constructs numerical tables in log enthalpy, a very small enthalpy value $H_\text{min}$ (of order $10^{-16}$) is used in the tabulated EOS files. Then, the stellar surface is defined as the location where the pressure $P=P_\text{min}$ and the enthalpy $H=H_\text{min}$. Interpolating the model's enthalpy by demanding that its value matches $H_\text{min}$ allows to pinpoint the surface's location. In addition, Figures~\ref{fig:2Drestmass_APR}, \ref{fig:2Drestmass_DD2} and \ref{fig:2Drestmass_MPA1} display the two-dimensional rest-mass density distributions for the most massive models in the meridional plane. 

Note that the surfaces and meridional density profiles are qualitatively similar for the Type C solutions obtained with $\{\lambda_1, \lambda_2\} = \{2.0, 0.5\}$ and $\{\lambda_1, \lambda_2\} = \{1.5, 0.5\}$. Both choices lead to a quasi-toroidal surface shape, typical for Type C solutions. For $\{\lambda_1, \lambda_2\} = \{2.0, 0.5\}$, we observe a stronger deformation close to the rotation axis than for $\{\lambda_1, \lambda_2\} = \{1.5, 0.5\}$, which is explained by the stronger differential rotation. 

For the Type A solutions obtained with $\{\lambda_1, \lambda_2\} = \{1.6, 1.0\}$, the surfaces of all models (most massive to least massive) are quite similar to each other (when coordinates are scaled by $r_e$). Even though the remnant models have significant oblateness, they still retain their quasi-spherical shape (the central density is also the maximum density). It is interesting to note that for these selected Type A remnants, the axis ratio $r_p/r_e$ ranges between $\sim 0.47-0.54$, whereas for the Type C remnants the corresponding range is considerably lower at $\sim 0.3-0.42$.

Recently, \citet{Kastaun_etal_2016} introduced a new measure that replaces density profiles, mass, and compactness in a way that can be used unambiguously for rapidly and differentially rotating merger remnants, without a clearly defined surface. Therefore, we stress that the profiles presented here serve simply as an indication about the different configurations possible for the different cold EOS employed and for the $\{\lambda_1, \lambda_2\}$ options we considered. In a realistic remnant with a hot envelope and mass-shedding, the density distribution would not terminate at the same radius as in our models and one would need to define an approximate surface shape, based e.g. on the location where the density drops to a certain fraction of the maximum density.

\begin{figure}
    \includegraphics[width=0.93\columnwidth]{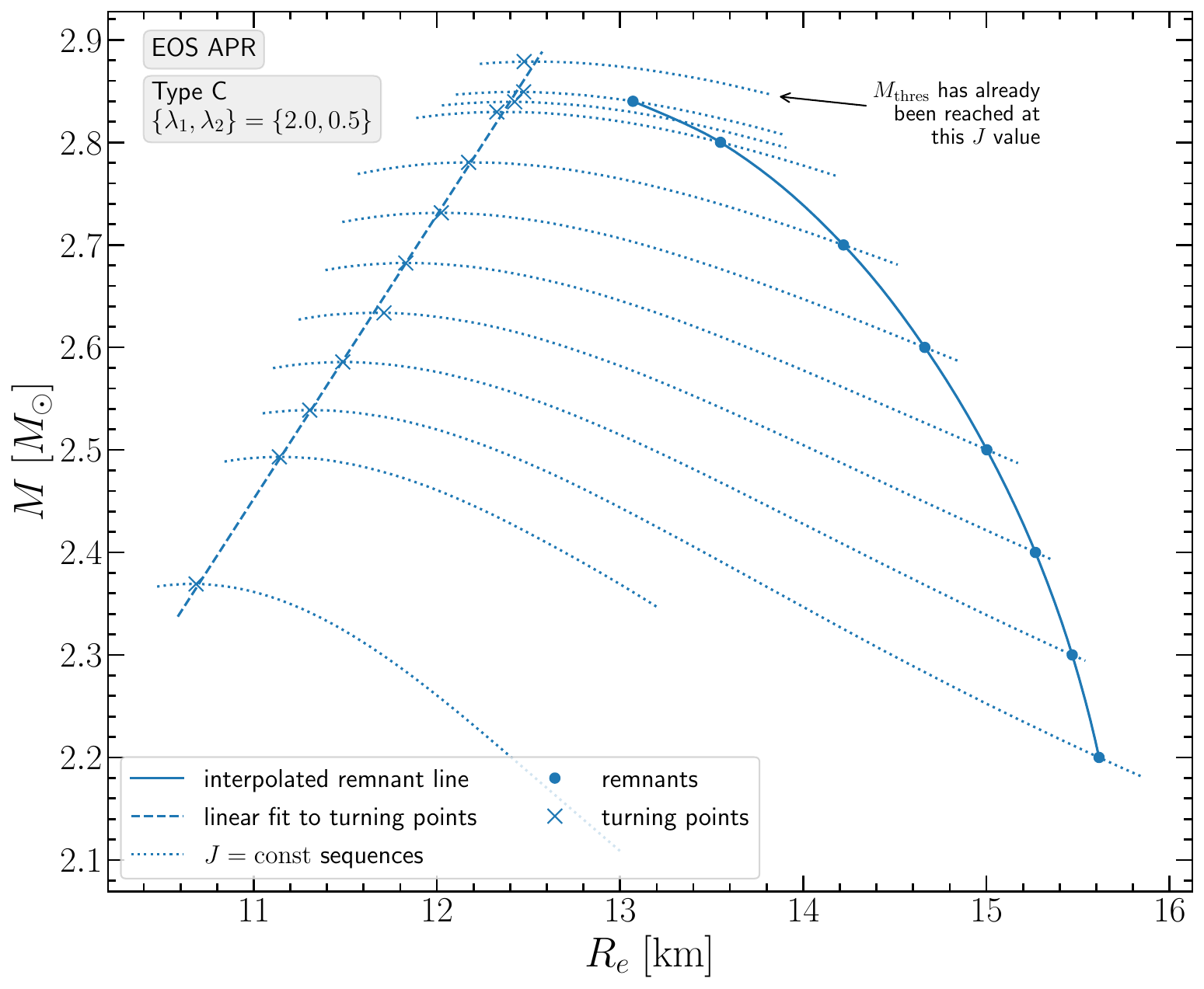}
    \caption{Gravitational mass $M$ vs. equatorial circumferential radius $R_e$ for models constructed with the Uryu+ rotation law, using $\{\lambda_1, \lambda_2\}= \{2.0, 0.5\}$, for the APR EOS. Each dotted line is a $J$-constant sequence and a cross indicates the turning point model. The dashed line is a linear fit through the turning point models. The filled circles represent the sequence of remnant models.}
    \label{fig:MvR_APR_remnants}
\end{figure}

\begin{figure}
    \includegraphics[width=0.93\columnwidth]{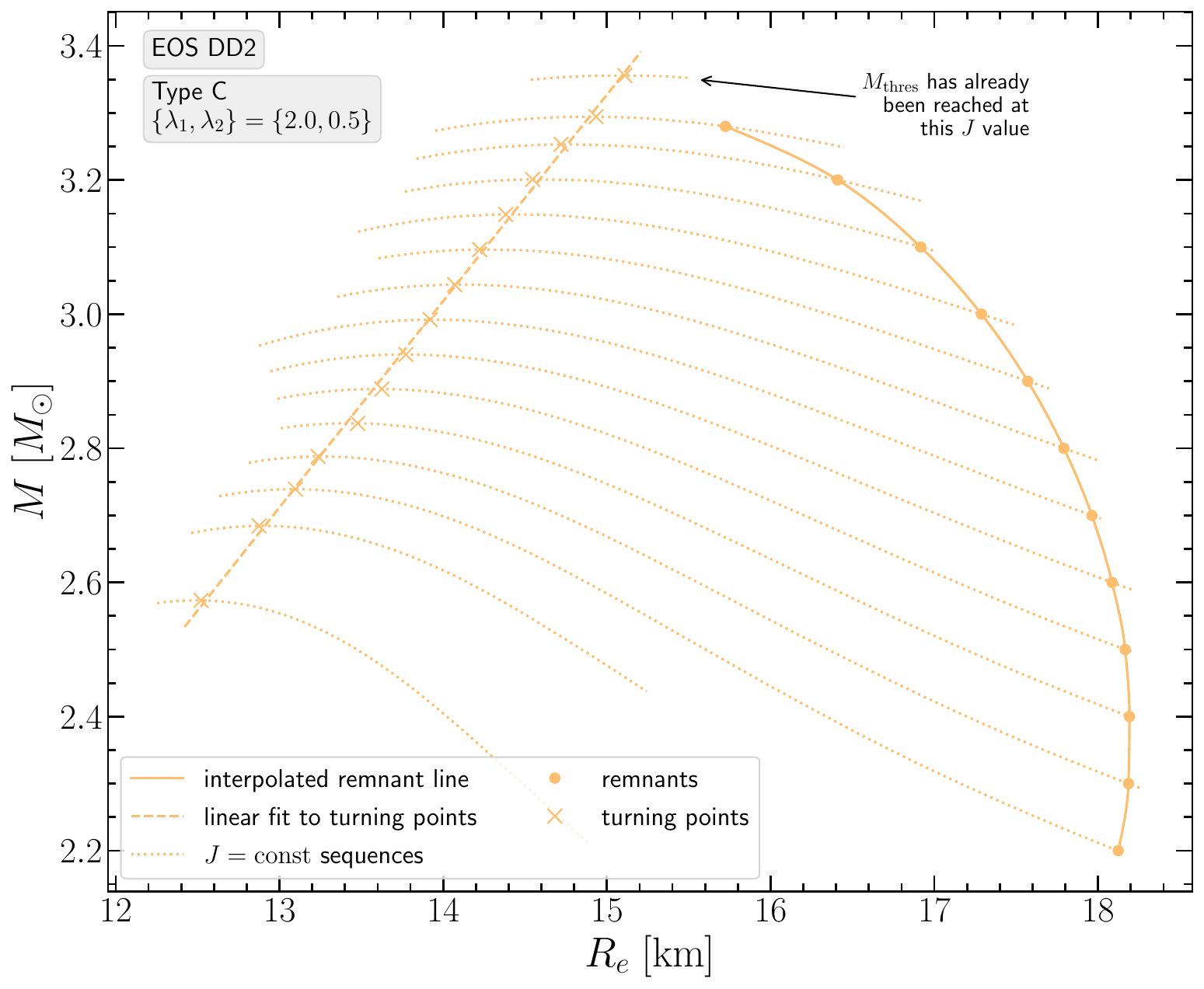}
    \caption{Same as Figure~\ref{fig:MvR_APR_remnants} for the DD2 EOS.}
    \label{fig:MvR_DD2_remnants}
\end{figure}

\begin{figure}
    \includegraphics[width=0.93\columnwidth]{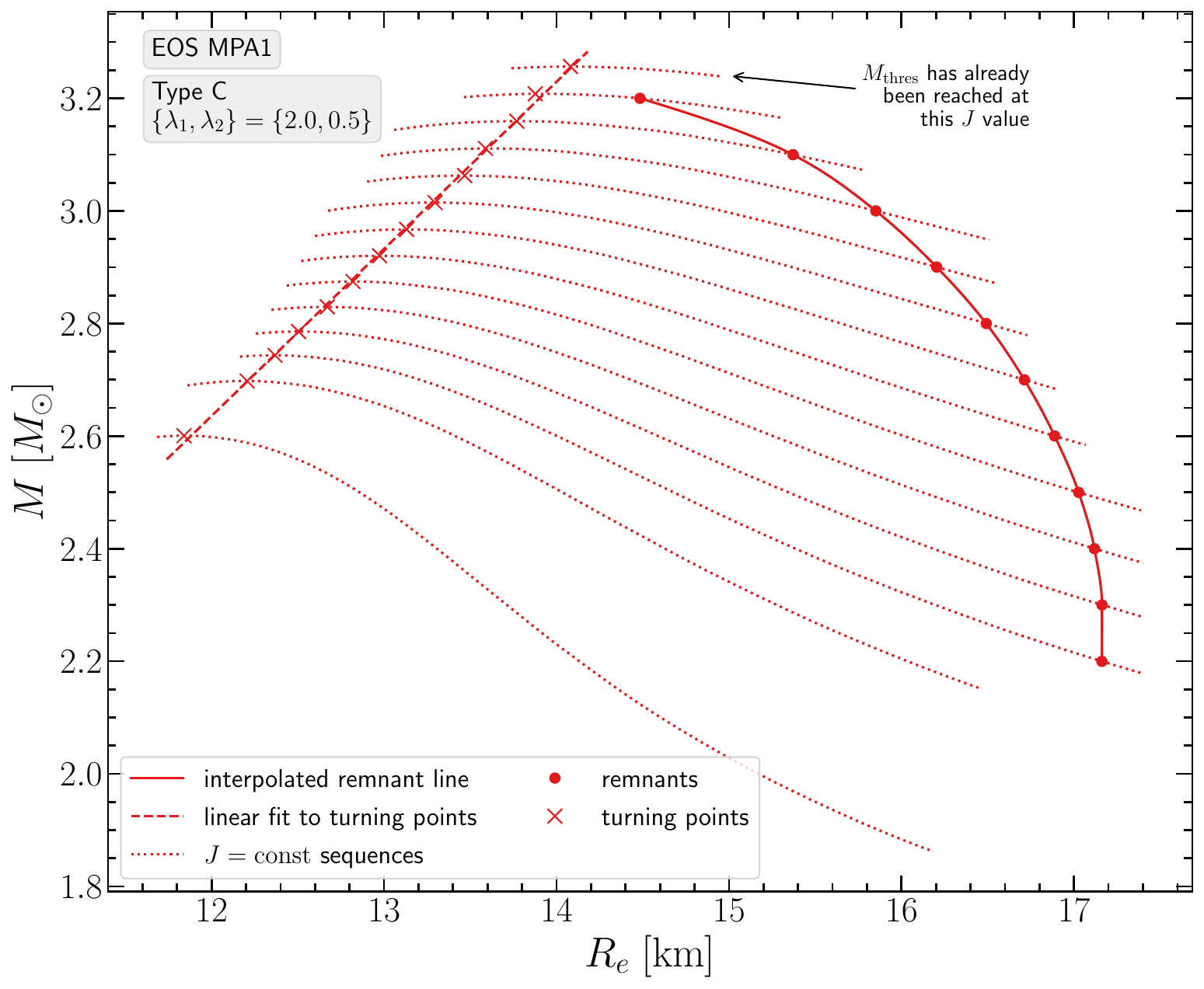}
    \caption{Same as Figure~\ref{fig:MvR_APR_remnants} for the MPA1 EOS.}
    \label{fig:MvR_MPA1_remnants}
\end{figure}

\begin{figure}
        \includegraphics[width=0.93\columnwidth]{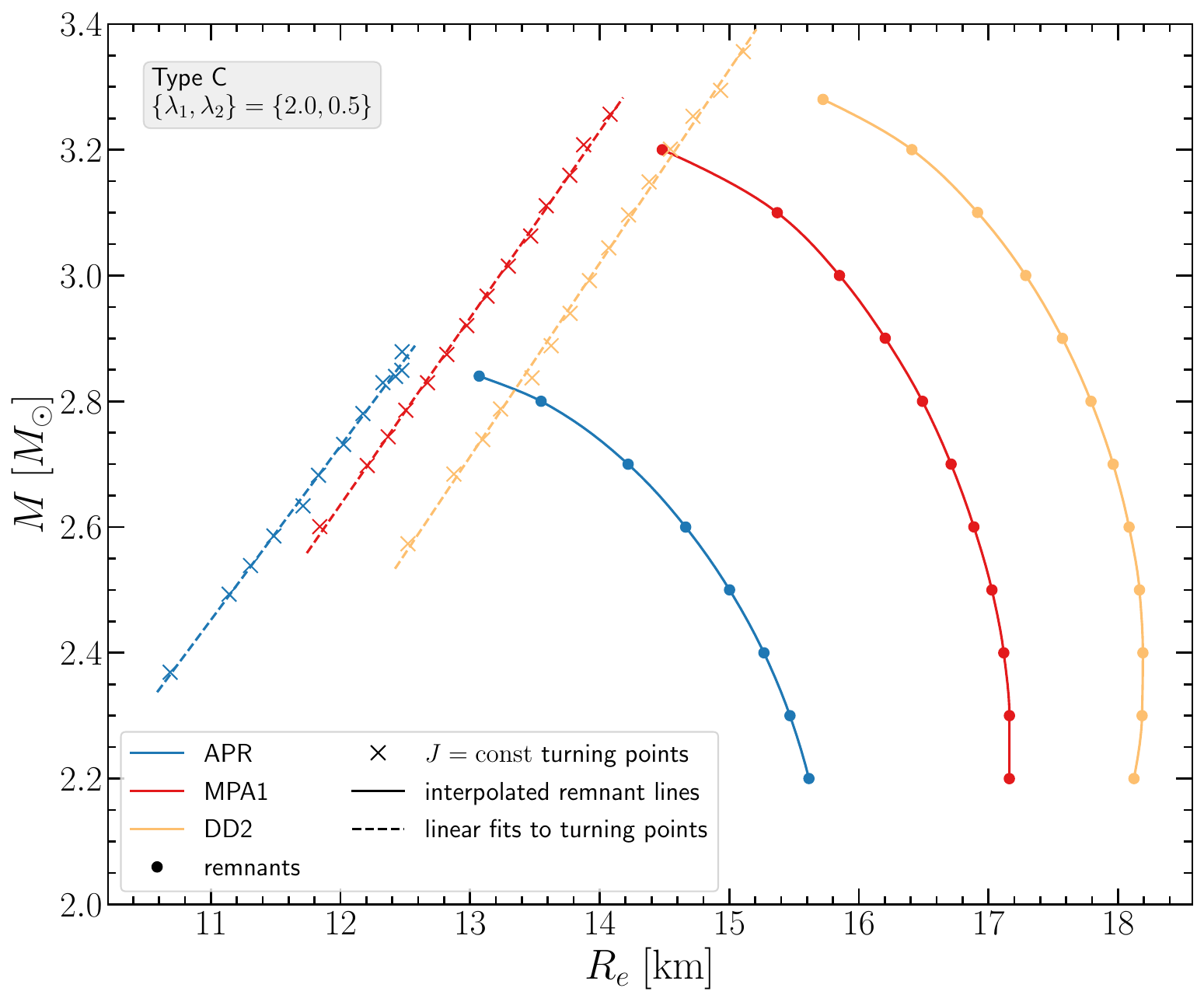}
    \caption{Same as Figure~\ref{fig:MvR_APR_remnants}, showing collectively remnants and turning points for all EOS considered.}
    \label{fig:MvR_all_remnants}
\end{figure}

\subsection{Mass vs. equatorial radius} 

Having presented the main characteristics of our models in the previous subsections, we further analyze our findings by constructing $M(R_e)$ plots (i.e. gravitational mass versus equatorial circumferential radius) for the remnant models. We focus on the choice $\{\lambda_1, \lambda_2\}= \{2.0, 0.5\}$ that represents the strongest differential rotation we consider, but note that a similar picture holds for the case $\{\lambda_1, \lambda_2\}= \{1.5, 0.5\}$.

Figures~\ref{fig:MvR_APR_remnants}, \ref{fig:MvR_DD2_remnants} and \ref{fig:MvR_MPA1_remnants} show $M(R_e)$ for the EOS APR, DD2 and MPA1 respectively. In each of these figures, remnant sequences are shown as dots (denoting the equilibrium models), connected by solid interpolated lines. The dotted lines represent $J$-constant sequences, turning points are marked by crosses and a linear fit (dashed line) approximates the turning point sequence. The annotation in these figures means that while the empirical relation \eqref{eq:LSGF20eqn1} by \citet{Lucca_etal_2021} provides a predicted $J_\text{merger}$ value for a desired target value $M_\text{tot}$, the intersection of the remnant sequence and the turning point sequence has already taken place. Therefore, the target $M_\text{tot}$ model for the specific $J_\text{merger}$ value predicted by \eqref{eq:LSGF20eqn1}, does not exist, since it would exceed the value of $M_\text{thres}$.

\begin{table}
        \centering
        \caption{Coefficients of the linear fits $M = a_1 R_e - b_1$ and their respective errors, for the turning point sequences with $\{\lambda_1, \lambda_2\}=\{2.0, 0.5\}$ of each EOS (Figures~\ref{fig:MvR_APR_remnants}, \ref{fig:MvR_DD2_remnants} and \ref{fig:MvR_MPA1_remnants}). The errors in the coefficients of the linear fits, $\delta a_1$ and $\delta b_1$, are calculated with the standard formulas of simple linear regression and correspond to uncertainties at the $1\sigma$ level.}
        \label{tab:MvR_TP_linear_fits_coeffs}
        \begin{tabular}{ccccc}
                \hline
                 EOS & $a_1$ & $b_1$ & $\delta a_1$ & $\delta b_1$\\
                \hline
                APR & 0.2769 & 0.5936 & 0.0048 & 0.0574\\
                DD2 & 0.3077 & 1.2876 & 0.0040 & 0.0564\\
				MPA1 & 0.2969 & 0.9268 & 0.0029 & 0.0382\\ 			             
                \hline        
        \end{tabular}
\end{table}

For the turning point sequences, Table~\ref{tab:MvR_TP_linear_fits_coeffs} lists the coefficients $a_1, b_1$ for a linear fit of the form
\begin{equation}
M = a_1 R_e - b_1,
\label{eq:MR_TP_linfit}
\end{equation}
as well as their respective errors (with uncertainties at the $1\sigma$ level). We find that the three slopes ($a_1$) are comparable, a fact which may be related to the universalities found by \citet{Bozzola_etal_2018} for turning point sequences. Only the slope of the fit for the APR EOS (the softest of the three) differs somewhat from the corresponding slope for the other two EOS. We note that the relative difference in slopes between the linear fits for the APR and DD2 EOS (the stiffest EOS we consider) is only $\sim10\%$, which is noteworthy given that equation \eqref{eq:MR_TP_linfit} correlates a bulk quantity (mass) with a local quantity (radius) that encodes more prominently a dependence on the EOS.

Collecting data from all EOS in a single $M(R_e)$ plot, Figure~\ref{fig:MvR_all_remnants}, one could extrapolate the remnant sequences so that they intersect with the turning point sequences and obtain the intersection $(M_\text{thres}, R_{e-\text{thres}})$. We note that the $M_\text{thres}$ values determined in this way are in good agreement with those reported in Table~\ref{tab:Mthres_eq_comparison}. Still, we regard the values in Table~\ref{tab:Mthres_eq_comparison} as better estimates for $M_\text{thres}^\text{eq}$, since they involve bulk quantities of the star, such as the angular momentum $J$ and the mass $M$, with direct input from numerical simulations, via \eqref{eq:LSGF20eqn1}, to determine $J$. Moreover, the precise determination of $R_{e-\text{thres}}$, in actual remnants is affected by the thermal properties of low-density material and one needs to resort to a particular definition, based on an isodensity surface, where the density has become a certain small fraction of the maximum density. The notion of a "bulk" region and the relative measure introduced in \citet{Kastaun_etal_2016} is relevant in this respect.

\subsection{Mass vs. equatorial compactness and a criterion for prompt collapse}

\begin{figure}
        \includegraphics[width=0.93\columnwidth]{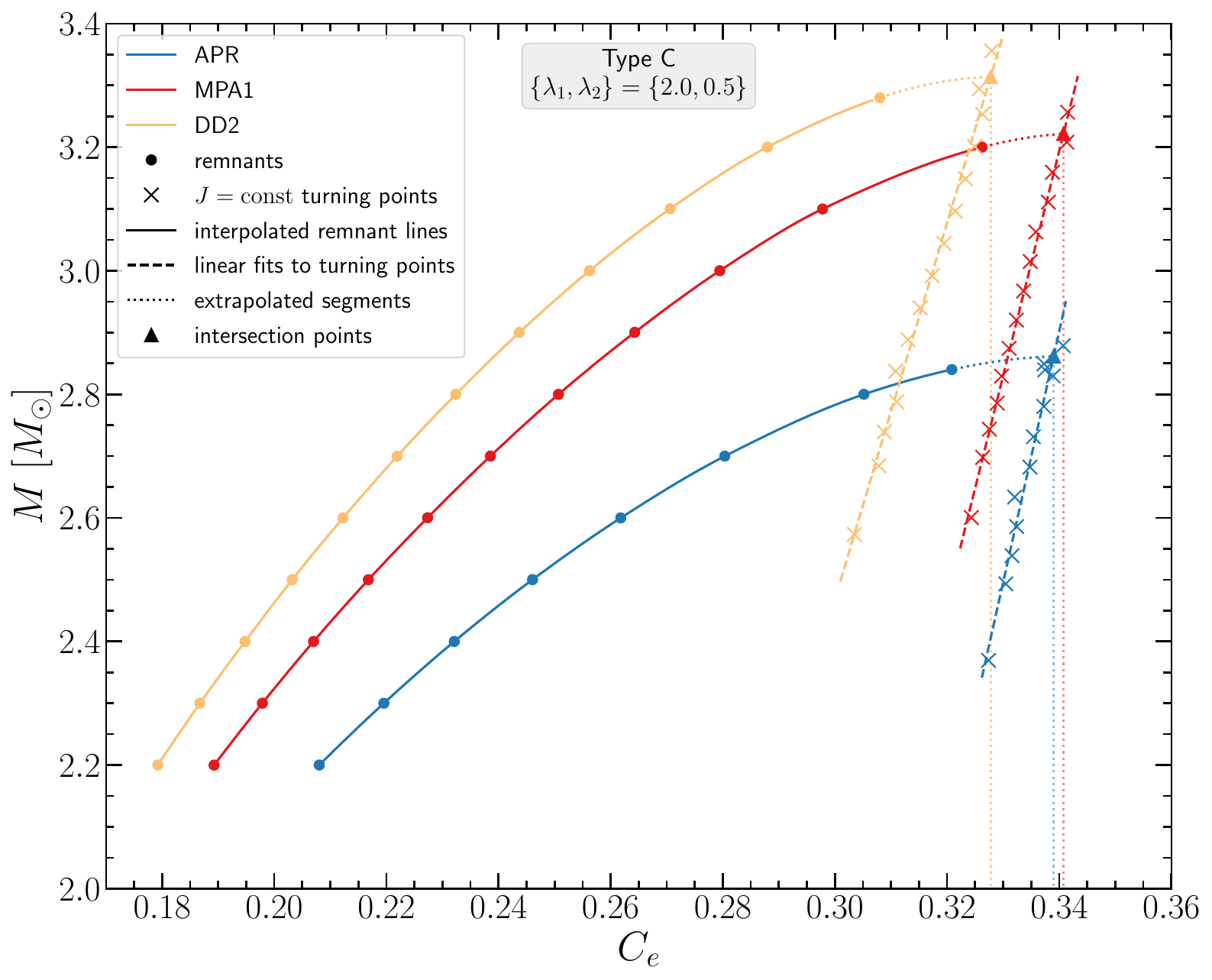}
    \caption{Gravitational mass $M$ vs. equatorial compactness $C_e$ for all EOS considered for the Type C models with $\{\lambda_1, \lambda_2\}= \{2.0, 0.5\}$.}
    \label{fig:MvC_all_EOS}
\end{figure}

The almost identical slope of the turning point lines in the $M(R_e)$ plot of Figure~\ref{fig:MvR_all_remnants}, implies that the threshold mass to collapse will be attained at about the same value of the ratio $M/R_e$, for all three EOS considered. In correspondence to the usual definition of the compactness of a nonrotating star $C=M/R$, we define the ratio $C_e = M/R_e$ as the {\it equatorial compactness}.

Figure~\ref{fig:MvC_all_EOS} displays the $M(C_e)$ relation for the remnant sequences and turning point sequences for the three EOS. The turning point sequences are practically straight lines and we list the coefficients (along with their $1\sigma$ errors) of linear fits of the form
\begin{equation}
M = a_2 C_e - b_2
\label{eq:MC_TP_linfit}
\end{equation}
in Table~\ref{tab:MvC_TP_linear_fits_coeffs}. Dotted lines are extrapolations\footnote{The remnant sequences are interpolated using the package \texttt{PchipInterpolator} from the \texttt{SciPy} library \citep{2020SciPy-NMeth}. This package incorporates the Piecewise Cubic Hermite Interpolating Polynomial (PCHIP) algorithm \citep{PCHIP_ref} and additionally provides an extrapolation option. The latter uses the monotonic cubic spline from the last interval of the data to find the value of the points in the extrapolated range.} of each remnant sequence to its intersection with the corresponding turning point sequence for the same EOS. The intersection of the two sequences for each EOS marks the value $C_{e-{\rm thres}}$ for the model at the threshold mass $M_{\rm thres}$. These values are also listed in Table~\ref{tab:MvC_TP_linear_fits_coeffs} along with the maximum compactness $C_{\rm max}^{\rm TOV}$ of stable nonrotating models, for the same EOS. 

At present, we only list three data points and it is not possible to decide whether the data imply a nearly constant $C_{e-{\rm thres}}\sim 1/3$ or a strong correlation between $C_{e-{\rm thres}}$ and $C_{\rm max}^{\rm TOV}$. A linear fit of the current data yields
\begin{equation}
    C_{e-{\rm thres}} \simeq 0.48 C_{\rm max}^{\rm TOV} + 0.184,
\end{equation}
but additional EOS will be required to clarify whether the equatorial compactness at the threshold mass to prompt collapse, $C_{e-{\rm thres}}$, is a universal value, or whether it is strongly correlated with the maximum compactness $C_{\rm max}^{\rm TOV}$ of stable nonrotating models. Either way, our results imply that a criterion for prompt collapse can be formulated, using the equatorial compactness of equilibrium models of BNS remnants. 

It is interesting that in simulations with the SFHo EOS by \citet{Kastaun_Ohme_2021} a ``bulk'' compactness of 0.31 for the remnant 1ms before collapse is reported, with $C_\text{max}^{\rm TOV}=0.295$ for the maximum mass TOV model for this specific EOS.
In \citet{Kastaun_etal_2016} it was found that the core of the remnant has a mass profile that resembles that of a TOV solution, with similar findings echoed by \citet{Ciolfi_etal_2017}. The latter study went on to suggest the conjecture that merger remnants that do not admit a TOV core equivalent\footnote{The TOV core equivalent, is defined as the “bulk” of a TOV star, meaning the region enclosed by the isodensity surface with the maximum compactness $C_V=M_b / R_V$, where $V$ is the total enclosed proper 3-volume by the isodensity surface, $M_b$ is the respective total baryon mass and $R_V$ is a volumetric radius defined through $V = \frac{4}{3} \pi R_V^3$. See \citet{Kastaun_etal_2016} for further details.} promptly collapse to a black hole. It will be interesting to explore these prospects with equilibrium modelling in future work, now that we have provided proof of concept that the Uryu+ law can capture the $\Omega \sim \omega$ feature in the core of remnants, as it has been observed in simulations.

\begin{table}
        \centering
        \caption{Coefficients of the linear fits $M = a_2 C - b_2$ and their respective errors, for the turning point sequences with $\{\lambda_1, \lambda_2\}=\{2.0, 0.5\}$ of each EOS (Figure~\ref{fig:MvC_all_EOS}). The errors in the coefficients of the linear fits, $\delta a_2$ and $\delta b_2$, are calculated with the standard formulas of simple linear regression and correspond to uncertainties at the $1\sigma$ level. Values for $C_{e-\text{thres}}$ at the intersection points are also listed, together with the corresponding values for $C_\text{max}^{\rm TOV}$ of the maximum mass TOV model for each EOS.}
        \label{tab:MvC_TP_linear_fits_coeffs}
        \begin{tabular}{ccccccc}
                \hline
                 EOS & $a_2$ & $b_2$ & $\delta a_2$ & $\delta b_2$ & $C_{e-\text{thres}}$ & $C_\text{max}^{\rm TOV}$\\
                \hline
                APR & 40.6218 & 10.9095 & 2.8194 & 0.9435 & 0.339 & 0.326\\
                DD2 & 30.4530 & 6.6678 & 0.8999 & 0.2854 & 0.328 & 0.300\\
				MPA1 & 36.3972 & 9.1811 & 0.7741 & 0.2579 & 0.341 & 0.321\\ 			             
                \hline        
        \end{tabular}
\end{table}

\section{Discussion and outlook}
\label{sec:discussion}

In this study, we followed up on our recent work \citep{Iosif_Stergioulas_2021} and expanded our investigation of relativistic equilibrium models, using the 4-parameter differential rotation law of \citet{Uryu_etal_2017} and cold, tabulated EOS of high-density matter. We constructed sequences of merger remnants, taking advantage of empirical relations for the angular momentum at merger, derived through numerical simulations. Comparing the rotational profiles $\Omega(r_c)$, the $\Omega(F)$ profiles, the stellar surfaces and the meridional rest mass density profiles of our remnant models between the three EOS we employed, we found that all respective profiles studied are qualitatively similar when the rotation law parameters are held fixed to different values. Furthermore, choices for the rotation law parameters that yield the \textit{same} Type of solutions (i.e. either Type A or C) also show qualitative similarities in profiles even when varying the parameter $\lambda_1$.

In addition, we constructed constant angular momentum sequences. From the intersection of the line connecting the turning points of $J$-constant sequences and the merger remnant sequence, we were able to reproduce the threshold mass to prompt collapse, $M_{\rm thres}$ with a relative difference of only $\sim 1\%$, compared to accurate binary neutron star merger simulations. We stress that this was achieved using the same rotation law, i.e. the same values for the rotation law parameters $\{\lambda_1, \lambda_2\}$ for all three EOS used in this study. 

Our investigation of the threshold mass to collapse, using equilibrium models, also points towards a possible connection between the equatorial compactness $C_e = M/R_e$ at the threshold and the maximum compactness of nonrotating models $C_\text{max}^\text{TOV}$. This lends support to the conjecture by \citet{Ciolfi_etal_2017}, that merger remnants collapse if their relatively cold and slowly rotating inner region does not admit a stable TOV equivalent. 

Another key prediction of binary neutron star merger simulations is a relatively slowly rotating inner region, where the angular velocity $\Omega$ (as measured by an observer at infinity) is mostly due to the frame dragging angular velocity $\omega$. In our investigation of the parameter space of the Uryu+ rotation law, we naturally find quasi-spherical (Type A) remnant models with this property. Both the density distribution and the angular velocity profile of these models have striking similarities with merger remnants produced in simulations (except for the low-density regions, since we neglect thermal effects).

In a forthcoming study, we plan to take the next step in this program and construct equilibrium models with finite-temperature EOS, using temperature and electron fraction profiles extracted from simulations of BNS mergers. This should allow us to isolate the effect of the thermal state of the remnant on the key properties that we discuss above. Based on \citet{Kaplan_etal_2014, Camelio_etal_2019, Camelio_etal_2020, 2020MNRAS.497.5480C} we do not expect dramatic deviations for bulk properties of the remnants (such as their mass and angular momentum) when thermal effects are included, whereas the corresponding angular velocity profiles could exhibit a larger effect depending on the different thermal treatments \citep{Raithel_etal_2021}. One can still expect an increased radius, due to the added thermal support (see e.g. \citealt{Koliogiannis_Moustakidis_2021}), and consequences on longer timescales, such as convective instabilities.

In our study, we construct models of merger remnants that have angular momentum equal to the angular momentum at the time of merger, as extracted from simulations. Due to the excitation of non-axisymmetric oscillations in the remnant (as well as non-axisymmetric features, such as spiral arms), the GW emission in the post-merger phase will result in a reduction of the angular momentum with time. The remnant's evolutionary path due to various dissipative processes taking place, could be tracked by constructing evolutionary sequences of equilibrium models. To a first approximation, one could neglect mass losses and construct evolutionary sequences keeping the rest mass of the remnant, $M_0$, fixed in time.

\begin{figure}
        \includegraphics[width=0.93\columnwidth]{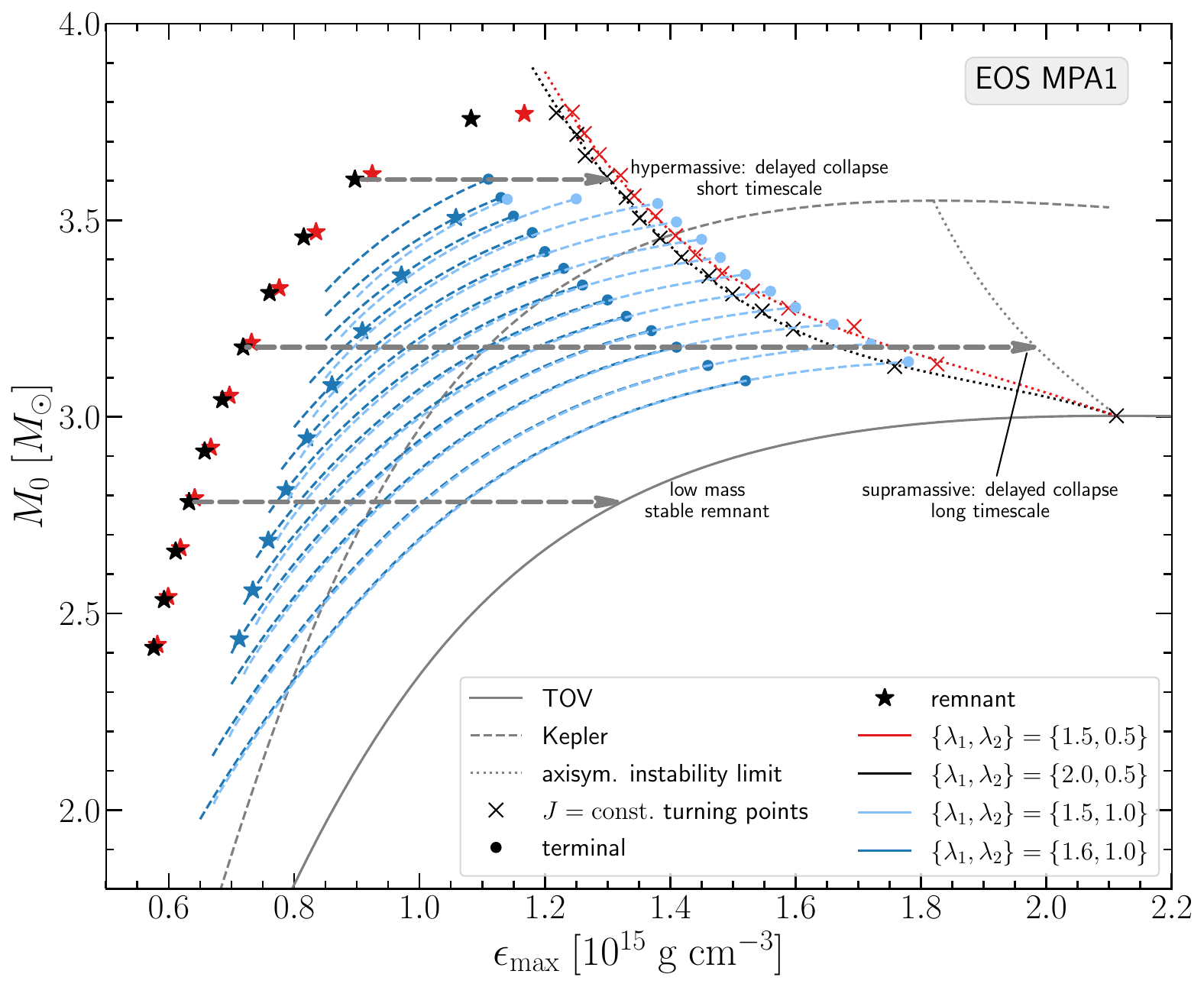}
    \caption{Rest mass $M_0$ vs. maximum energy density $\epsilon_\mathrm{max}$ for the MPA1 EOS. The different sequences are as in Fig. \ref{fig:MPA1_Mass_emax_TypeA}. The grey dashed arrows indicate possible evolutionary paths of merger remnants as they lose angular momentum, if mass loss is neglected.}
    \label{fig:M0_v_emax_MPA1_remnant_evolution}
\end{figure}

With this viewpoint and following the related discussion in \citet{Kaplan_etal_2014}, we show in Figure~\ref{fig:M0_v_emax_MPA1_remnant_evolution} the rest mass $M_0$ versus the maximum energy density $\epsilon_\text{max}$ for the MPA1 EOS and the rotation law with $\{\lambda_1, \lambda_2\}=\{1.6, 1.0\}$. Analogously to Figure~\ref{fig:MPA1_Mass_emax_TypeA}, the blue dashed lines represent constant angular momentum sequences, with sequences for $J=\{3.0, 3.5, 4.0\}$ as well as those reported in Table~\ref{tab:MPA1_remnants_physical_quantities}.

We note that no remnant models are located for these lower $J$-constant sequences, since their lower masses (implying a binary total mass $M_\text{tot} < 2.2 M_\odot$) would make them irrelevant for this study. The arrows in Figure~\ref{fig:M0_v_emax_MPA1_remnant_evolution} approximate the evolutionary path of remnants as they lose angular momentum (assuming constant rest mass). A remnant will evolve towards larger densities and higher compactness, as its angular momentum decreases. In addition, the rotation law may also evolve in time. Three representative evolutionary paths are shown, resulting in a delayed collapse to a black hole on a short timescale (when the remnant is hypermassive); a delayed collapse on a longer timescale (when the remnant is supramassive and can still exist as a uniformly rotating model); and a stable remnant (when its mass is smaller than the maximum mass for nonrotating models). 

The description of post-merger remnants as quasi-equilibrium models has applications in the interpretation of the post-merger GW spectrum; in the study of the threshold mass to prompt collapse; in the construction of universal or empirical relations between properties of the remnants and properties of nonrotating models; and in modeling of processes taking place on longer timescales that are relevant for multi-messenger follow-up studies of GW detections \citep{ciolfi_etal_2021_multimessenger, rosati_etal_2021_synergies}. We plan to elaborate further on these aspects in future work.

\begingroup
\setlength{\tabcolsep}{4.2pt}
\begin{table*}
        \centering
        \caption{Physical quantities for equilibrium remnant models constructed with the APR EOS and different choices of rotation law parameters. The columns show the polar to equatorial axis ratio $r_p/r_e$, the central energy density $\epsilon_c$, the maximum energy density $\epsilon_\text{max}$, the gravitational mass $M$, the rest mass $M_0$, the angular momentum $J$, the ratio of the rotational kinetic energy $T$ over the absolute value of the gravitational binding energy $|W|$, the angular velocity at the rotation axis $\Omega_c$, the maximum value of angular velocity $\Omega_\text{max}$, the angular velocity at the equator $\Omega_e$, the angular velocity of a free particle in circular orbit at the equator $\Omega_K$, the circumferential radius $R_e$, the coordinate radius $r_e$ at the equator and the 3-dimensional general relativistic virial index GRV3.}
        \label{tab:APR_remnants_physical_quantities}
        \begin{tabular}{cccccccccccccccc}
                \hline
                 $r_p/r_e$ & $\epsilon_c (\times 10^{15})$ & $\epsilon_\text{max} (\times 10^{15})$ & $M$ & $M_0$ & $J$ & $T/|W|$ & $\Omega_c$ & $\Omega_\text{max}$ & $\Omega_e$ & $\Omega_K$ & $R_e$ & $r_e$ & GRV3\\
                \{$\lambda_1$, $\lambda_2$\} & $[\mathrm{g \, cm^{-3}}]$ & $[\mathrm{g \, cm^{-3}}]$ & $[M_\odot]$ & $[M_\odot]$ & $[\frac{G M_{\odot}^2} {c}]$ & & [rad/ms] & [rad/ms] & [rad/ms] & [rad/ms] & [km] & [km] & $(\times 10^{-5})$\\
                \hline
                \{2.0, 0.5\} \\
                0.2981 & 0.543 & 0.7788 & 2.20 & 2.4297 & 4.3365 & 0.2253 & 9.9765 & 19.9530 & 4.9882 & 9.1037 & 15.6150 & 11.8159 & 7.6519 \\
                0.2952 & 0.563 & 0.8138 & 2.30 & 2.5558 & 4.6733 & 0.2258 & 10.4436 & 20.8873 & 5.2218 & 9.4001 & 15.4686 & 11.4730 & 6.4827 \\
                0.2950 & 0.592 & 0.8556 & 2.40 & 2.6847 & 5.0105 & 0.2257 & 10.9829 & 21.9658 & 5.4914 & 9.7420 & 15.2677 & 11.0720 & 7.6786 \\
                0.2976 & 0.632 & 0.9074 & 2.50 & 2.8175 & 5.3469 & 0.2250 & 11.6224 & 23.2451 & 5.8112 & 10.1461 & 15.0024 & 10.6012 & 6.8382 \\
                0.3027 & 0.685 & 0.9736 & 2.60 & 2.9540 & 5.6820 & 0.2239 & 12.4015 & 24.8030 & 6.2007 & 10.6314 & 14.6637 & 10.0506 & 6.3562 \\
                0.3100 & 0.758 & 1.0659 & 2.70 & 3.0956 & 6.0176 & 0.2223 & 13.4186 & 26.8376 & 6.7093 & 11.2489 & 14.2197 & 9.3826 & 7.4427 \\
                0.3204 & 0.879 & 1.2256 & 2.80 & 3.2440 & 6.3517 & 0.2203 & 15.0118 & 30.0244 & 7.5059 & 12.1789 & 13.5483 & 8.4607 & 8.7865 \\
                0.3261 & 0.976 & 1.3619 & 2.84 & 3.3058 & 6.4839 & 0.2194 & 16.2264 & 32.4557 & 8.1132 & 12.8548 & 13.0703 & 7.8583 & 7.9165 \\
				\{1.5, 0.5\} \\
				0.3806 & 0.720 & 0.7911 & 2.20 & 2.4383 & 4.3365 & 0.2229 & 10.6528 & 15.9793 & 5.3264 & 9.2112 & 15.4348 & 11.6363 & 7.8407 \\
				0.3795 & 0.748 & 0.8280 & 2.30 & 2.5648 & 4.6734 & 0.2233 & 11.1264 & 16.6896 & 5.5632 & 9.5197 & 15.2823 & 11.2862 & 6.3160 \\
				0.3807 & 0.784 & 0.8722 & 2.40 & 2.6946 & 5.0105 & 0.2231 & 11.6700 & 17.5050 & 5.8350 & 9.8752 & 15.0766 & 10.8789 & 7.9935 \\
				0.3843 & 0.831 & 0.9276 & 2.50 & 2.8286 & 5.3469 & 0.2221 & 12.3134 & 18.4701 & 6.1567 & 10.2962 & 14.8070 & 10.4015 & 7.1972 \\
				0.3897 & 0.892 & 0.9985 & 2.60 & 2.9658 & 5.6820 & 0.2206 & 13.0935 & 19.6402 & 6.5467 & 10.7975 & 14.4685 & 9.8486 & 7.1570 \\
				0.3970 & 0.978 & 1.0986 & 2.70 & 3.1081 & 6.0175 & 0.2187 & 14.1200 & 21.1800 & 7.0600 & 11.4392 & 14.0226 & 9.1754 & 7.4033 \\
				0.4067 & 1.130 & 1.2790 & 2.80 & 3.2566 & 6.3517 & 0.2165 & 15.7837 & 23.6757 & 7.8919 & 12.4311 & 13.3304 & 8.2262 & 8.4743 \\
				0.4119 & 1.281 & 1.4642 & 2.84 & 3.3179 & 6.4839 & 0.2156 & 17.2826 & 25.9241 & 8.6413 & 13.2718 & 12.7567 & 7.5147 & 9.6980 \\
                \{1.6, 1.0\} \\
				0.4669 & 1.0128 & 1.0128 & 2.20 & 2.4567 & 4.3365 & 0.2052 & 6.4676 & 10.3482 & 6.4676 & 7.9280 & 16.8809 & 13.1466 & 9.4359 \\
				0.4813 & 1.0583 & 1.0583 & 2.30 & 2.5862 & 4.6733 & 0.2056 & 6.7609 & 10.8175 & 6.7609 & 8.5470 & 16.2804 & 12.3306 & 9.2157 \\
				0.4930 & 1.1154 & 1.1154 & 2.40 & 2.7188 & 5.0105 & 0.2051 & 7.0683 & 11.3092 & 7.0683 & 9.1070 & 15.7973 & 11.6330 & 9.5772 \\
				0.5051 & 1.1883 & 1.1883 & 2.50 & 2.8551 & 5.3468 & 0.2038 & 7.4161 & 11.8658 & 7.4161 & 9.6956 & 15.3167 & 10.9335 & 9.8606 \\
				0.5185 & 1.2862 & 1.2862 & 2.60 & 2.9961 & 5.6820 & 0.2018 & 7.8342 & 12.5347 & 7.8342 & 10.3614 & 14.7943 & 10.1838 & 11.8614 \\
				0.5334 & 1.4320 & 1.4320 & 2.70 & 3.1421 & 6.0175 & 0.1996 & 8.3892 & 13.4227 & 8.3892 & 11.1812 & 14.1770 & 9.3242 & 12.3628 \\    
                \hline
        \end{tabular}
\end{table*}
\endgroup

\begingroup
\setlength{\tabcolsep}{4.2pt}
\begin{table*}
        \centering
        \caption{Physical quantities for equilibrium remnant models constructed with the DD2 EOS and different choices of rotation law parameters. The different quantities are defined as in Table~\ref{tab:APR_remnants_physical_quantities}.}
        \label{tab:DD2_remnants_physical_quantities}
        \begin{tabular}{cccccccccccccccc}
                \hline
                 $r_p/r_e$ & $\epsilon_c (\times 10^{15})$ & $\epsilon_\text{max} (\times 10^{15})$ & $M$ & $M_0$ & $J$ & $T/|W|$ & $\Omega_c$ & $\Omega_\text{max}$ & $\Omega_e$ & $\Omega_K$ & $R_e$ & $r_e$ & GRV3\\
                \{$\lambda_1$, $\lambda_2$\} & $[\mathrm{g \, cm^{-3}}]$ & $[\mathrm{g \, cm^{-3}}]$ & $[M_\odot]$ & $[M_\odot]$ & $[\frac{G M_{\odot}^2} {c}]$ & & [rad/ms] & [rad/ms] & [rad/ms] & [rad/ms] & [km] & [km] & $(\times 10^{-5})$\\
                \hline
                \{2.0, 0.5\} \\
                0.3208 & 0.3812 & 0.4983 & 2.20 & 2.4009 & 4.4513 & 0.2172 & 7.6442 & 15.2884 & 3.8221 & 7.3691 & 18.1234 & 14.3799 & 8.3706 \\
                0.3094 & 0.3789 & 0.5101 & 2.30 & 2.5198 & 4.8401 & 0.2198 & 7.8595 & 15.7190 & 3.9298 & 7.4853 & 18.1858 & 14.2529 & 6.9913 \\
                0.3015 & 0.3807 & 0.5240 & 2.40 & 2.6406 & 5.2273 & 0.2215 & 8.0980 & 16.1960 & 4.0490 & 7.6254 & 18.1922 & 14.0690 & 7.7354 \\
                0.2952 & 0.3847 & 0.5396 & 2.50 & 2.7630 & 5.6204 & 0.2228 & 8.3557 & 16.7114 & 4.1778 & 7.7787 & 18.1660 & 13.8509 & 10.1183 \\
                0.2921 & 0.3935 & 0.5581 & 2.60 & 2.8877 & 6.0109 & 0.2234 & 8.6459 & 17.2919 & 4.3230 & 7.9591 & 18.0857 & 13.5768 & 6.9961 \\
                0.2911 & 0.4060 & 0.5795 & 2.70 & 3.0146 & 6.4027 & 0.2236 & 8.9700 & 17.9401 & 4.4850 & 8.1617 & 17.9628 & 13.2582 & 6.9080 \\
                0.2922 & 0.4230 & 0.6051 & 2.80 & 3.1443 & 6.7954 & 0.2234 & 9.3389 & 18.6779 & 4.6694 & 8.3936 & 17.7916 & 12.8879 & 7.8284 \\
                0.2952 & 0.4447 & 0.6359 & 2.90 & 3.2768 & 7.1887 & 0.2229 & 9.7627 & 19.5254 & 4.8813 & 8.6588 & 17.5710 & 12.4645 & 6.3214 \\
                0.2999 & 0.4726 & 0.6745 & 3.00 & 3.4125 & 7.5817 & 0.2220 & 10.2658 & 20.5315 & 5.1329 & 8.9707 & 17.2880 & 11.9735 & 6.1169 \\
                0.3065 & 0.5100 & 0.7266 & 3.10 & 3.5524 & 7.9741 & 0.2209 & 10.8964 & 21.7932 & 5.4482 & 9.3560 & 16.9163 & 11.3854 & 7.7065 \\
                0.3145 & 0.5630 & 0.8044 & 3.20 & 3.6962 & 8.3673 & 0.2196 & 11.7581 & 23.5171 & 5.8791 & 9.8656 & 16.4095 & 10.6486 & 7.2057 \\
                0.3226 & 0.6410 & 0.9276 & 3.28 & 3.8156 & 8.6800 & 0.2188 & 12.9607 & 25.9233 & 6.4803 & 10.5450 & 15.7235 & 9.7506 & 7.2275 \\
				\{1.5, 0.5\} \\
				0.3962 & 0.4716 & 0.5025 & 2.20 & 2.4074 & 4.4513 & 0.2142 & 8.1481 & 12.2221 & 4.0740 & 7.4195 & 17.9799 & 14.2399 & 7.0193 \\
				0.3884 & 0.4773 & 0.5151 & 2.30 & 2.5270 & 4.8401 & 0.2171 & 8.3792 & 12.5687 & 4.1896 & 7.5467 & 18.0227 & 14.0929 & 8.6483 \\
				0.3833 & 0.4860 & 0.5299 & 2.40 & 2.6485 & 5.2272 & 0.2190 & 8.6292 & 12.9438 & 4.3146 & 7.6968 & 18.0142 & 13.8936 & 7.0557 \\
				0.3799 & 0.4970 & 0.5468 & 2.50 & 2.7723 & 5.6203 & 0.2203 & 8.9002 & 13.3503 & 4.4501 & 7.8635 & 17.9698 & 13.6556 & 8.5267 \\
				0.3784 & 0.5110 & 0.5663 & 2.60 & 2.8973 & 6.0109 & 0.2209 & 9.1956 & 13.7934 & 4.5978 & 8.0505 & 17.8832 & 13.3750 & 9.3444 \\
				0.3789 & 0.5290 & 0.5895 & 2.70 & 3.0256 & 6.4027 & 0.2210 & 9.5275 & 14.2913 & 4.7638 & 8.2655 & 17.7494 & 13.0433 & 6.7115 \\
				0.3808 & 0.5510 & 0.6168 & 2.80 & 3.1562 & 6.7955 & 0.2206 & 9.9002 & 14.8504 & 4.9501 & 8.5067 & 17.5734 & 12.6666 & 7.3657 \\
				0.3839 & 0.5778 & 0.6495 & 2.90 & 3.2889 & 7.1888 & 0.2199 & 10.3245 & 15.4868 & 5.1623 & 8.7788 & 17.3533 & 12.2428 & 7.5190 \\
				0.3887 & 0.6130 & 0.6916 & 3.00 & 3.4258 & 7.5817 & 0.2188 & 10.8354 & 16.2531 & 5.4177 & 9.1050 & 17.0641 & 11.7429 & 6.8216 \\
				0.3946 & 0.6600 & 0.7482 & 3.10 & 3.5656 & 7.9741 & 0.2176 & 11.4737 & 17.2106 & 5.7369 & 9.5030 & 16.6920 & 11.1526 & 7.5437 \\
				0.4020 & 0.7330 & 0.8376 & 3.20 & 3.7105 & 8.3673 & 0.2161 & 12.3874 & 18.5812 & 6.1937 & 10.0535 & 16.1580 & 10.3829 & 7.4784 \\
				0.4090 & 0.8670 & 1.0083 & 3.28 & 3.8301 & 8.6800 & 0.2155 & 13.8873 & 20.8311 & 6.9437 & 10.9019 & 15.3236 & 9.3187 & 8.2883 \\
                \{1.6, 1.0\} \\
                0.4828 & 0.6261 & 0.6261 & 2.20 & 2.4207 & 4.4513 & 0.1992 & 5.0343 & 8.0549 & 5.0343 & 6.5162 & 19.4227 & 15.7321 & 8.5887 \\
				0.4829 & 0.6451 & 0.6451 & 2.30 & 2.5421 & 4.8400 & 0.2019 & 5.1666 & 8.2666 & 5.1666 & 6.7141 & 19.3079 & 15.4256 & 9.8640 \\
				0.4854 & 0.6669 & 0.6670 & 2.40 & 2.6658 & 5.2273 & 0.2034 & 5.3104 & 8.4966 & 5.3104 & 6.9468 & 19.1243 & 15.0469 & 8.9839 \\
				0.4885 & 0.6916 & 0.6916 & 2.50 & 2.7912 & 5.6203 & 0.2045 & 5.4627 & 8.7404 & 5.4627 & 7.1879 & 18.9252 & 14.6504 & 10.0100 \\
				0.4933 & 0.7209 & 0.7209 & 2.60 & 2.9192 & 6.0109 & 0.2048 & 5.6300 & 9.0080 & 5.6300 & 7.4581 & 18.6776 & 14.2023 & 8.8487 \\
				0.4990 & 0.7556 & 0.7556 & 2.70 & 3.0495 & 6.4027 & 0.2045 & 5.8134 & 9.3014 & 5.8134 & 7.7484 & 18.4016 & 13.7228 & 9.1569 \\
				0.5058 & 0.7978 & 0.7978 & 2.80 & 3.1827 & 6.7953 & 0.2037 & 6.0192 & 9.6308 & 6.0192 & 8.0671 & 18.0894 & 13.2030 & 10.1405 \\
				0.5134 & 0.8510 & 0.8510 & 2.90 & 3.3189 & 7.1888 & 0.2026 & 6.2561 & 10.0098 & 6.2561 & 8.4222 & 17.7347 & 12.6358 & 9.5439 \\
				0.5222 & 0.9226 & 0.9226 & 3.00 & 3.4589 & 7.5818 & 0.2010 & 6.5434 & 10.4695 & 6.5434 & 8.8369 & 17.3132 & 11.9945 & 10.9882 \\
				0.5322 & 1.0304 & 1.0304 & 3.10 & 3.6029 & 7.9740 & 0.1992 & 6.9248 & 11.0797 & 6.9248 & 9.3564 & 16.7807 & 11.2302 & 11.8057 \\	    
                \hline
        \end{tabular}
\end{table*}
\endgroup

\begingroup
\setlength{\tabcolsep}{4.2pt}
\begin{table*}
        \centering
        \caption{Physical quantities for equilibrium remnant models constructed with the MPA1 EOS and different choices of rotation law parameters. The different quantities are defined as in Table~\ref{tab:APR_remnants_physical_quantities}.}
        \label{tab:MPA1_remnants_physical_quantities}
        \begin{tabular}{cccccccccccccccc}
                \hline
                 $r_p/r_e$ & $\epsilon_c (\times 10^{15})$ & $\epsilon_\text{max} (\times 10^{15})$ & $M$ & $M_0$ & $J$ & $T/|W|$ & $\Omega_c$ & $\Omega_\text{max}$ & $\Omega_e$ & $\Omega_K$ & $R_e$ & $r_e$ & GRV3\\
                \{$\lambda_1$, $\lambda_2$\} & $[\mathrm{g \, cm^{-3}}]$ & $[\mathrm{g \, cm^{-3}}]$ & $[M_\odot]$ & $[M_\odot]$ & $[\frac{G M_{\odot}^2} {c}]$ & & [rad/ms] & [rad/ms] & [rad/ms] & [rad/ms] & [km] & [km] & $(\times 10^{-5})$\\
                \hline
                \{2.0, 0.5\} \\
               	0.3065 & 0.4260 & 0.5761 & 2.2 & 2.4131 & 4.4180 & 0.2205 & 8.4137 & 16.8274 & 4.2068 & 7.9797 & 17.1615 & 13.3944 & 14.4367 \\
				0.2985 & 0.4285 & 0.5925 & 2.3 & 2.5346 & 4.7867 & 0.2222 & 8.6843 & 17.3686 & 4.3421 & 8.1391 & 17.1622 & 13.2038 & 13.8202 \\
				0.2934 & 0.4350 & 0.6110 & 2.4 & 2.6578 & 5.1565 & 0.2232 & 8.9821 & 17.9643 & 4.4911 & 8.3202 & 17.1186 & 12.9678 & 13.4671 \\
				0.2911 & 0.4465 & 0.6324 & 2.5 & 2.7835 & 5.5263 & 0.2236 & 9.3157 & 18.6314 & 4.6578 & 8.5283 & 17.0270 & 12.6816 & 13.8714 \\
				0.2915 & 0.4641 & 0.6573 & 2.6 & 2.9123 & 5.8965 & 0.2234 & 9.6920 & 19.3842 & 4.8460 & 8.7668 & 16.8878 & 12.3448 & 9.5482 \\
				0.2936 & 0.4861 & 0.6855 & 2.7 & 3.0430 & 6.2685 & 0.2230 & 10.1113 & 20.2226 & 5.0556 & 9.0301 & 16.7122 & 11.9693 & 8.2679 \\
				0.2975 & 0.5140 & 0.7187 & 2.8 & 3.1769 & 6.6425 & 0.2221 & 10.5906 & 21.1815 & 5.2953 & 9.3294 & 16.4912 & 11.5439 & 7.1198 \\
				0.3042 & 0.5510 & 0.7608 & 2.9 & 3.3153 & 7.0144 & 0.2207 & 11.1621 & 22.3244 & 5.5811 & 9.6864 & 16.2037 & 11.0456 & 8.3531 \\
				0.3121 & 0.5970 & 0.8154 & 3.0 & 3.4568 & 7.3878 & 0.2190 & 11.8549 & 23.7102 & 5.9274 & 10.1095 & 15.8508 & 10.4744 & 8.6592 \\
				0.3224 & 0.6610 & 0.8970 & 3.1 & 3.6037 & 7.7617 & 0.2169 & 12.7936 & 25.5878 & 6.3968 & 10.6687 & 15.3706 & 9.7608 & 7.3993 \\
				0.3367 & 0.7900 & 1.0824 & 3.2 & 3.7578 & 8.1352 & 0.2144 & 14.6102 & 29.2249 & 7.3051 & 11.6946 & 14.4834 & 8.5922 & 8.1130 \\
				\{1.5, 0.5\} \\
				0.3866 & 0.5380 & 0.5819 & 2.2 & 2.4205 & 4.4179 & 0.2178 & 8.9666 & 13.4499 & 4.4833 & 8.0489 & 17.0026 & 13.2377 & 15.8326 \\
				0.3816 & 0.5485 & 0.5991 & 2.3 & 2.5424 & 4.7867 & 0.2196 & 9.2482 & 13.8723 & 4.6241 & 8.2177 & 16.9911 & 13.0347 & 12.0241 \\
				0.3788 & 0.5625 & 0.6187 & 2.4 & 2.6666 & 5.1565 & 0.2207 & 9.5562 & 14.3342 & 4.7781 & 8.4102 & 16.9350 & 12.7851 & 12.1648 \\
				0.3784 & 0.5805 & 0.6414 & 2.5 & 2.7934 & 5.5264 & 0.2210 & 9.8964 & 14.8445 & 4.9482 & 8.6295 & 16.8339 & 12.4881 & 12.2902 \\
				0.3795 & 0.6020 & 0.6671 & 2.6 & 2.9221 & 5.8966 & 0.2208 & 10.2697 & 15.4046 & 5.1349 & 8.8718 & 16.6965 & 12.1527 & 10.9684 \\
				0.3824 & 0.6280 & 0.6969 & 2.7 & 3.0541 & 6.2685 & 0.2200 & 10.6889 & 16.0334 & 5.3445 & 9.1453 & 16.5173 & 11.7716 & 8.2365 \\
				0.3866 & 0.6590 & 0.7321 & 2.8 & 3.1888 & 6.6425 & 0.2189 & 11.1632 & 16.7448 & 5.5816 & 9.4523 & 16.2976 & 11.3458 & 7.3561 \\
				0.3927 & 0.6980 & 0.7763 & 2.9 & 3.3274 & 7.0144 & 0.2173 & 11.7225 & 17.5838 & 5.8613 & 9.8135 & 16.0180 & 10.8541 & 7.8691 \\
				0.4004 & 0.7485 & 0.8350 & 3.0 & 3.4700 & 7.3878 & 0.2153 & 12.4110 & 18.6165 & 6.2055 & 10.2491 & 15.6659 & 10.2810 & 9.1894 \\
				0.4099 & 0.8240 & 0.9246 & 3.1 & 3.6177 & 7.7617 & 0.2129 & 13.3604 & 20.0406 & 6.6802 & 10.8311 & 15.1810 & 9.5587 & 7.8279 \\
				0.4228 & 1.0200 & 1.1673 & 3.2 & 3.7707 & 8.1352 & 0.2105 & 15.5006 & 23.2512 & 7.7503 & 12.0506 & 14.1554 & 8.2317 & 9.7803 \\	
                \{1.6, 1.0\} \\
                0.4842 & 0.7127 & 0.7127 & 2.2 & 2.4349 & 4.4180 & 0.2026 & 5.5146 & 8.8233 & 5.5146 & 7.1943 & 18.1611 & 14.4404 & 14.5244 \\
				0.4872 & 0.7343 & 0.7343 & 2.3 & 2.5591 & 4.7867 & 0.2041 & 5.6713 & 9.0741 & 5.6713 & 7.4547 & 17.9811 & 14.0644 & 13.5309 \\
				0.4917 & 0.7588 & 0.7588 & 2.4 & 2.6855 & 5.1565 & 0.2048 & 5.8389 & 9.3422 & 5.8389 & 7.7359 & 17.7683 & 13.6528 & 12.4246 \\
				0.4974 & 0.7872 & 0.7872 & 2.5 & 2.8143 & 5.5264 & 0.2047 & 6.0192 & 9.6308 & 6.0192 & 8.0385 & 17.5258 & 13.2088 & 11.9574 \\
				0.5045 & 0.8206 & 0.8206 & 2.6 & 2.9460 & 5.8966 & 0.2040 & 6.2158 & 9.9452 & 6.2158 & 8.3660 & 17.2534 & 12.7314 & 11.2757 \\
				0.5125 & 0.8604 & 0.8604 & 2.7 & 3.0807 & 6.2686 & 0.2028 & 6.4319 & 10.2911 & 6.4319 & 8.7184 & 16.9557 & 12.2248 & 11.8631 \\
				0.5215 & 0.9088 & 0.9088 & 2.8 & 3.2183 & 6.6425 & 0.2012 & 6.6744 & 10.6790 & 6.6744 & 9.1027 & 16.6285 & 11.6841 & 11.0840 \\
				0.5322 & 0.9712 & 0.9712 & 2.9 & 3.3599 & 7.0144 & 0.1991 & 6.9594 & 11.1351 & 6.9594 & 9.5434 & 16.2476 & 11.0829 & 12.1694 \\
				0.5443 & 1.0582 & 1.0582 & 3.0 & 3.5061 & 7.3878 & 0.1967 & 7.3166 & 11.7066 & 7.3166 & 10.0705 & 15.7915 & 10.3956 & 13.0393 \\   
                \hline
        \end{tabular}
\end{table*}
\endgroup

\section*{Acknowledgements}

We thank Andreas Bauswein, John Friedman, Dimitrios Papadopoulos and Christos Tsagas for helpful discussions and comments. We also thank the anonymous referee for providing useful comments that improved the final manuscript. PI gratefully acknowledges support by a Virgo-EGO Scientific Forum (VESF) PhD fellowship. NS acknowledges support by the COST actions CA16214 “PHAROS”, CA16104 “GWVerse” and CA18108 “QG-MM”. The authors gratefully acknowledge the Italian Instituto Nazionale di Fisica Nucleare (INFN), the French Centre National de la Recherche Scientifique (CNRS) and the Netherlands Organization for Scientific Research, for the construction and operation of the Virgo detector and the creation and support of the EGO consortium. 

\section*{Data Availability}

The data underlying this article are available in the article.



\bibliographystyle{mnras}
\bibliography{references}




\appendix

\section{Comparison of angular momentum empirical relations}
\label{appendix}

In this appendix, we present a more detailed comparison for the empirical relations \eqref{BS17-Jempirical} and \eqref{eq:LSGF20eqn1} discussed in Section~\ref{sec:remn_seq}, that were used to estimate the angular momentum at the time of merger.

As a first step, the \citeauthor{Lucca_etal_2021} relation \eqref{eq:LSGF20eqn1} is cast into the form used in \citeauthor{Bauswein_Stergioulas_2017}, i.e. it is expressed as a function of the total binary mass $M_\text{tot}$ (for equal mass binaries, $M_\text{NS} = M_\text{tot}/2$) 
\begin{equation}
    \frac{c J_{\text{merger}}}{G {M_\odot}^2} = \frac{a_1}{2} \left( \frac{R_\text{NS}}{\frac{G M_\odot}{c^2}} \right) \frac{M_\text{tot}}{M_\odot} + a_2 \left( \frac{R_\text{NS}}{\frac{G M_\odot}{c^2}} \right)^2 \, .
    \label{eq:appdx_LSGF20eqn1_step3}
\end{equation}

\begin{figure}
        \centering   
        \includegraphics[width=0.93\columnwidth]{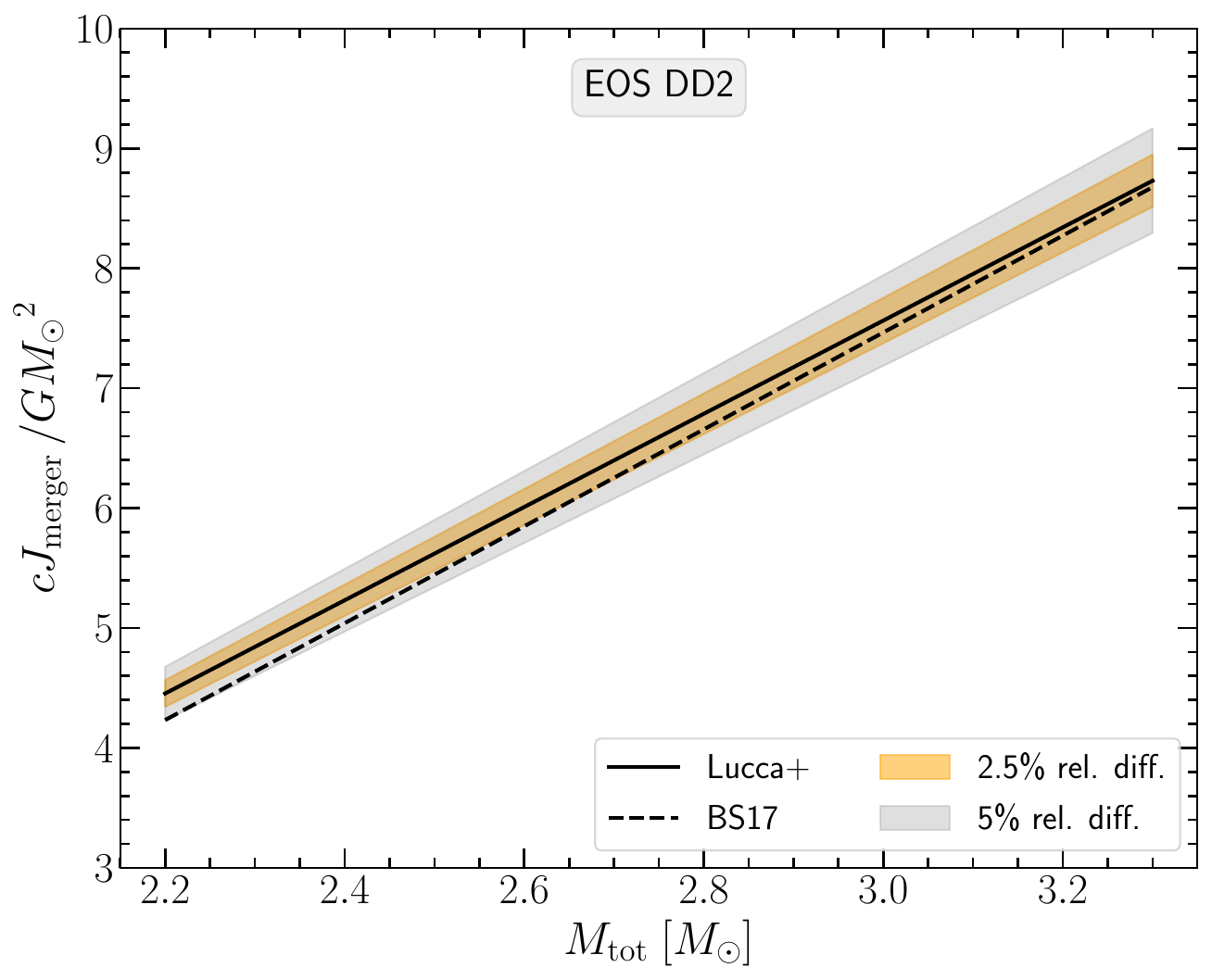}
    \caption{Comparison of empirical relations for the angular momentum at the time of merger, $J_\text{merger}$ vs. the total binary mass, $M_\text{tot}$, for the DD2 EOS. The solid line represents the empirical relation from \citet{Lucca_etal_2021} cast into the form used in \citeauthor{Bauswein_Stergioulas_2017}, i.e equation \eqref{eq:appdx_LSGF20eqn1_step3}, while the dashed line represents the empirical relation from \citet{Bauswein_Stergioulas_2017}, i.e. equation \eqref{BS17-Jempirical}. The orange and grey filled areas denote the relative difference bands at the 2.5\% and 5\% level respectively.}
    \label{fig:J_Mtot_DD2}
\end{figure}

In Figure~\ref{fig:J_Mtot_DD2} the two angular momentum empirical relations \eqref{BS17-Jempirical} and \eqref{eq:appdx_LSGF20eqn1_step3} are shown for the DD2 EOS. An $M_\text{tot}$ ranging from $2.2 M_\odot$ to $3.3 M_\odot$ is assumed. Note that the threshold mass for prompt collapse to a black hole for this particular EOS is $M_\text{thres} \simeq 3.325 M_\odot$ as reported from numerical simulations by \citet[Supplementary Material]{Bauswein_etal_2020}. In equation \eqref{eq:appdx_LSGF20eqn1_step3} we have set $R_\text{NS} = 13.1$ km, which is a typical circumferential radius value for progenitor neutron stars constructed with the DD2 EOS (see Figure~\ref{fig:M_R_plot}). The two curves show very good agreement, 
deviating for the most part at the 2.5\% level and at the 5\% level only for very low mass remnants.

To investigate further the two angular momentum empirical formulas, we consider six zero-temperature, tabulated EOS, namely APR, DD2 and MPA1 already mentioned in Section~\ref{sec:eos} with the addition of SFHx \citep{SFHx_EOS}, SLy4 \citep{Douchin_Haensel_2001} and WFF1 EOS \citep{WFF_EOS}. For each EOS, we consider a remnant mass range starting from $2.2 M_\odot$ and increasing with a step of $0.1 M_\odot$ up to approximately the threshold mass value $M_\text{thres}$, as the latter is reported in \citet{Bauswein_etal_2020}. Constructing the corresponding nonrotating progenitor models with $M_\text{NS}=M_\text{tot} / 2$, we then use the empirical relations \eqref{BS17-Jempirical} and \eqref{eq:LSGF20eqn1} to get predictions for the value of $J_\text{merger}$ from each formula. 

\begin{figure} 
        \centering   
        \includegraphics[width=0.93\columnwidth]{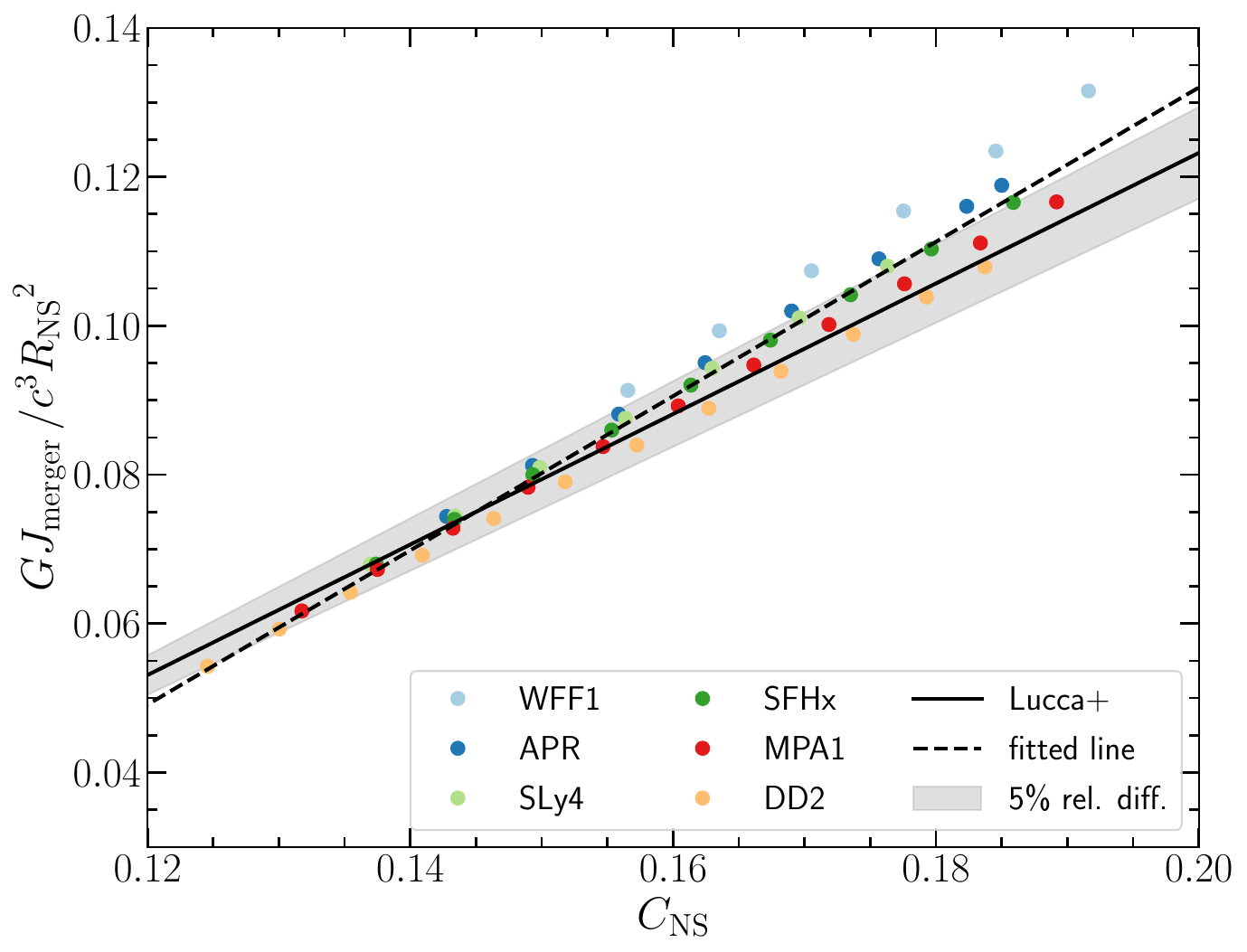}
    \caption{Angular momentum at the time of merger, $J_\text{merger}$, normalized by the TOV progenitor neutron star circumferential radius squared, $R_\text{NS}^2$ vs. the corresponding compactness $C_\text{NS}$ (equal mass binaries are assumed). The filled circles correspond to the data presented in Tables~\ref{tab:Jmerger_comp_APR}--\ref{tab:Jmerger_comp_WFF1}, i.e. they represent $J_\text{merger}$ predictions (normalized by $R_\text{NS}^2$) following the empirical formula in \citet{Bauswein_Stergioulas_2017} for remnants with different EOS and their correlation with the respective TOV progenitors' compactness $C_\text{NS}$. The dashed line is a linear fit to the filled circles data, i.e equation \eqref{eq:fittedline_BS17_in_LSGF_form}, the solid line represents the empirical formula of \citet{Lucca_etal_2021}, i.e. equation \eqref{eq:LSGF20eqn1}, and the grey filled area the 5\% relative difference band.}
    \label{fig:J_norm_R2_v_C}
\end{figure}

Tables~\ref{tab:Jmerger_comp_APR}, \ref{tab:Jmerger_comp_DD2}, \ref{tab:Jmerger_comp_MPA1}, \ref{tab:Jmerger_comp_SFHX}, \ref{tab:Jmerger_comp_SLy} and \ref{tab:Jmerger_comp_WFF1} report the $J_\text{merger}$ estimates from the \citet{Bauswein_Stergioulas_2017} and \citet{Lucca_etal_2021} expressions. For the EOS DD2, MPA1, SFHx and SLy4 excellent agreement is noticed, as the relative difference ranges from $\sim0.5\%$ up to $\sim5\%$. For the APR EOS, somewhat larger deviations ($>6\%$) are noticed for remnants with masses over $2.6 M_\odot$, with the \citet{Bauswein_Stergioulas_2017} approximate formula predicting larger angular momentum values. The picture is similar for the WFF1 EOS, with the deviations augmented to the 10\% level. 

In Figure~\ref{fig:J_norm_R2_v_C}, we compare the two empirical relations again, but this time focusing on the form of the \citeauthor{Lucca_etal_2021} relation. Specifically, using the equilibrium progenitor models' calculated properties and assuming that the $J_\text{merger}$ values follow equation \eqref{BS17-Jempirical}, we investigate the correlation between $GJ_{\text{merger}} \big|_{BS17}  \, / c^3 R^2_{\text{NS}}$ and $C_\text{NS}$. A linear fit of the same form as equation \eqref{eq:LSGF20eqn1} is calculated
\begin{equation}
    \frac{GJ_{\text{merger}} \big|_{BS17} }{c^3 R^2_{\text{NS}}} = b_1 C_{\text{NS}} + b_2 \, ,
    \label{eq:fittedline_BS17_in_LSGF_form}
\end{equation}
with $ b_1 =  1.0358 \pm 0.0265$ and $ b_2 =  -\left(7.517 \pm 0.429 \right) \times 10^{-2} $ ($1\sigma$ credible level). Note that (as expected) the errors in the coefficients of \eqref{eq:LSGF20eqn1} are smaller, compared to our equilibrium model "hybrid" expression \eqref{eq:fittedline_BS17_in_LSGF_form}. That said, both expressions are in very good agreement (at the 5\% level) for the most part. Larger deviations are noticed between the two, only for the softer EOS, but these are gradually disfavored by the tightening astrophysical and nuclear constraints on the EOS of high density matter (see discussion in Section~\ref{sec:eos}).

Overall, the agreement between the empirical formulas from \citet{Bauswein_Stergioulas_2017} and \citet{Lucca_etal_2021} at a level of a few per cent is notable, given that (i) they are independent analyses of different numerical simulations datasets and (ii) the formula appearing in \citeauthor{Bauswein_Stergioulas_2017} was a benchmark for the DD2 EOS with an added approximate treatment to describe other EOS. In fact, since one cannot be certain that the specifics of the angular momentum extraction were performed in an identical way between the two works, the actual agreement could prove to be even better. In any case, an accuracy of a few per cent is adequate to get useful insights into merger remnants properties with equilibrium models.

\begin{table}
	\centering
	\caption{Angular momentum predictions from two different empirical relations, for the APR EOS. The total binary mass $M_\text{tot}$, the  progenitor masses $M_\text{NS} = M_\text{tot}/2$, the equatorial circumferential radii $R_\text{NS}$ and the values of compactness $C_\text{NS}$ are shown. The $J_\text{merger}$ columns correspond to angular momentum predictions from the empirical relations reported in \citet{Bauswein_Stergioulas_2017} (abbreviated as BS17) and \citet{Lucca_etal_2021} (abbreviated as Lucca+). The last column reports the absolute values of the relative difference between the $J_\text{merger}$ predictions. Note that $M_\text{tot}$ is terminated approximately at the $M_\text{thres}$ value, as the latter is reported in \citet{Bauswein_etal_2020}.}
	\label{tab:Jmerger_comp_APR}
	\begin{tabular}{ccccccc}
	\hline
	$M_\text{tot}$ & $M_\text{NS}$ & $R_\text{NS}$ & $C_\text{NS}$ & $\frac{cJ_\text{merger}}{G M_\odot^2}$ & $\frac{cJ_\text{merger}}{G M_\odot^2}$  & rel. diff. \\
	$[M_\odot]$ & $[M_\odot]$ & [km] & & [BS17] & [Lucca+] & [\%] \\
	\hline
	2.20 & 1.10 & 11.3774 & 0.143 & 4.4165 & 4.3365 & 1.85 \\
	2.30 & 1.15 & 11.3735 & 0.149 & 4.8203 & 4.6733 & 3.14 \\
	2.40 & 1.20 & 11.3689 & 0.156 & 5.2248 & 5.0105 & 4.28 \\
	2.50 & 1.25 & 11.3634 & 0.162 & 5.6290 & 5.3469 & 5.28 \\
	2.60 & 1.30 & 11.3567 & 0.169 & 6.0324 & 5.6821 & 6.17 \\
	2.70 & 1.35 & 11.3482 & 0.176 & 6.4372 & 6.0175 & 6.97 \\
	2.80 & 1.40 & 11.3379 & 0.182 & 6.8417 & 6.3517 & 7.71 \\
	2.84 & 1.42 & 11.3331 & 0.185 & 7.0021 & 6.4838 & 7.99 \\			
	\hline        
    \end{tabular}
\end{table}

\begin{table}
	\centering
	\caption{Same as Table~\ref{tab:Jmerger_comp_APR} for the DD2 EOS.}
	\label{tab:Jmerger_comp_DD2}
	\begin{tabular}{ccccccc}
	\hline
	$M_\text{tot}$ & $M_\text{NS}$ & $R_\text{NS}$ & $C_\text{NS}$ & $\frac{cJ_\text{merger}}{G M_\odot^2}$ & $\frac{cJ_\text{merger}}{G M_\odot^2}$ & rel. diff. \\
	$[M_\odot]$ & $[M_\odot]$ & [km] & & [BS17] & [Lucca+] & [\%] \\
	\hline
	2.20 & 1.10 & 13.0385 & 0.125 & 4.2314 & 4.4513 & 4.94 \\
	2.30 & 1.15 & 13.0594 & 0.130 & 4.6361 & 4.8401 & 4.21 \\
	2.40 & 1.20 & 13.0795 & 0.135 & 5.0379 & 5.2273 & 3.62 \\
	2.50 & 1.25 & 13.0985 & 0.141 & 5.4448 & 5.6203 & 3.12 \\
	2.60 & 1.30 & 13.1161 & 0.146 & 5.8481 & 6.0109 & 2.71 \\
	2.70 & 1.35 & 13.1320 & 0.152 & 6.2519 & 6.4027 & 2.36 \\
	2.80 & 1.40 & 13.1460 & 0.157 & 6.6561 & 6.7954 & 2.05 \\
	2.90 & 1.45 & 13.1578 & 0.163 & 7.0607 & 7.1888 & 1.78 \\
	3.00 & 1.50 & 13.1672 & 0.168 & 7.4648 & 7.5818 & 1.54 \\
	3.10 & 1.55 & 13.1743 & 0.174 & 7.8683 & 7.9740 & 1.33 \\
	3.20 & 1.60 & 13.1784 & 0.179 & 8.2734 & 8.3673 & 1.12 \\
	3.28 & 1.64 & 13.1795 & 0.184 & 8.5960 & 8.6800 & 0.97 \\			         
	\hline        
    \end{tabular}
\end{table}

\begin{table}
	\centering
	\caption{Same as Table~\ref{tab:Jmerger_comp_APR} for the MPA1 EOS.}
	\label{tab:Jmerger_comp_MPA1}
	\begin{tabular}{ccccccc}
	\hline
	$M_\text{tot}$ & $M_\text{NS}$ & $R_\text{NS}$ & $C_\text{NS}$ & $\frac{cJ_\text{merger}}{G M_\odot^2}$ & $\frac{cJ_\text{merger}}{G M_\odot^2}$ & rel. diff. \\
	$[M_\odot]$ & $[M_\odot]$ & [km] & & [BS17] & [Lucca+] & [\%] \\
	\hline
	2.2 & 1.10 & 12.3276 & 0.132 & 4.3011 & 4.4180 & 2.65 \\
	2.3 & 1.15 & 12.3498 & 0.137 & 4.7060 & 4.7867 & 1.68 \\
	2.4 & 1.20 & 12.3711 & 0.143 & 5.1109 & 5.1565 & 0.88 \\
	2.5 & 1.25 & 12.3913 & 0.149 & 5.5146 & 5.5264 & 0.21 \\
	2.6 & 1.30 & 12.4102 & 0.155 & 5.9177 & 5.8966 & 0.36 \\
	2.7 & 1.35 & 12.4277 & 0.160 & 6.3217 & 6.2685 & 0.85 \\
	2.8 & 1.40 & 12.4437 & 0.166 & 6.7273 & 6.6425 & 1.28 \\
	2.9 & 1.45 & 12.4579 & 0.172 & 7.1300 & 7.0144 & 1.65 \\
	3.0 & 1.50 & 12.4702 & 0.178 & 7.5342 & 7.3878 & 1.98 \\
	3.1 & 1.55 & 12.4805 & 0.183 & 7.9387 & 7.7616 & 2.28 \\
	3.2 & 1.60 & 12.4885 & 0.189 & 8.3431 & 8.1352 & 2.56 \\	
	\hline        
    \end{tabular}
\end{table}

\begin{table}
	\centering
	\caption{Same as Table~\ref{tab:Jmerger_comp_APR} for the SFHx EOS.}
	\label{tab:Jmerger_comp_SFHX}
	\begin{tabular}{ccccccc}
	\hline
	$M_\text{tot}$ & $M_\text{NS}$ & $R_\text{NS}$ & $C_\text{NS}$ & $\frac{cJ_\text{merger}}{G M_\odot^2}$ & $\frac{cJ_\text{merger}}{G M_\odot^2}$  & rel. diff. \\
	$[M_\odot]$ & $[M_\odot]$ & [km] & & [BS17] & [Lucca+] & [\%] \\
	\hline
	2.2 & 1.10 & 11.8206 & 0.137 & 4.3573 & 4.3801 & 0.52 \\
	2.3 & 1.15 & 11.8451 & 0.143 & 4.7619 & 4.7342 & 0.58 \\
	2.4 & 1.20 & 11.8649 & 0.149 & 5.1649 & 5.0877 & 1.52 \\
	2.5 & 1.25 & 11.8832 & 0.155 & 5.5700 & 5.4439 & 2.32 \\
	2.6 & 1.30 & 11.8965 & 0.161 & 5.9736 & 5.7989 & 3.01 \\
	2.7 & 1.35 & 11.9073 & 0.167 & 6.3775 & 6.1543 & 3.63 \\
	2.8 & 1.40 & 11.9140 & 0.173 & 6.7813 & 6.5092 & 4.18 \\
	2.9 & 1.45 & 11.9174 & 0.180 & 7.1856 & 6.8640 & 4.69 \\
	3.0 & 1.50 & 11.9151 & 0.186 & 7.5901 & 7.2173 & 5.17 \\	
	\hline        
    \end{tabular}
\end{table}

\begin{table}
	\centering
	\caption{Same as Table~\ref{tab:Jmerger_comp_APR} for the SLy4 EOS.}
	\label{tab:Jmerger_comp_SLy}
	\begin{tabular}{ccccccc}
	\hline
	$M_\text{tot}$ & $M_\text{NS}$ & $R_\text{NS}$ & $C_\text{NS}$ & $\frac{cJ_\text{merger}}{G M_\odot^2}$ & $\frac{cJ_\text{merger}}{G M_\odot^2}$ & rel. diff. \\
	$[M_\odot]$ & $[M_\odot]$ & [km] & & [BS17] & [Lucca+] & [\%] \\
	\hline
	2.2 & 1.10 & 11.8588 & 0.137 & 4.3833 & 4.3831 & 0.01 \\
	2.3 & 1.15 & 11.8421 & 0.143 & 4.7886 & 4.7341 & 1.15 \\
	2.4 & 1.20 & 11.8238 & 0.150 & 5.1916 & 5.0820 & 2.16 \\
	2.5 & 1.25 & 11.8033 & 0.156 & 5.5957 & 5.4292 & 3.07 \\
	2.6 & 1.30 & 11.7800 & 0.163 & 6.0007 & 5.7756 & 3.90 \\
	2.7 & 1.35 & 11.7537 & 0.170 & 6.4039 & 6.1181 & 4.67 \\
	2.8 & 1.40 & 11.7239 & 0.176 & 6.8078 & 6.4588 & 5.40 \\	
	\hline        
    \end{tabular}
\end{table}

\begin{table}
	\centering
	\caption{Same as Table~\ref{tab:Jmerger_comp_APR} for the WFF1 EOS.}
	\label{tab:Jmerger_comp_WFF1}
	\begin{tabular}{ccccccc}
	\hline
	$M_\text{tot}$ & $M_\text{NS}$ & $R_\text{NS}$ & $C_\text{NS}$ & $\frac{cJ_\text{merger}}{G M_\odot^2}$ & $\frac{cJ_\text{merger}}{G M_\odot^2}$  & rel. diff. \\
	$[M_\odot]$ & $[M_\odot]$ & [km] & & [BS17] & [Lucca+] & [\%] \\
	\hline
	2.2 & 1.10 & 10.3754 & 0.157 & 4.5093 & 4.2029 & 7.29 \\
	2.3 & 1.15 & 10.3840 & 0.164 & 4.9126 & 4.5119 & 8.88 \\
	2.4 & 1.20 & 10.3911 & 0.171 & 5.3178 & 4.8225 & 10.27 \\
	2.5 & 1.25 & 10.3970 & 0.178 & 5.7222 & 5.1326 & 11.49 \\
	2.6 & 1.30 & 10.4010 & 0.185 & 6.1260 & 5.4420 & 12.57 \\
	2.7 & 1.35 & 10.4034 & 0.192 & 6.5297 & 5.7512 & 13.54 \\
	\hline        
    \end{tabular}
\end{table}

\bsp    
\label{lastpage}
\end{document}